\documentclass[11pt]{article}
\usepackage[dvips]{graphicx}
\usepackage{amsmath,amssymb,epsfig,float}
\usepackage{subfigure}
\usepackage{subfloat}

\newcommand{\nn}{\nonumber}
\newcommand{\be}{\begin{equation}}
\newcommand{\ee}{\end{equation}}
\newcommand{\ba}{\begin{array}}
\newcommand{\bqa}{\begin{eqnarray}}
\newcommand{\eqa}{\end{eqnarray}}
\newcommand{\cO}{{\cal O}}
\newcommand{\mL}{\mathcal{L}}

\newcommand{\ket}{\,\rangle}
\newcommand{\bra}{\langle \,}

\begin{document}
\title{\bf Resonances from meson-meson scattering in $U(3)$ $\chi$PT }

\author{ {\large Zhi-Hui Guo$^{1,2}$\thanks{guo@um.es} and J.~A. Oller$^2$\thanks{oller@um.es}} \vspace{.3cm}\\ 
$^1${\it {\it Department of Physics. Hebei Normal University. 050016 Shijiazhuang. China}}\\
$^2${\it {\it Departamento de F\'{\i}sica. Universidad de Murcia. E-30071 Murcia. Spain}}
}
\vspace{1.3cm}

\date{}
\maketitle

\begin{abstract}
In this work, the complete one loop calculation of meson-meson scattering amplitudes 
within $U(3) \otimes U(3)$ chiral perturbation theory with explicit resonance states is carried out for the first time. 
Partial waves are unitarized from the perturbative calculation employing a non-perturbative approach 
 based on the  N/D method. 
Once experimental data are reproduced in a satisfactory way we then study the resonance properties, such as the pole 
positions, corresponding residues and  their $N_C$ behaviors. The resulting $N_C$ dependence is the first one in 
the literature that takes into account the fact that the $\eta_1$ becomes the ninth Goldstone boson in the chiral 
limit for  large $N_C$. Within this scheme the vector resonances studied, $\rho(770)$, 
$K^*(892)$ and $\phi(1020)$, follow an $N_C$ trajectory in agreement with their standard $\bar{q}q$ interpretation. The scalars 
$f_0(1370)$, $a_0(1450)$ and $K^*(1430)$ also have for large $N_C$ a $\bar{q}q$ pole position trajectory and 
all of them tend to a bare octet of scalar resonances around 1.4~GeV. The $f_0(980)$ tends asymptotically to the bare pole position 
of a singlet scalar resonance around 1~GeV. The $\sigma$, $\kappa$ and $a_0(980)$ scalar resonances have a very different $N_C$ 
behavior. The case of the $\sigma$ resonance is analyzed with special detail.  
\end{abstract}


\section{Introduction}

 
A completely reliable theory to describe the physics in the intermediate energy region (resonance region) of QCD 
is still missing nowadays, since neither the low energy effective theory of QCD, namely chiral 
perturbation theory ($\chi$PT) 
nor perturbative QCD  is valid in this region. 
The appearance of  very broad resonances, such as $\sigma$ and $\kappa$~\cite{pdg}, 
makes the discussion even more difficult and thus poses a great challenge for theorists. 
Several attempts have been done to address the question by implementing non-perturbative 
methods inspired from the S-matrix theory \cite{chewbook,martin,eden}, typically on the $\chi$PT results. 
 One has approaches like the  Inverse Amplitude Method (IAM)
~\cite{truong,dobado,dobado97,pavon} to establish the unitarized amplitudes, 
 others are based on the N/D equations \cite{oller97,oller99prd,chew60,oller08prl,igi},  
Roy equations~\cite{leutwyler06,descotes06,madrid}, or on building  specific dispersion relations 
for the scattering amplitudes~\cite{zheng01,zheng04}, etc. 
 Although the different approaches determine values for the pole positions of the broad resonances $\sigma$ and $\kappa$ 
in good agreement between each other, their nature is still a controversial subject. 
One way to get insight into them is to track the $N_C$ trajectories of resonance poles in the 
complex plane. All the previous works along this research line 
\cite{pelaez04,pelaez06prl,sun07,guo,guo2,arriola1} are performed within $SU(3)$ 
or  $SU(2)$  $\chi$PT \cite{gasserleu}. In the former case the degrees of freedom correspond to the lightest octet of 
pseudoscalar mesons, pions $\pi$, kaons $K$ and the  isoscalar  $\eta_8$, while the singlet pseudoscalar $\eta_1$ 
is considered as a heavy field buried in the chiral counterterms. In the case of $SU(2)$ $\chi$PT only pions 
are taken as degrees of freedom. In Ref.~\cite{broniowski} the scalar-isoscalar states, including the 
$\sigma$, are investigated within the large-$N_C$ Regge approach.

$U(3)\otimes U(3)$ chiral symmetry in QCD is broken because of quantum effects that violate the conservation of the 
singlet axial vector current by the  $U_A(1)$ anomaly \cite{bardeen,fuji,adler}. 
As a result the singlet pseudoscalar $\eta_1$ is not a pseudo-Goldstone boson \cite{wein23}. Nevertheless, 
from the large $N_C$ QCD point of view, the quark loop responsible for the $U_A(1)$ anomaly \cite{bardeen} is $1/N_C$ suppressed, 
thus indicating that the $\eta_1$ becomes the ninth pseudo-Goldstone boson in the large $N_C$ limit~\cite{ua1innc,wit79}. 
Hence it is necessary to include the $\eta_1$ meson as a dynamical degree of freedom if one attempts to discuss 
the $N_C$ trajectories of various resonances. 
 This important fact was lacking in previous studies and is one of the main motivations of our current work.

Thanks to the large $N_C$ QCD, the singlet $\eta_1$ field can be conveniently incorporated into the 
effective field theory by enlarging the number of degrees of freedom of the theory from the octet of pseudo-Goldstone 
bosons to the nonet, which is usually called $U(3)$ $\chi$PT \cite{oriua}.  
 One advantage of foremost importance in our work is to use  $U(3)$ $\chi PT$ for considering the running with $N_C$ 
of resonance poles because then one has a framework that conceptually admits to consider the 
large $N_C$ limit in $\chi PT$. E.g. it has then the proper number of degrees of freedom.  

 Compared with  $SU(3)$ $\chi$PT,
 employed in the previous studies \cite{pelaez04,pelaez06prl,sun07,guo,guo2,arriola1}, 
 a novel ingredient in $U(3)$ $\chi$PT
is the $\eta_1$ mass term from the $U_A(1)$ anomaly, which has nothing to do with the current quark masses. 
The appearance of this new scale in $\chi$PT could totally breakdown the well celebrated chiral power counting.  
The introduction of the $1/N_C$ expansion in $U(3)$ $\chi$PT fixes the problem, since the singlet 
mass squared $M_0^2$ behaves as $\cO(1/N_C)$ in the large $N_C$ limit. Thus, in $U(3)$ $\chi$PT there are three expansion
parameters: momentum, quark masses and $1/N_C$, giving rise to a joint triple expansion $\delta \sim p^2 \sim m_q \sim 1/N_C$. 
However, one should bear in mind that the value of the singlet mass $M_0$, mainly determined from 
 the masses of the physical states  $\eta$ and  $\eta'$, is not a small quantity (for a 
review see Ref.~\cite{feldmann}). 
In connection with this problem Ref.\cite{borasoy02npa} used    
the infrared regularization method  (IR), which
is proposed to cure chiral violating terms from loop corrections in 
 baryon chiral perturbation theory in the low energy region~\cite{becher99epjc,alarcon},
 to investigate the $\eta'$ physics within $U(3)$ $\chi$PT. 
In IR the basic one loop scalar integral is divided into
the infrared singular and regular parts. The regular part is not considered  in IR 
because its  effects are absorbed into the low energy constants
of the theory. One only needs to consider the infrared singular part. 
However, contrary to baryon chiral perturbation theory, 
nothing prevents the appearance of large $\eta'$ masses in the vertices from derivatives acting 
on the external $\eta'$ fields, while this is 
not an issue  in baryon chiral perturbation theory because of baryon number conservation. 
 By the same token, in meson $\chi$PT  the total energy squared in the center of mass frame (CM) is not 
restricted to be around one, two, etc $\eta'$ masses, which could indeed spoil the validity of a loop 
calculation within IR.  Although there are some special processes, 
such as $\eta' \to \eta \pi \pi$~\cite{borasoy02npa}, which happens to 
be covered by the constrained energy region of IR, 
in general one can not naively apply IR in the pure meson sector. 

In the present discussion, we employ  standard dimensional regularization, 
as in conventional $SU(3)$ $\chi$PT \cite{gasserleu},
 already applied in $U(3)$ $\chi$PT to calculate the pseudoscalar weak decay constants in Ref.~\cite{kaiser98}.
Then, the triple chiral expansion scheme is necessary for preserving the power counting. 
Investigations within $U(3)$ $\chi$PT along this line  mainly focused on 
the construction of higher order Lagrangians~\cite{herrera97,kaiser00} and 
the  tree level and one-loop calculations for $\eta-\eta'$ mixing~\cite{u3chpteta,kaiser98}. 
An important contribution of our work 
is to offer the first complete one-loop calculation in $U(3)$ $\chi$PT for meson-meson scattering,  
including both the loop diagrams contributing to the pseudo-Goldstone masses 
and decay constants, as well as the genuine ones of  meson-meson scattering. 
A preliminary version of this calculation is given in Ref.~\cite{mongolia}. 

 Previous analyses from other groups only considered  
tree level amplitudes or performed partial one-loop calculations \cite{borasoy03prd,zhou11prd,jamin}. 
 The  $\eta'\to \eta\pi\pi$ decay was studied in detail. Ref.~\cite{borasoy02npa} performed a one-loop calculation 
within IR, and Ref.~\cite{kubis09} undertook a two-loop calculation  within 
the framework of non-relativistic field theory. However, in this decay  
the three-momenta of the pions are not  small compared with their mass for some region of the kinematics.   
Recently, Ref.~\cite{escribano10}  discussed the same process  within the triple expansion scheme 
by considering  tree level amplitudes and part of the $s$-channel loops, resumming 
$\pi\pi$ final state interactions.

Since one of the main interests of our work is to study resonance properties, we will explicitly include bare 
resonance fields in the Lagrangian within the framework of 
resonance chiral theory~\cite{ecker89npb,ecker89plb}. 
However, including resonances as  dynamical degrees of freedom is not enough to guarantee that one  can 
safely apply the perturbative results up to the resonance region, since the loops of 
pseudo-Goldstone bosons, specially the unitarity  or the $s$-channel loops, start to 
play an important role around the energy region where resonances emerge. Thus, a proper way 
to resum  unitarity loops is crucial to study resonance properties. We use in this work  the 
method provided in Ref.~\cite{oller99prd} to accomplish this resummation. Notice also that 
not all the resonances that stem in our study correspond to bare fields because new resonances 
come out when strong enough interactions are resummed.

The present paper is organized as follows. Section \ref{introlag} is devoted 
to the introduction of the relevant chiral Lagrangian. The structure of the 
perturbative scattering amplitudes is elaborated in Section \ref{pertamp}, 
which is followed by the partial wave projection and its unitarization in 
Section \ref{pwamp}. The phenomenological discussion, including  the fit quality and  
 features of resonances, such as  masses, widths, residues and $N_C$ behavior, is given in detail in Section \ref{pheno}. We conclude in Section \ref{conclusion}.

\section{Relevant Chiral Lagrangian}\label{introlag}

The chiral Lagrangian at leading order in $U(3)$ $\chi$PT reads \cite{oriua} 
\begin{eqnarray} \label{lolagrangian}
\mL_{\chi}=\frac{ F^2}{4}\langle u_\mu u^\mu \rangle+
\frac{F^2}{4}\langle \chi_+ \rangle
+ \frac{F^2}{3}M_0^2 \ln^2{\det u}\,,
\end{eqnarray}
where $\langle \ldots \rangle$ denotes the trace in flavor space and 
the last term corresponds to the $U_A(1)$ anomaly $\eta_1$ mass term. 
The definitions for the chiral building blocks are   
\begin{eqnarray}
  u_\mu &=& i u^+ D_\mu U u^+  \,, \nn\\
 \chi_+ &=& u^+ \chi u^+ + u \chi^+ u \,,\nn\\
U &=&  u^2 = e^{i\frac{ \sqrt2\Phi}{ F}} \,, \nn\\
D_\mu U &=& \partial_\mu U - i r_\mu U + i U l_\mu \,, \nn\\
\chi &=& 2 B (s + i p) \,,
\end{eqnarray}
where $r_\mu\,, l_\mu \,, s\,, p$ stand for external sources and 
 the pseudo-Goldstone bosons are collected in the matrix  
\begin{equation}\label{phi1}
\Phi \,=\, \left( \begin{array}{ccc}
\frac{1}{\sqrt{2}} \pi^0+\frac{1}{\sqrt{6}}\eta_8+\frac{1}{\sqrt{3}} \eta_1 & \pi^+ & K^+ \\ \pi^- &
\frac{-1}{\sqrt{2}} \pi^0+\frac{1}{\sqrt{6}}\eta_8+\frac{1}{\sqrt{3}} \eta_1   & K^0 \\  K^- & \bar{K}^0 &
\frac{-2}{\sqrt{6}}\eta_8+\frac{1}{\sqrt{3}} \eta_1 
\end{array} \right)\,.
\end{equation}
In addition, $F$ is the axial decay constant of the pseudo-Goldstone bosons in the simultaneous chiral and large $N_C$ limit. 
In the same limit the parameter $B$ is related to the quark condensate through 
$\langle 0|\bar{q}^iq^j|0\rangle =- F^2 B\delta^{ij}$.  
 The explicit chiral symmetry breaking is implemented by taking the vacuum expectation 
values of the scalar external source field $s = {\rm diag}(m_u,m_d,m_s)$, 
 with $m_q$ the light quark masses. 
Throughout we always work in the isospin limit, i.e. taking $m_u = m_d$.

We follow the framework in Ref.\cite{ecker89npb} to include bare resonance fields, 
where the interaction terms of pseudo-Goldstone bosons and resonances are invariant under    
 chiral symmetry \cite{callen} and the discrete symmetries of charge conjugation $C$ and parity $P$. 
Due to the presence of heavy resonance states, the momentum 
expansion is not valid any more in this theory. 
Nevertheless the $1/N_C$ expansion provides another guided principle to construct the Lagrangian. 
 The generic $N_C$ leading structure of the interacting operator in resonance chiral theory has only one 
flavor trace and in a schematic way can be written as 
\begin{equation} 
 {\cal O}_i \, \sim \, \langle \, R_1 R_2 \ldots R_j \, \chi^{(n)} \, \rangle \, , 
\end{equation}
where $\chi^{(n)}$ denotes the chiral building block with  chiral order $\cO(p^n)$,
that only incorporates the pseudo-Goldstone bosons and the external source fields, and  
$R_i$ stands for the resonance fields. 
Although in the resonance chiral theory the interacting terms with  
higher chiral orders ($ n \geq 4$) are not suppressed in general, 
many of them are absent if one invokes the short distance constraints 
from QCD in large $N_C$ \cite{ecker89plb} and  does not allow fine tuning between different 
operators~\cite{cirigliano06npb}. 
To be practical in the phenomenological discussion 
we restrict ourselves by including the full set of operators with $n\leq 2$. 
Concerning the number of resonance states in the interacting vertices, 
the most relevant ones in meson-meson scattering consist of one single 
resonance field. 

 In regards to the operators attached  only to the interactions between the 
pseudo-Goldstone bosons beyond leading order, it is commonly believed 
that they encode the high energy dynamics of the underlying theory, which 
could be represented by the heavier states, such as resonances. 
It has been shown that at $\cO(p^4)$  in $SU(3)$ $\chi$PT   
the low energy constants (LECs) are saturated in good approximation by the lowest multiplet of resonances 
and thus no additional pieces of LECs seem to be typically needed in the theory~\cite{ecker89npb,ecker89plb}. 

In the present work,  we exploit this assumption on the saturation of the LECs by the lightest resonances, so that, 
 instead of local chiral terms contributing to meson-meson scattering we take the tree level exchanges of  the scalar and vector resonances.  In this way we keep all local contributions to meson-meson scattering up to and including ${\cal O}(\delta^3)$, 
while also generating higher order ones. We also calculate in addition  the one-loop contributions  that count one order higher in $\delta$. 

 The relevant resonance operators for the interactions with the pseudo-Goldstone bosons read~\cite{ecker89npb} 
\begin{eqnarray}\label{lagscalar}
\mL_{S}&=& \,\,\,\,\,c_d\bra S_8 u_\mu u^\mu \ket + c_m \bra S_8 \chi_+ \ket
 + \widetilde{c}_d S_1 \bra u_\mu u^\mu \ket
+ \widetilde{c}_m  S_1 \bra  \chi_+ \ket
\nonumber \\ & & 
+ \hat{c}_d \bra S_9 u_\mu \ket\bra u_\mu \ket 
+ \hat{c}_m S_1 \ln^2{\det u}\,,
\end{eqnarray}
\begin{eqnarray}\label{lagvector}
\mL_{V}= \frac{i G_V}{2\sqrt2}\langle V_{\mu\nu}[ u^\mu, u^\nu]\rangle \,,
\end{eqnarray}
where the resonance states are collected in the building block $R =S, V$,  
\begin{eqnarray}
S_1 & =& \sigma_1\,,\nonumber\\
\label{s1s8}
S_{8} &=& \left(
                                        \begin{array}{ccc}
                                          \frac{a_0^0}{\sqrt2}+\frac{\sigma_8}{\sqrt6} & a_0^+ & \kappa^{+}  \\
                                          a_0^- & -\frac{a_0^0}{\sqrt2}+\frac{\sigma_8}{\sqrt6}  & \kappa^{0}  \\
                                          \kappa^{-} & \bar{\kappa}^{0} & -\frac{2\sigma_8}{\sqrt6} \\
                                        \end{array}
                                       \right) \,,\nn \\
S_9&=&S_8+\frac{1}{\sqrt{3}}S_1~,\nn\\
V_{\mu\nu}&=& \left(
                                        \begin{array}{ccc}
                              \frac{\rho_0}{\sqrt2}+\frac{1}{\sqrt{6}}\omega_8 + \frac{1}{\sqrt{3}} \omega_1 & \rho^+ & K^{*+}  \\
                         \rho^- & -\frac{\rho_0}{\sqrt2}+\frac{1}{\sqrt{6}}\omega_8+ \frac{1}{\sqrt{3}}\omega_1   & K^{*0}  \\
                K^{*-} & \bar{K}^{*0} & -\frac{2}{\sqrt{6}}\omega_8+ \frac{1}{\sqrt3}\omega_1 \\
                                        \end{array}
                                       \right)_{\mu\nu} \,.
\end{eqnarray}
The corresponding kinetic terms for resonance states read~\cite{ecker89npb}
\begin{eqnarray}\label{kinerv}
\mathcal{L}_{\rm kin}^V&=&-{1\over 2} \bra \nabla^\lambda
V_{\lambda\mu}\nabla_\nu V^{\nu\mu}
-{1\over 2}M^2_V V_{\mu\nu}V^{\mu\nu} \ket \,, 
\end{eqnarray}
\begin{eqnarray}\label{kiners}
\mathcal{L}_{\rm kin}^S&=&{1\over 2} \bra \nabla^\mu S_8 \nabla_\mu S_8
-M^2_{S_8} S_8^{2} \ket +{1\over 2}\big(\partial^\mu S_{1}
\partial_\mu S_{1}-M^2_{S_1} S_1^{2}\big)\,,
\end{eqnarray}
where 
\begin{eqnarray}
\nabla_\mu R &=& \partial_\mu R + [\Gamma_\mu, R]\,, \quad R = V, S, \nonumber\\
\Gamma_\mu  &=& \frac{1}{2}\bigg[ u^\dagger (\partial_\mu- i\,r_\mu) u + u (\partial_\mu- i\,l_\mu) u^\dagger \bigg]\,.
\end{eqnarray}
Compared with Ref.~\cite{ecker89npb} two additional $1/N_C$ suppressed
operators appear in ${\cal L}_S$ due to the inclusion of the singlet $\eta_1$. These are  the monomials proportional to $\hat{c}_d$ and  $\hat{c}_m$ 
in Eq.~\eqref{lagscalar}. From the exchange of the scalar resonances these terms give rise to tree-level meson-meson 
contributions that are higher order, at least ${\cal O}(\delta^4)$. 
  In addition, these new operators  mainly contribute to processes 
involving the $\eta'$ meson\footnote{The term with $\hat{c}_m$ is purely proportional to $\eta_1^2$ while that with $\hat{c}_d$ 
requires at least one $\eta_1$, which mainly becomes an $\eta'$ because the $\eta_1$ contribution to the $\eta$ is suppressed by 
the pseudoscalar mixing angle, as it is shown later.} and since we deal  with  experimental 
data  related to $\pi$, K and $\eta$ in the present discussion, as shown explicitly below, 
 the states with  $\eta'$ only enter through an indirect way. 
So their effects are rather tiny in the current discussion and we discard these two new 
terms throughout. We have checked that if included our results barely change.
For the remaining parameters of scalar resonances, 
instead of imposing the large $N_C$ relations to the couplings and masses of the octet and singlet, 
such as $\widetilde{c}_{d\,,m} = c_{d\,,m}/ \sqrt3$ and $M_{\sigma_1} = M_{\sigma_8}$~\cite{ecker89npb}, 
we free them in our discussion. 
In this way, we consider  effects beyond leading 
order of $1/N_C$ in the scalar resonance Lagrangian implicitly.

The antisymmetric tensor formalism is used to describe the vector resonance since, as  
it is demonstrated in Ref.\cite{ecker89plb}, only in this way one does not need to 
include extra terms in the pseudo-Goldstone Lagrangian to fulfill the QCD short distance constraints.  
For the vector resonances $\omega$ and $\phi$, we assume ideal mixing throughout 
\begin{eqnarray}
\omega_1 = \sqrt{\frac{2}{3}} \omega - \sqrt{\frac{1}{3}} \phi \,, \qquad
\omega_8 = \sqrt{\frac{2}{3}} \phi + \sqrt{\frac{1}{3}} \omega \,,
\end{eqnarray}
and we do not include any $1/N_C$ 
suppressed operators in this respect. Nonetheless, we employ different bare masses for the $\rho(770)$ and $K^*(892)$ 
in order to obtain a good fit to data.

Before finishing this section, we introduce the last pieces of the chiral Lagrangian 
involving only pseudo-Goldstone bosons at ${\cal O}(\delta)$ \cite{kaiser00}
\begin{align}\label{deflam2}
\mathcal{L}_{\Lambda} & = \Lambda_1\frac{F^2}{12}D_\mu\psi D^\mu\psi -i\Lambda_2\frac{F^2}{12} \bra U^+ \chi - \chi^+ U \ket \psi ~,\nn\\
\psi&=-i\ln \det U~,\nn\\
D_\mu\psi&=\partial_\mu \psi-2\langle a_\mu\rangle~,
\end{align}
with $a_\mu=(r_\mu-l_\mu)/2$. 
The inclusion of such operators does not improve our fits to data indeed. However, we take 
into account the monomial proportional to $\Lambda_2$  to bring our prediction 
for the masses of $\eta$ and $\eta'$ at their physical values \cite{pdg}. 
This is necessary in the present work, since in the fit all of the 
pseudo-Goldstone masses have their physical values, 
while we need to use our prediction for the masses when we discuss the $N_C$ dependence 
of our results, in  particular the movement of resonance poles with $N_C$. 
Thus, it is necessary to match our prediction for the masses of the pseudoscalars as a function of $N_C$ with their 
physical values for $N_C =3$. Precisely the $\Lambda_2$ term leads to an important contribution 
to the $\eta$ and $\eta'$ masses and in the triple expansion scheme its chiral order 
is  lower  than the one of chiral loops. 
Though this operator contributes to  meson-meson scattering as well, 
it mainly contributes to the processes involving $\eta'$ and the 
inclusion of this term barely affects the global fit. 
 Unlike the $\Lambda_2$ operator the monomial proportional to  $\Lambda_1$ in Eq.~\eqref{deflam2} only contributes to the masses 
and the scattering amplitudes in an indirect way,
i.e. through the normalization of the $\eta'$ field, and its influence in the global fit 
is tiny. Indeed, if we include this counterterm in our fits to data the resulting fitted value tends to vanish. 
As a result, we do not consider any further the $\Lambda_1$ term neither in scattering nor for the $ \eta-\eta'$ mixing and masses. 

Finally, we want to point  out that  there is no double counting problem 
by having both the resonance Lagrangians in Eqs.\eqref{lagscalar}-\eqref{kiners}
 and the local pseudo-Goldstone operator $\Lambda_2$, 
since the $\Lambda_2$ term can be only generated by integrating out the excited pseudo scalar resonances, 
instead of the scalar and vector ones considered here.

\section{Structure of the scattering amplitudes}\label{pertamp}

Even in the leading order Lagrangian  Eq.(\ref{lolagrangian}), 
the flavor eigenstates $\eta_8$ and $\eta_1$ are not mass eigenstates 
and we use the angle $\theta$ to describe the mixing of $\eta_8$ and $\eta_1$ at this order
\begin{eqnarray}\label{deflomixing}
\eta_8 &=&  c_\theta \overline{\eta}+ s_\theta\overline{\eta}'\,,\nonumber \\
\eta_1 &=& -s_\theta\overline{\eta} + c_\theta\overline{\eta}'\,,
\end{eqnarray}
with $c_\theta=\cos\theta$ and $ s_\theta=\sin\theta$.
 In our notation  $\overline{\eta}$ and  $\overline{\eta}'$ are the fields that diagonalize the quadratic 
terms of the Lagrangian Eq.~\eqref{lolagrangian}. The $\eta-\eta'$ mixing at leading order is discussed in Appendix~\ref{app.B}, Eqs.~\eqref{defmetab2}-\eqref{deftheta0}.
The differences between  $\overline{\eta}\,,\overline{\eta}'$ and the physical states $\eta\,,\eta'$ are
caused by higher order operators, including  loops, and can be treated perturbatively within the triple expansion scheme.

\begin{figure}[h]
\begin{center}
\includegraphics[angle=0, width=0.5\textwidth]{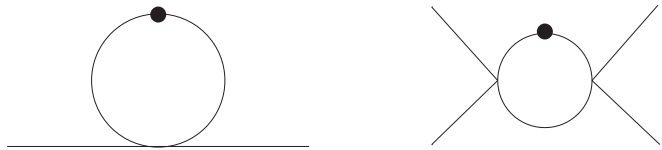}
\caption{ The dot denotes the mixing of $\eta_8$ and $\eta_1$ at leading order, which is proportional to $m_K^2-m_\pi^2$. }
\label{figmassinsert}
\end{center}
\end{figure}

Next we calculate the contributions beyond the leading order to the scattering amplitudes in terms of the 
$\overline{\eta}$ and $\overline{\eta}'$ fields, while for the leading order terms to meson-meson scattering, stemming 
from ${\cal L}_\chi$ Eq.~\eqref{lolagrangian}, one has to take care of the 
full $ \eta-\eta'$ mixing (see below).
We also  point out that for the calculation of  the basic amplitudes it is more 
reasonable to use the  $\overline{\eta}$ and $\overline{\eta}'$ fields than the  $\eta_1$ and $\eta_8$ ones. This is because the 
insertion of the leading order mixing of $\eta_8$ and $\eta_1$ does not increase the order of a diagram, 
as illustrated in Fig.~\ref{figmassinsert} for a one-loop contribution. 
 The leading order mixing is 
proportional to $m_K^2-m_\pi^2$, and is always accompanied by the inclusion of one extra  $\eta_8$ or $\eta_1$ propagator 
that compensates the chiral power of the vertex. As a result 
 loop diagrams with an arbitrary number of insertions of  $\eta_8 - \eta_1$ mixing vertices have the same order. 
On the contrary, the mixing of $\overline{\eta} \,, \overline{\eta}'$ only receives contribution 
from higher orders, which guarantees that  diagrams with insertions from  the 
$\overline{\eta}-\overline{\eta}'$ mixing are indeed suppressed.

\begin{figure}[h]
\begin{center}
\includegraphics[angle=0, width=0.6\textwidth]{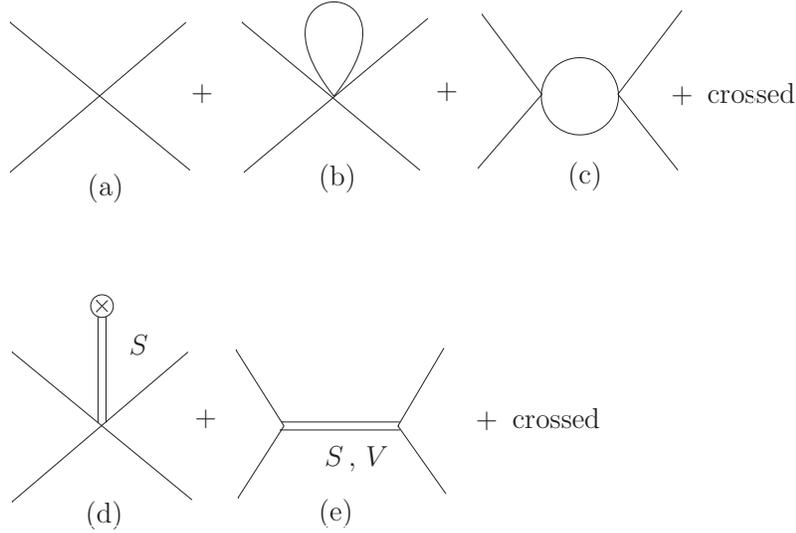}
\caption{ Relevant  Feynman diagrams in the scattering amplitudes up to one-loop order. In digram (d) the coupling of the scalar resonances with the vacuum is indicated by a cross. }
\label{Fig.sc.diagram}
\end{center}
\end{figure}

\begin{figure}[h]
\begin{center}
\includegraphics[angle=0, width=0.5\textwidth]{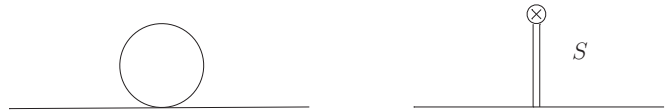}
\caption{ Relevant Feynman diagrams for the pseudoscalar self-energy. }
\label{Fig.selfenergy}
\end{center}
\end{figure}

\begin{figure}[h]
\begin{center}
\includegraphics[angle=0, width=0.5\textwidth]{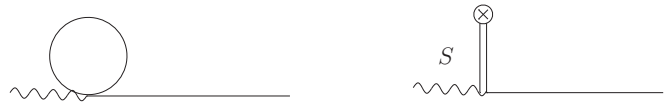}
\caption{ Relevant Feynman diagrams for the pseudoscalar decay constants. 
 The wiggly line corresponds to the axial-vector external source.}
\label{Fig.Fp}
\end{center}
\end{figure}

The calculation of the scattering amplitudes comprises the contact vertex from ${\cal L}_\chi$ Eq.\eqref{lolagrangian}, 
the one loop and resonance exchange graphs, 
 as illustrated in Fig.~\ref{Fig.sc.diagram}. In addition,  the mass and wave function renormalization terms, displayed 
in Fig.~\ref{Fig.selfenergy}, should also be included. The latter ones  only affect the tree level scattering 
amplitudes from Eq.~(\ref{lolagrangian}) up to the order we attempt to calculate, 
as the contributions from the resonances and the loop diagrams are already beyond the leading order. 
The decay constants of the pseudo-Goldstone bosons, illustrated in Fig.~\ref{Fig.Fp}, 
also give rise to one higher order contributions by rewriting  $F$ in terms of the physical 
ones in the leading order scattering amplitudes. We express the scattering amplitudes employing one single physical 
decay constant $F_\pi$ throughout. The relation between $F$ and $F_\pi$ is given in Eq.~\eqref{ffpi}.

Another subtle contribution to the scattering amplitudes is related to the $\eta-\eta'$ mixing, as commented above. 
We recall that $\overline{\eta}\,, \overline{\eta}'$ result from the diagonalization of 
$\eta_8\,, \eta_1$ at leading order, Eq.~\eqref{deflomixing}. After including the higher order contributions from 
resonances and chiral loops, $\overline{\eta}\,, \overline{\eta}'$ will mix again and 
the physical states $\eta\,, \eta'$ can be obtained by diagonalizing $\overline{\eta}\,, \overline{\eta}'$. 
This extra diagonalization process  contributes relevant pieces to the scattering amplitudes 
through the leading order results from Eq.~(\ref{lolagrangian}). 
Thus it is necessary to work out the $\overline{\eta}-\overline{\eta}'$ mixing at the one loop level
by calculating the diagrams in Fig.~\ref{Fig.selfenergy}. However, for the already next-to-leading order 
contributions and higher, namely, those amplitudes obtained from the exchange of resonances  and involving loops, we do not need to 
distinguish between $\overline{\eta}\,, \overline{\eta}'$ and $\eta\,, \eta'$, 
as their differences only cause higher order effects that are beyond our current consideration.

We parameterize the higher order $\overline{\eta}-\overline{\eta}'$ mixing as
\begin{eqnarray}\label{defmixingpara}
\mL&=& \frac{1 + \delta_{\overline{\eta}} }{2}\partial_\mu \overline{\eta} \partial^\mu\overline{\eta}
+\frac{1+ \delta_{\overline{\eta}'} }{2}\partial_\mu \overline{\eta}' \partial^\mu \overline{\eta}'
+\delta_k\, \partial_\mu \overline{\eta} \partial^\mu \overline{\eta}'
\nonumber \\ & &
-\frac{m_{\overline{\eta}}^2 + \delta_{m_{\overline{\eta}}^2} }{2} \overline{\eta}\, \overline{\eta}
- \frac{m_{\overline{\eta}'}^2 + \delta_{m_{\overline{\eta}'}^2 } }{2} \overline{\eta}' \overline{\eta}'
- \delta_{m^2} \,\overline{\eta}\, \overline{\eta}' \,.
\end{eqnarray}
where $m_{\overline{\eta}}$ and $m_{\overline{\eta}'}$ defined in Eqs.(\ref{defmetab2}) and (\ref{defmetaPb2}) 
stand for the leading order masses of $\overline{\eta}$
and $\overline{\eta}'$ respectively, while the different $\delta_i$, given in Eq.~\eqref{mix.del.1}, 
contain the higher order contributions. 
The physical eigenstates $\eta$ and $\eta'$ are related with the $\overline{\eta}$ and $\overline{\eta}'$ fields by diagonalizing 
canonically the quadratic terms in Eq.~\eqref{defmixingpara} 
in the following way
\begin{eqnarray} \label{etatoetabar}
 \left(
 \begin{array}{c}
 \eta   \\
 \eta'  \\
 \end{array}
 \right) = \left(
                                        \begin{array}{cc}
                                          \cos\theta_\delta  & -\sin\theta_\delta  \\
                                           \sin\theta_\delta & \cos\theta_\delta  \\
                                        \end{array}
    \right) \left(
    \begin{array}{cc}
 1+\frac{\delta_{\overline{\eta}}}{2}\,\, &
 \,\, \frac{\delta_{k}}{2}  \\
\frac{\delta_{k}}{2}\,\, &
\,\, 1+\frac{\delta_{\overline{\eta}'}}{2}  \\
                                        \end{array}
    \right)
      \left(
       \begin{array}{c}
       \overline{\eta}   \\
       \overline{\eta}' 
       \end{array}
        \right)\,.
\end{eqnarray}
We calculate the leading order contributions, corresponding to the diagram (a) of Fig.~\ref{Fig.sc.diagram},
 in terms of the physical $\eta$ and $\eta'$ fields, so that we can directly use the physical values for the masses. 
Up to the order considered in our calculation for meson-meson scattering, namely, up to ${\cal O}(\delta^3)$ for local tree 
level contributions and up to ${\cal O}(\delta^4)$ for the one-loop graphs,  we can invert Eq.~\eqref{etatoetabar} in a 
perturbative way. Then we have 
\begin{eqnarray} \label{etabartoeta}
 \left(
 \begin{array}{c}
  \overline{\eta}   \\
 \overline{\eta}'   \\
 \end{array}
 \right) = \left(
    \begin{array}{cc}
 1-\frac{\delta_{ \overline{\eta}}}{2} \,\, &\,\,
  -\frac{\delta_{k}}{2}   \\
-\frac{\delta_{k}}{2}  \,\,
&\,\, 1-\frac{\delta_{ \overline{\eta}'}}{2}   \\
                                        \end{array}
    \right)  \left(
                                        \begin{array}{cc}
                                          \cos\theta_\delta  & \sin\theta_\delta  \\
                                           -\sin\theta_\delta & \cos\theta_\delta  \\
                                        \end{array}
    \right)
      \left(
       \begin{array}{c}
       \eta   \\
       \eta' \\
       \end{array}
        \right)\,,
\end{eqnarray}
where $\theta_\delta$ is determined through
\begin{eqnarray}
\tan\theta_\delta = \frac{\hat{\delta}_{m^2}}{m_{\eta'}^2- \hat{m}_{\eta}^2}\,,
\end{eqnarray}
with
\begin{eqnarray}\label{mixnlo}
\hat{m}_{\eta}^2 &=& m_{\overline{\eta}}^2 + \delta_{m_{\overline{\eta}}^2} - m_{\overline{\eta}}^2\, \delta_{\overline{\eta}}\,,
\nonumber \\
\hat{m}_{\eta'}^2 &=& m_{\overline{\eta}'}^2 + \delta_{m_{\overline{\eta}'}^2} - m_{\overline{\eta}'}^2\, \delta_{\overline{\eta}'}\,,
\nonumber \\
\hat{\delta}_{m^2} &=& \delta_{m^2}- \frac{1}{2} \delta_k ( m_{\overline{\eta}}^2+m_{\overline{\eta}'}^2) \,,
\nonumber \\
2 m_{\eta'}^2 &=& \hat{m}_{\eta}^2+ \hat{m}_{\eta'}^2 + \sqrt{ (\hat{m}_{\eta}^2- \hat{m}_{\eta'}^2)^2 + 4 \hat{\delta}_{m^2}^2 }~,
\nonumber \\
2 m_{\eta}^2 &=& \hat{m}_{\eta}^2+ \hat{m}_{\eta'}^2 - \sqrt{ (\hat{m}_{\eta}^2 -\hat{m}_{\eta'}^2)^2 + 4 \hat{\delta}_{m^2}^2 }~.
\end{eqnarray}
In the previous equations $\theta_\delta$ and $\delta_i$  are originated by the higher order contributions
from the resonances and one loop graphs. Up to the precision we consider, 
it is safe to take only linear terms in the $\delta_i$ so that $\cos {\theta_\delta} = \sqrt{1 - \sin^2{\theta_\delta}} \simeq 1 $.
From the relations between $\eta \,,\eta'$ and $\overline{\eta}\,,\overline{\eta}'$, given in 
Eqs.(\ref{etatoetabar}) and (\ref{etabartoeta}), there are two ways to proceed for the calculation of  
the physical scattering amplitudes. One method, the one that we follow, consists of writing 
the physical amplitudes by expressing  $\overline{\eta}$ and
$\overline{\eta}'$ in terms of the physical fields $\eta$ and $\eta'$ in Eq.(\ref{lolagrangian}), and then calculate the tree level amplitudes.
In this way, all the masses from the kinematics are the physical ones, since they correspond 
to the  $\eta$ and $\eta'$ fields. 
Another method is to determine the amplitudes with $\eta$ and $\eta'$ 
by using Eq.(\ref{etatoetabar}) as a linear superposition of those calculated in terms of the bar fields. 
 However, 
one should bear in mind that for this case the masses of $\overline{\eta}$ and $\overline{\eta}'$ from 
the kinematics are those of the physical $\eta$ and $\eta'$ fields to which they 
are attached \cite{jamin}.

The explicit expressions for the mass and wave function renormalization,   
mixing parameters entering in Eq.(\ref{defmixingpara}),  pion decay constant, and  scattering amplitudes are given in the Appendices \ref{app.B}--\ref{app.D}.

\section{Partial wave amplitude and its unitarization}\label{pwamp}

Once  the perturbative amplitudes are calculated from $U(3)$ $\chi$PT, as done in the previous section, we can proceed to
perform the partial wave projections of the isospin amplitudes 
and construct the corresponding unitarized amplitudes as well. 
The  amplitudes $T^I$ with well-defined isospin $I$ from different processes are derived by assigning 
the following phase convention to the pseudoscalars in connection with the isospin basis states. This 
convention is consistent with the one taken 
in Eq.~\eqref{phi1}, 
\begin{eqnarray} 
&&| \eta \ket =  | 0\, 0 \ket\,, \qquad | \eta' \ket =  | 0\, 0 \ket\,, \nonumber \\ &&
| \pi^+ \ket = - | 1\,\, 1 \ket, \quad | \pi^- \ket = | 1\,\, -1 \ket, \quad| \pi^0 \ket =  | 1\,\, 0 \ket\,, \nonumber \\&&
| K^+ \ket = - | \frac{1}{2}\,\, \frac{1}{2} \ket \,,  
| K^0 \ket =  | \frac{1}{2}\,\, -\frac{1}{2} \ket \,,
| \bar{K}^0 \ket = | \frac{1}{2}\,\, \frac{1}{2} \ket \,,  
| K^- \ket =  | \frac{1}{2}\,\, -\frac{1}{2} \ket \,, 
\end{eqnarray}
where $|I\,I_3\rangle$ is a state with isospin $I$ and third component $I_3$. 

For $\pi\pi \to \pi\pi$ scattering there are three isospin amplitudes,
 $I = 0, 1, 2$. They read   
\begin{eqnarray}
T^0(s,t,u) &=& 3\,A(s,t,u) + A(t,s,u) + A(u,t,s) \,, \nonumber \\ 
T^1(s,t,u) &=&  A(t,s,u) - A(u,t,s) \,, \nonumber \\ 
T^2(s,t,u) &=& A(t,s,u) + A(u,t,s) \,,
\end{eqnarray}
where $A(s,t,u)$ stands for the process $\pi^+\pi- \to \pi^0 \pi^0$ and 
$s, t, u$ are the standard Mandelstam variables. These equations, and other similar ones that follow,
 are obtained by invoking crossing symmetry \cite{morgan}.

For $K\pi \to K \pi$ and $\pi\pi \to K \bar{K}$, the different isospin amplitudes can also be expressed in terms of one single 
amplitude, again by using crossing symmetry. One then has
\begin{eqnarray}
T^{\frac{3}{2}}(s,t,u) &=& T_{K^+\pi^+ \to K^+\pi^+  }(s,t,u)\,,\nonumber \\
T^{\frac{1}{2}}(s,t,u) &=& \frac{3}{2} T^{\frac{3}{2}}(u,t,s) - \frac{1}{2} T^{\frac{3}{2}}(s,t,u) \,,\nonumber \\
T^{0}(s,t,u) &=& \sqrt{\frac{3}{2}}\,\bigg[ T^{\frac{3}{2}}(t,s,u) + T^{\frac{3}{2}}(u,s,t) \bigg] \,,\nonumber \\
T^{1}(s,t,u) &=&  T^{\frac{3}{2}}(u,s,t) - T^{\frac{3}{2}}(t,s,u)\,.
\end{eqnarray}

For $\pi\pi \to \eta\eta  $ and $\pi\eta \to \pi \eta$, we have
\begin{eqnarray}
T^0(s,t,u) &=& -\sqrt{3}\,C(s,t,u) \,, \nonumber \\ 
T^1(s,t,u) &=&  C(t,s,u) \,,  
\end{eqnarray}
with $C(s,t,u) = T_{\pi^0\pi^0 \to \eta\eta}(s,t,u)$. 
For $\pi\pi \to \eta\eta', \eta'\eta'$ and $\pi\eta(\eta') \to \pi \eta'$, analogous formulas 
emerge  and the only difference is that  $C(s,t,u)$ is redefined accordingly to the scattering  
process. 

The reactions $K\pi\to K\eta$ and $K \eta'$ are pure $I=1/2$ and are given by
\begin{eqnarray}
T^{\frac{1}{2}}(s,t,u)=-\sqrt{3}T_{K^+\pi^0\to K^+\eta}(s,t,u)
\end{eqnarray}
and similarly when the $\eta'$ is in the final state. In terms of them one 
also has  the  scattering amplitudes for the $t$-crossed processes  $K\bar{K}\to \pi\eta$ and $\pi\eta'$, which is pure 
 $I=1$. It reads 
\begin{eqnarray}
T^{1}(s,t,u)=-\sqrt{2}T_{K^+\pi^0\to K^+\eta}(t,s,u)~,
\end{eqnarray}
and analogously for $K\bar{K}\to \pi\eta'$.

 There are two isospin amplitudes in $K\bar{K} \to K\bar{K}$, that can be expressed as
\begin{eqnarray}
T^0(s,t,u) &=& 2 D(s,t,u) + D(t,s,u) \,, \nonumber \\ 
T^1(s,t,u) &=& D(t,s,u) \,,  
\end{eqnarray}
with $D(s,t,u)= T_{K^+ K^- \to K^0 \bar{K}^0}(s,t,u)$. Notice that here we have derived 
both isospin amplitudes in terms of only one, while previous works used both 
$T_{K^+ K^- \to K^0 \bar{K}^0}(s,t,u)$ and $T_{K^+ K^- \to K^+ \bar{K}^-}(s,t,u)$ 
\cite{guerrero,prdlong,prlshort}.

For $K\bar{K} \to\eta\eta$ and $K \eta \to K \eta$, the amplitudes involved read
\begin{eqnarray}
T^0(s,t,u) &=& -\sqrt{2} E(s,t,u) \,  \,, \nonumber \\ 
T^{\frac{1}{2}}(s,t,u) &=& E(t,s,u) \,.  
\end{eqnarray}
with $E(s,t,u)=T_{K^0\bar{K}^0 \to\eta\eta}(s,t,u)$. For $K\bar{K} \to \eta\eta', \eta'\eta'$ and $K\eta(\eta') \to K \eta$ and $ K\eta'$, 
the formulas are 
analogous and the only difference is that one needs to redefine $E(s,t,u)$ for the corresponding 
process. 

All the processes $\eta^{(')}\eta^{(')}\to \eta^{(')}\eta^{(')}$, with $\eta^{(')}$ either an $\eta$ or an $\eta'$, are  $I=0$ and no crossing relation can be invoked to reduce the number of amplitudes needed, one for each process.

Next we can perform the partial wave projection for the isospin amplitudes 
and the convention we use is 
\begin{eqnarray} \label{defpwp}
T_J^{I}(s)=\frac{1}{2 (\sqrt2)^N}\int^{1}_{-1} d x \, P_J(x)\, T^{I}\big(s, t(x), u(x)\, \big)\,,
\end{eqnarray}
where $x=\cos\varphi$, with $\varphi$ the scattering angle between the incoming and outgoing particles in 
the CM, and  $P_J(x)$ stands for the Legendre polynomial. In the previous equation
 $(\sqrt2)^N$ is a symmetry factor to account for
identical particle states, such as $\pi \pi$ (with isospin 0, 1 or 2), $\eta\eta$ and $\eta'\eta'$. It corresponds to the so-called unitary normalization of Ref.~\cite{oller97}. E.g. $N=1$ for $\pi\pi\to K\bar{K}$ and $N=2$ for $\pi\pi \to \eta\eta$, and so on.

The unitarization method we follow
was developed in Ref.~\cite{oller99prd} and is based on the  $N/D$ method~\cite{chew60}. 
The essential of this approach is to separate the crossed-channel cuts (LHC) and the right-hand cut (RHC), due to unitarity,   in two different functions, $N(s)$ and $g(s)$. The former does not contain the 
two-particle unitarity cut but it has the LHC, while the later contains only the 
two-particle unitarity cut and not the LHC. A more detailed account of this unitarization 
method can be found in Refs.~\cite{oller99prd,plbkn,higgs,lacour}. In the latter reference 
an explicit integral equation for the function  $N(s)$ in nucleon-nucleon scattering is deduced. 

Let us consider first 
the elastic case, its generalization to the coupled channel case is straightforward and given below.
 Because of unitarity above the two-particle threshold $s_{th}$ a two-meson partial wave amplitude  $T_J^I(s)$, 
with well-defined isospin $I$ and angular momentum $J$, fulfills     
\begin{eqnarray} \label{utim2}
&& {\rm Im} T_J^I(s) =  T_J^I(s)\, \rho(s) \, {T_J^I}(s)^{*} 
 \Rightarrow  {\rm Im} T_J^I(s)^{-1} = 
 -\rho(s) \,.
\end{eqnarray}
  In the previous equation 
\begin{align}\label{defrho}
 \rho(s) & =  \frac{\sqrt{[s-(m_a + m_b)^2][s-(m_a - m_b)^2]}}{16\pi s}  \nn\\
  & = \frac{q}{8\pi\sqrt{s}}~,
\end{align}
with $m_a$ and $m_b$ the masses of the two particles in the state and $q$ the CM three-momentum. 
 Eq.~\eqref{utim2} implies that the imaginary part of the inverse of a partial wave is known. 
This can be used 
to write down a dispersion relation of the inverse of $T_J^I(s)$ taking as integration contour 
a circle of infinity radius that engulfs the right hand cut \cite{oller99prd}. It then results 
\begin{eqnarray}\label{defnd}
{T_J^{I}}(s)^{-1} = N_J^I(s)^{-1}+ g(s)~,
\end{eqnarray}
so that the function of $N_J^I(s)$, by construction, 
does not contain the RHC and 
$g(s)$ results from the known discontinuity along this cut, Eq.~\eqref{utim2}. In this way $g(s)$ is given 
by the following dispersion relation,
\begin{eqnarray}\label{defgfuncionds}
 g(s) = g(s_0) - \frac{s-s_0}{\pi}\int_{s_{\rm th}}^{\infty} \frac{\rho(s')}{(s'-s)(s'-s_0)} d s'\,.
\end{eqnarray} 
with 
\begin{eqnarray}
 {\rm Im} \,g(s) = -\rho(s)~,
\end{eqnarray}
for $ s> s_{\rm th} = (m_a + m_b)^2$. The dispersive integral in Eq.~\eqref{defgfuncionds} can be represented by the typical two-point one loop function~\cite{oller99prd}
\begin{eqnarray}\label{defgfuncionloop}
x_{\pm}&=&\frac{s+m_a^2-m_b^2}{2s}\pm\frac{1}{-2s}\sqrt{ -4 s (m_a^2-i0^+)+(s+m_a^2-m_b^2)^2}~,\nonumber\\
16\pi^2 g(s) &=&  a_{SL}(\mu)+\log\frac{m_b^2}{\mu^2} 
-x_+\log\frac{x_+-1}{x_+}
-x_- \log\frac{x_--1}{x_-} \,.
\end{eqnarray}

For the case of equal mass scattering, 
$g(s)$ reduces to the simple form
\begin{equation}
16\pi^2\, g(s) =  a_{SL}(\mu) + \log{\frac{m^2}{\mu^2}} - \sigma(s) \log{\frac{\sigma(s)-1}{\sigma(s)+1}}\,,
\end{equation}
with 
\begin{equation}\label{defsigfunc}
 \sigma(s) = \sqrt{1 - \frac{4 m^2}{s}}\,.
\end{equation}

In order to determine the interacting kernel $N_J^I(s)$ we proceed as explained in Refs.~\cite{plbkn,higgs}, so 
that we match the unitarized amplitude Eq.(\ref{defnd}) to the perturbative calculation, which 
is performed up to the one-loop level.\footnote{Eq.~\eqref{defnd} is valid below the first threshold of multi-particle 
states. Note that such states are further suppressed in the $\delta$ counting beyond our present one-loop calculation for the 
interaction kernel (they at least imply a two-loop calculation) and we do not consider them any longer.}
 In this way,  Eq.~\eqref{defnd} should be  expanded up to 
one power of $g(s)$. It then results for   $N_J^I(s)$,  
\begin{eqnarray}\label{defnfunction}
N_J^I(s) = {T_J^I}(s)^{\rm (2)+Res+Loop}+ T_J^I(s)^{(2)}  \,\, g(s) \, \, T_J^I(s)^{(2)} \,. 
\end{eqnarray}
Here $T_J^I(s)^{\rm (2)+ Res + Loop}$ stands for the partial wave amplitude from the 
perturbative calculation, with the superscripts ${\rm(2),~ Res}$ and Loop  
denoting the tree level amplitude from Eq.~(\ref{lolagrangian}), resonance exchanges and 
loop contributions, respectively, depicted in Fig.~\ref{Fig.sc.diagram}. The 
wave-function renormalization and $\eta-\eta'$ mixing contributions are included in Res and Loop. 

It can be easily checked that  $N_J^I(s)$ from Eq.~\eqref{defnfunction}  does not contain the RHC. This is due to the fact that the perturbative partial wave amplitudes satisfy unitarity perturbatively   
\begin{eqnarray}
 {\rm Im}\, T_J^I(s)^{\rm (2)+Res+Loop} &=& 
T_J^I(s)^{(2)}  \,\, \rho(s) \, \, T_J^I(s)^{(2)} 
\nonumber \\ 
&=& -T_J^I(s)^{(2)}  \,\,  {\rm Im} \,g(s) \, \, T_J^I(s)^{(2)} \,, 
\end{eqnarray}
for $ s> s_{\rm th}$. 
 Regarding the LHC contribution,  we would like to emphasize that it is perturbatively 
treated and collected in the $N_J^I(s)$ function 
defined in Eq.(\ref{defnfunction}), unlike the RHC that 
we take into account non-perturbatively. 

If the amplitude $T_J^I(s)$ has a zero in the complex plane, e.g. an Adler zero that appears in the subthreshold region 
of some partial S-waves as a consequence of chiral symmetry \cite{adlerzero}, this is accounted for in Eq.~\eqref{defnd} 
by the corresponding zero of $N_J^I(s)$. The zeroes on the real axis correspond to poles of the inverse of the amplitude, the so-called Castillejo-Dalitz-Dyson poles \cite{cdd}. As discussed in detail in Ref.~\cite{oller99prd} when the LHC contributions are neglected $\hat{T}_J^I(s)=T_J^I(s)/q^{2J}(s)$ can be written as
\begin{align}
\hat{T}_J^I(s)&=\frac{1}{{\hat{D}_J^I(s)}}~,\nn\\
\hat{D}_J^I(s)&=-\frac{(s-s_0)^{J+1}}{\pi}\int_{s_{th}}^\infty ds' \frac{q^{2J}(s')\rho(s')}{(s'-s)(s'-s_0)^{J+1}}
+\sum_{m=0}^J a_m s^m + \sum_i^M \frac{R_i}{s-s_i}~.
\label{nd.paper}
\end{align}
In the previous equation $J+1$ subtractions at $s_0$ have been taken, being the $a_m$, $m=1,\ldots,J$ the corresponding 
subtraction constants. The last sum accounts for the presence of poles in the inverse of the partial wave, with $s_i$ and 
$R_i$ their locations and residues, respectively. For the S-waves 
one has the location of non-trivial poles due to the Adler zeroes. For a given partial wave its position fixes one of the $s_i$ pole positions in the right-hand-side of Eq.~\eqref{nd.paper}. In the presence of LHC, $\hat{D}_J^I(s)=N_J^I(s)^{-1}+g(s)$ by comparing the previous equation with Eq.~\eqref{defnd}, the zero in 
$N_J^I(s)$ is moved according to the chiral expansion, Eq.~\eqref{defnfunction},  from its leading position given by 
$T_J^I(s)^{(2)}$.  In the case of the IAM one has to modify the standard formula \cite{truong,dobado,dobado97} to account properly for the Adler zero region \cite{rios}, otherwise one has a double zero instead of a simple Adler zero and a spurious pole in the partial wave amplitude in the same subthreshold region \cite{rios,pennington}.

The previous formalism can be easily generalized to  scattering processes with multiple coupled channels  
by employing a matrix notation. In this way,  $T_J^I(s)$, $N_J^I(s)$ and $g(s)\to g_J^I(s)$ are now matrices and 
Eq.~\eqref{defnd} still holds. Since phase space is diagonal then the matrix $g_J^I(s)$ is also diagonal, with its matrix elements given by Eq.~\eqref{defgfuncionloop}, evaluated with the appropriate masses for the corresponding channel.  The matrix $T_J^I(s)$ is symmetric, as required by time reversal invariance \cite{martin}, 
which then implies from Eq.~\eqref{defnd} that $N_J^I(s)$ is also symmetric. 

Explicitly, for the $IJ=00$ case there are five channels and the corresponding matrices read    
\begin{eqnarray}\label{n00}
N^{0}_0(s) \,=\, \left( \begin{array}{ccccc}
N_{\pi \pi \to \pi \pi} & N_{\pi \pi \to K \bar{K} } & N_{\pi \pi \to \eta \eta} 
& N_{\pi \pi \to \eta \eta'}  & N_{\pi \pi \to \eta' \eta'}  \\
N_{\pi \pi \to K \bar{K}} & N_{K \bar{K} \to K \bar{K} } & N_{K \bar{K} \to \eta \eta} 
& N_{K \bar{K} \to \eta \eta'}  & N_{K \bar{K} \to \eta' \eta'}
\\
N_{\pi \pi \to \eta \eta} & N_{ K \bar{K} \to \eta \eta } & N_{\eta \eta\to \eta \eta} 
& N_{\eta \eta \to \eta \eta'}  & N_{\eta \eta \to \eta' \eta'}
\\
N_{\pi \pi \to \eta \eta'} & N_{ K \bar{K} \to  \eta \eta' } & N_{ \eta \eta \to  \eta \eta'} 
& N_{ \eta \eta'\to \eta \eta'} & N_{ \eta \eta' \to \eta' \eta'}  \\
N_{\pi \pi \to  \eta' \eta'} & N_{ K \bar{K} \to \eta' \eta' } & N_{\eta \eta \to  \eta' \eta'}
& N_{ \eta \eta' \to  \eta' \eta'}  & N_{ \eta' \eta' \to \eta' \eta'}
\end{array} \right)\,,
\end{eqnarray}
\begin{eqnarray}\label{g00}
g^{0}_0(s) \,=\, \left( \begin{array}{ccccc}
g_{\pi \pi} & 0 & 0 &0&0 \\
0 & g_{K \bar{K}} & 0&0&0  \\
0 & 0 & g_{\eta \eta}&0&0  \\
0 & 0 & 0 & g_{\eta \eta'}&0  \\
0 & 0 &0&0& g_{\eta' \eta'}  \\
\end{array} \right)\,.
\end{eqnarray}

For the $IJ=1\, 0$ channel, the matrices are    
\begin{eqnarray}\label{n10}
N^{1}_0(s) \,=\, \left( \begin{array}{ccc}
N_{\pi \eta \to \pi\eta} & N_{ \pi\eta \to K \bar{K} } & N_{\pi\eta \to \pi \eta'}  \\
N_{ \pi\eta \to K \bar{K}}  & N_{K \bar{K} \to K \bar{K} }  & N_{K \bar{K}\to \pi \eta' }  \\
N_{\pi\eta \to \pi \eta'}  & N_{K \bar{K}\to \pi \eta'}  & N_{ \pi \eta' \to \pi \eta' }
\end{array} \right)\,,
\end{eqnarray}
\begin{eqnarray}\label{g10}
g^{1}_0(s) \,=\, \left( \begin{array}{ccc}
g_{ \pi\eta} & 0 & 0  \\
0 & g_{K \bar{K}} & 0  \\
0 & 0 & g_{ \pi\eta'}  \\
\end{array} \right)\,.
\end{eqnarray}

For the $IJ=\frac{1}{2}\,\, 0$ channel, it results   
\begin{eqnarray}\label{n1d20}
N^{\frac{1}{2}}_0(s) \,=\, \left( \begin{array}{ccc}
N_{K \pi \to K \pi} & N_{K \pi \to K \eta } & N_{K \pi \to K \eta'}  \\
N_{K \pi \to K \eta }  & N_{K \eta \to K \eta }  & N_{K \eta \to K \eta' }  \\
N_{K \pi \to K \eta' }  & N_{K \eta \to K \eta' }  & N_{K \eta' \to K \eta' }
\end{array} \right)\,,
\end{eqnarray}
\begin{eqnarray}\label{g1d20}
g^{\frac{1}{2}}_0(s) \,=\, \left( \begin{array}{ccc}
g_{K \pi} & 0 & 0  \\
0 & g_{K \eta} & 0  \\
0 & 0 & g_{K \eta'}  \\
\end{array} \right)\,.
\end{eqnarray}
Similar results hold for the $IJ=\frac{1}{2}\,\, 1$ channel.

For the $IJ=1\, 1$ quantum numbers  we have  
\begin{eqnarray}\label{n11}
N^{1}_1(s) \,=\, \left( \begin{array}{cc}
N_{\pi \pi \to \pi\pi} & N_{ \pi\pi \to K \bar{K} }  \\
N_{ \pi\pi \to K \bar{K}}  & N_{K \bar{K} \to K \bar{K} }  
\end{array} \right)\,,
\end{eqnarray}
\begin{eqnarray}\label{g11}
g^{1}_1(s) \,=\, \left( \begin{array}{cc}
g_{ \pi\pi} & 0   \\
0 & g_{K \bar{K}}  \\
\end{array} \right)\,.
\end{eqnarray}

Let us consider now the purely elastic channels where $N_J^I(s)$ and $g_J^I(s)$ are just functions, not matrices. 
For $IJ=\frac{3}{2}\,\, 0$, one has
\begin{eqnarray}\label{n3d20}
N^{\frac{3}{2}}_0(s) \,= N_{K \pi \to K \pi}\,,
\end{eqnarray}
\begin{eqnarray}\label{g3d20}
g^{\frac{3}{2}}_0(s) \,= g_{K \pi}\,.
\end{eqnarray}

For $IJ=2\,\, 0$,  
\begin{eqnarray}\label{n20}
N^{2}_0(s) \,= N_{\pi \pi \to \pi \pi}\,,\nonumber
\end{eqnarray}
\begin{eqnarray}\label{g20}
g^{2}_0(s) \,= g_{\pi\pi}\,.
\end{eqnarray}
Finally, the appropriate functions for  $IJ=0\,\, 1$ are   
\begin{eqnarray}\label{n01}
N^{0}_1(s) \,= N_{K \bar{K} \to K \bar{K}}\,,\nonumber
\end{eqnarray}
\begin{eqnarray}\label{g01}
g^{0}_1(s) \,= g_{K\bar{K}}\,.
\end{eqnarray}

After having the unitarized scattering amplitude from  Eq.~\eqref{defnd}, 
the S-matrix for the $IJ$ channel, $S_J^I(s)$, can be defined straightforwardly in matrix notation  
\begin{equation}
 S_J^I = 1 + 2 i \sqrt{\rho_J^I(s)}\cdot T_J^I(s)\cdot \sqrt{\rho_J^I(s)}\,.  
\end{equation}
with $\rho_J^I(s)=-\text{Im}g_J^I(s)$.
 From the matrix elements of the S-matrix we can read out the phase shifts $\delta_{kk}$ and $\delta_{kl}$, with $k\neq l$, since
\begin{align}
S_{ k k} &= |S_{  k k}| e^{2 i \delta_{ k k}}\,,\nn\\
S_{ k l} &= |S_{  k l}| e^{ i \delta_{ k l}}\,.
\end{align}

\section{Discussion and results}\label{pheno}

In this section we first discuss the fit to experimental data in order to fix the free parameters 
in our approach. Later we discuss the associated spectroscopy and its properties under variation of $N_C$. 

\subsection{Fit Quality}

\begin{figure}[H]
\begin{center}
\includegraphics[angle=0, width=1.\textwidth]{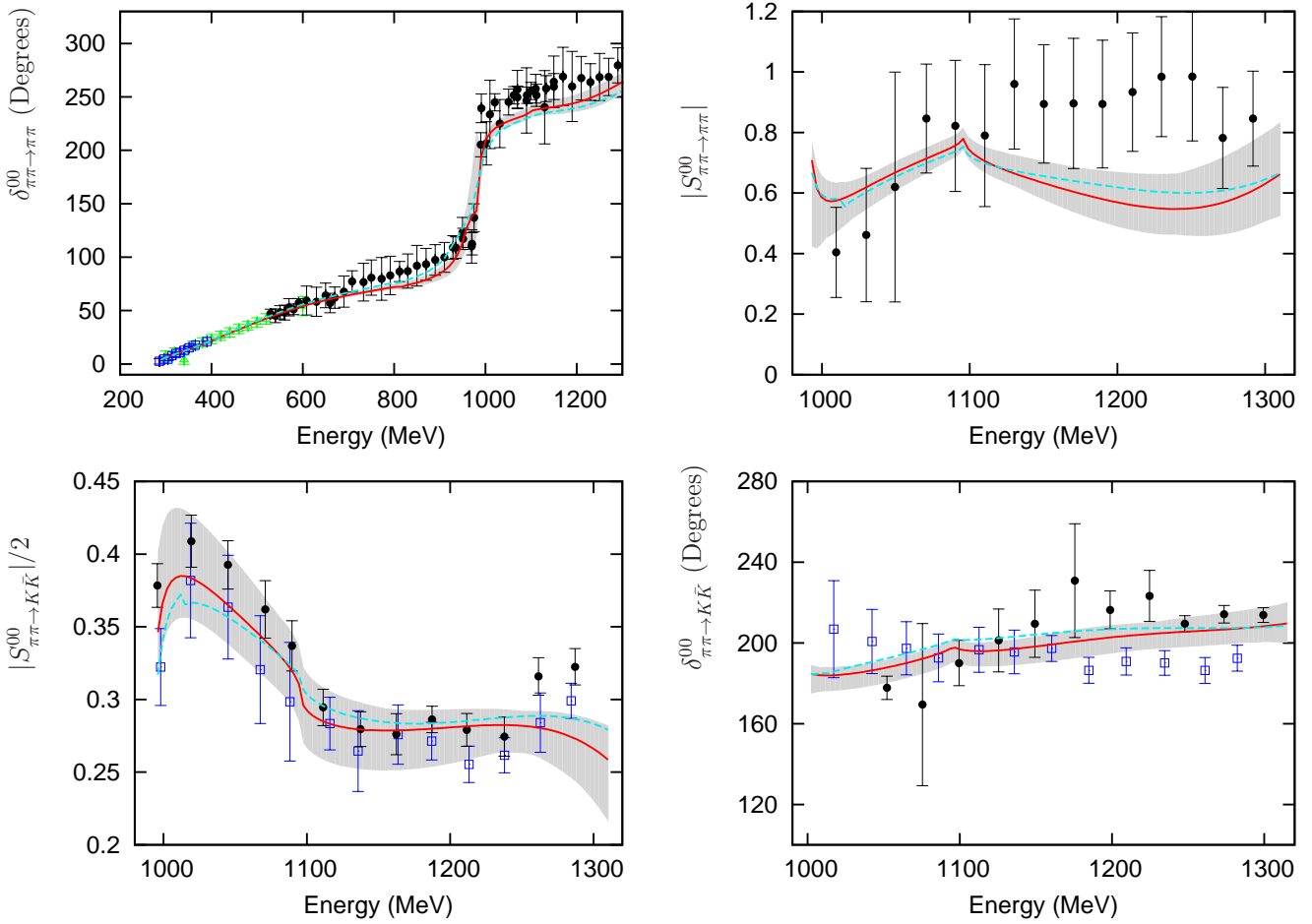}
\caption{(Color online.) Plots of the fit to the $IJ = 00$ case. From top to bottom and left to right: 
the phase shifts of $\pi\pi \to \pi\pi$ ($\delta_{\pi\pi\to\pi\pi}^{00}$), 
the modulus of the S-matrix element for $\pi\pi \to \pi\pi$ ($|S_{\pi\pi\to\pi\pi}^{00}|$), 
the half modulus of the S-matrix element for $\pi\pi \to K\bar{K}$ ($|S_{\pi\pi\to K\bar{K}}^{00}|/2$) 
and the phase shifts of $\pi\pi \to K\bar{K}$ ($\delta_{\pi\pi\to K\bar{K}}^{00}$).
$\delta_{\pi\pi\to\pi\pi}^{00}$ data correspond to Ref.~\cite{frogratt77npb} (triangle in green), \cite{na48} (square in blue) and the average data from Refs.~\cite{ochs74,refpipiphase,kaminski97zpc} (circle in black), as employed in Ref.~\cite{oller99prd}.  
 $|S_{\pi\pi \to \pi\pi}^{00}|$ is from Ref.~\cite{ochs74}.
$|S_{\pi\pi \to K \bar{K}}^{00}|/2$ is from Refs.~\cite{cohen80prd} (square in blue) and \cite{martin79npb} (circle in black). 
 The phase shifts 
$\delta_{\pi\pi \to K \bar{K}}^{00}$ are from Refs.~\cite{cohen80prd} (square in blue), \cite{etkin} (circle in black). 
The solid (red) line corresponds to the best fit, Eq.~\eqref{fitresult}, while the error bands are presented by the shadowed area. Finally, the more constrained 
fit of Eq.~\eqref{newfit} is given by the dashed line. 
 This notation also applies in Figs.~\ref{fitfig2} and \ref{fitfig3}.  
}
\label{fitfig1}
\end{center}
\end{figure}

We perform the fit for the $IJ = 00$ channel 
up to $\sqrt{s}=1300$~MeV. The inclusion of the $\eta'$ meson is not enough to guarantee that  
our current calculation can be applied higher in the energy region. There are good 
phenomenological reasons to expect that the $4\pi$ state plays an important at this 
energy level and its influence can not be simply neglected~\cite{oller08prl}.  
The observables fitted are shown in Fig.~\ref{fitfig1} and correspond to the elastic 
$\pi\pi\to \pi\pi$ phase shifts, $\delta^{00}_{\pi\pi\to\pi\pi}$, the elasticity parameter, $|S^{00}_{\pi\pi\to \pi\pi}|$,  
and the phase and one half  of the modulus of the S-matrix element for the inelastic process $\pi\pi\to K\bar{K}$  above the 
 $K\bar{K}$  threshold, $\delta^{00}_{\pi\pi\to K\bar{K}}$ and $\frac{1}{2}|S^{00}_{\pi\pi\to K \bar{K}}|$, respectively.  For references to the experimental data see Fig.~\ref{fitfig1}.

\begin{figure}[H]
\begin{center}
\includegraphics[angle=0, width=1.0\textwidth]{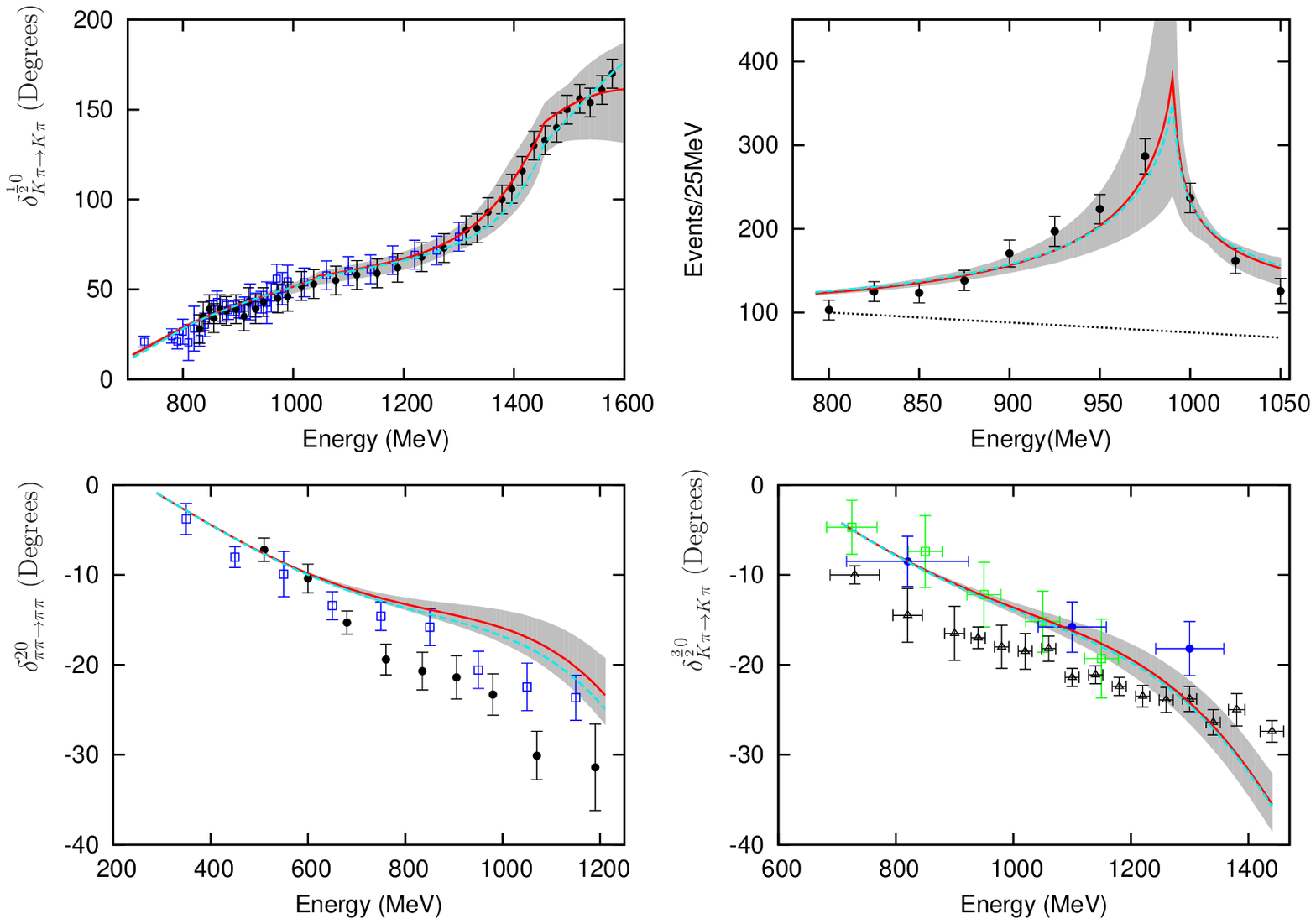}
\caption{(Color online.) From top to bottom and left to right: 
phase shifts of $K \pi \to K\pi$ with $IJ= \frac{1}{2}\,0$ ($\delta_{K \pi \to K\pi}^{\frac{1}{2} 0}$), 
$\pi\eta$ event distribution with $IJ=1\,0$, 
phase shifts of $\pi\pi \to \pi\pi$ with $IJ= 2\,0$ ($\delta_{\pi\pi \to \pi\pi}^{2\,0}$) and 
phase shifts of $K \pi \to K\pi$ with $IJ= \frac{3}{2}\,0$ ($\delta_{K\pi \to K\pi}^{\frac{3}{2}0}$). 
$\delta_{K \pi \to K\pi}^{\frac{1}{2}0}$ corresponds to the average data from 
Refs.~\cite{mercer71npb,estrabooks78npb,bingham72npb} (square in blue), as used in Ref.~\cite{oller99prd},
and Ref.\cite{aston88npb} (circle in black).  
Data for the  $\pi\eta$ event distribution are from Ref.~\cite{armstrong84zpc} and  
the dotted line corresponds to the background \cite{oller99prd}. 
$\delta_{\pi\pi \to \pi\pi}^{20}$  is from 
Refs.~\cite{hoogland77npb} (square in blue) and  \cite{losty74npb} (circle in black).
The experimental data for $\delta_{K \pi \to K\pi}^{\frac{3}{2}0}$ are taken from 
Refs.~\cite{bakker70npb} (square in green), \cite{cho70plb} (circle in blue) and \cite{estrabooks78npb} (triangle in black). 
 For the meaning of lines see Fig.~\ref{fitfig1}.} 
\label{fitfig2}
\end{center}
\end{figure}

For the $IJ = \frac{1}{2}\,0$ channel, since there is no significant 
inelasticity above the $K\eta'$ threshold, we fit the data up to $\sqrt{s}=1600$~MeV. 
Although for higher energies one already has the influence of 
the $K^*_0(1950)$ resonance \cite{jamin} which we have not included.  
The observables in the fit are the $K\pi\to K\pi$ phase shifts $\delta^{\frac{1}{2}0}_{K\pi\to K\pi}$.
The reference to experimental data is given in Fig.~\ref{fitfig2}.

For the $IJ = 1 \,0$ channel, there are no available  data for scattering 
up to now and it is typical to use a $\pi\eta$ event distribution from other 
production processes to obtain information on the $a_0(980)$ resonance.  
As the production mechanism is not under good theoretical control in our work, 
we decide to fit the data around the  resonance region up to 1050~MeV, as explained below. 
The experiment data for the $\pi\eta$ distribution are illustrated in Fig.~\ref{fitfig2}.

For the exotic channels with  $IJ = \frac{3}{2} \,0$ and $IJ = 2 \,0$,  
we fit the phase shifts of $K\pi \to K\pi$ and $\pi\pi \to \pi\pi$, respectively. 
 The references for the experimental data are given 
in Fig.~\ref{fitfig2}. 

\begin{figure}[ht]
\begin{center}
\includegraphics[ angle=0,width=1.0\textwidth]{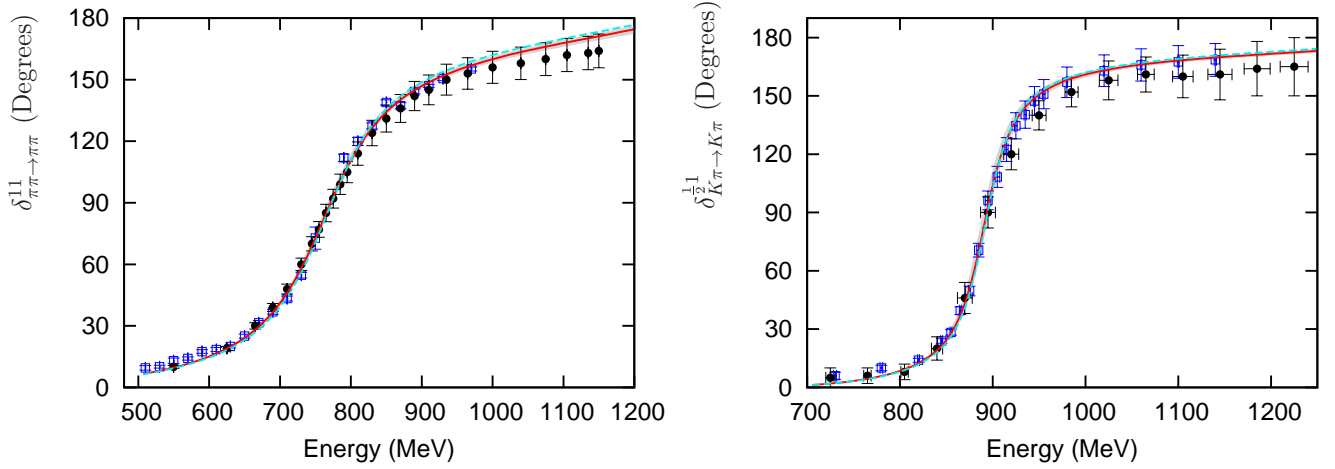}
\caption{(Color online.) Plots for the vector channels from the fits, Eqs.~\eqref{fitresult} and \eqref{newfit}. The left panel is for the $IJ=11$ channel 
and the right one is for $IJ=\frac{1}{2}1$. 
The data of phase shifts for $\pi\pi \to \pi\pi$ with $IJ=11$, $\delta_{\pi\pi \to \pi\pi}^{11}$ are from 
Refs.~\cite{lindenbaum92plb} (square in blue) and \cite{estrabooks74npb} (circle in black). 
The data of phase shifts for $K\pi \to K\pi$ with $IJ=\frac{1}{2}1$, $\delta_{K\pi \to K\pi}^{\frac{1}{2}1}$ are from 
Refs.~\cite{mercer71npb} (circle in black) and \cite{estrabooks78npb} (square in blue). For the meaning of lines see Fig.~\ref{fitfig1}.
} 
\label{fitfig3}
\end{center}
\end{figure}

The observables we fit in $IJ = 1 \,1$ and $IJ = \frac{1}{2} \,1$ are the phase shifts  
of $\pi\pi \to \pi\pi$ and $K\pi \to K\pi$, $\delta^{11}_{\pi\pi\to\pi\pi}$ and $\delta^{\frac{1}{2}1}_{K\pi\to K\pi}$, in order. They are fitted  up to 1200~MeV, as we only include the lowest 
multiplet of vector resonances in the Lagrangian~Eq.(\ref{lagvector}). 
The references to experimental data can be found in Fig.~\ref{fitfig3}.

Before moving to  the details of the phenomenological discussion, let us first comment on the 
free parameters in our theory. Apart from the couplings in 
Eqs.\eqref{lagscalar}, \eqref{lagvector} and \eqref{deflam2}, 
there are the bare mass parameters of the resonances,  
the $U_A(1)$ anomaly mass $M_0$ in Eq.(\ref{lolagrangian}) and the subtraction 
constants $a_{SL}$ defined in Eq.(\ref{defgfuncionloop})  for different channels. 
One way to reduce the number of the subtraction constants is to impose 
the isospin symmetry for them \cite{lambda}. In this way,  $a_{ SL}$ of  
$g_{\pi\pi}$ in  $IJ = 0 \,0$ and in $ IJ = 2 \, 0$ are the same. Similarly, 
$g_{K \bar{K}}$ in  $IJ = 0 \,0$ and $ IJ = 1 \, 0$ are also equal. This also applies   
to $g_{K \pi}$ in  $IJ = \frac{1}{2} \,0$ and in $ IJ = \frac{3}{2} \, 0$. 
In principle, the subtraction constants in the remaining channels are free. 
 Nevertheless, if one further assumes  $U(3)$ symmetry within each orbital angular momentum   
the values of different subtraction constants should be also equal. For this result one should extend 
the proof given in Appendix A of Ref.~\cite{lambda} in the $SU(3)$ limit to the $U(3)$ case, that holds for equal quark masses 
 at leading order in large $N_C$. Notice that in this limit the quarks behave equivalently so that the equality of 
the subtraction constants in $SU(3)$ also implies that they are equal for the $U(3)$ case.
We indeed exploit this feature as much as we can in the numerical discussion. 
We summarize the subtraction constants used in the fits discussed below:
\begin{eqnarray}
&& a_{ SL}^{00} = a_{ SL}^{00\,,\, \pi\pi} = a_{ SL}^{00\,,\, K\bar{K}} = a_{ SL}^{00\,,\, \eta\eta}
 = a_{ SL}^{00\,,\, \eta\eta'} = a_{ SL}^{00\,,\, \eta'\eta'} = a_{ SL}^{2 0 \,,\, \pi\pi} \,, \nonumber \\ &&
a_{ SL}^{\frac{1}{2} \,0} = a_{ SL}^{\frac{1}{2} \,0\,,\, K\pi} 
= a_{ SL}^{\frac{1}{2} \,0\,,\, K\eta} = a_{ SL}^{\frac{1}{2} \,0 \,,\, K\eta'}
= a_{ SL}^{\frac{3}{2} \,0\,,\, K\pi}  \nonumber \\ &&
 a_{ SL}^{1\,0 \,,\, \pi\eta'} = a_{SL}^{1\,0\,,\, K\bar{K}} =  a_{ SL}^{0\,0\,,\, K\bar{K}}\,,\nonumber  \\
&& a_{ SL}^{1 \,0\,,\, \pi\eta}\,,
\end{eqnarray}
and all of the subtraction constants in the vector channels are set equal to $a_{ SL}^{0 0}$.

In order to fit the $\pi\eta$ mass distribution we need to introduce two additional parameters  
to parameterize the production mechanism, ${\cal N}$ and $c$, that enter in the expression 
\begin{eqnarray}
\frac{d N_{\pi\eta}}{ d E_{\pi\eta}} 
= q_{\pi\eta}\, {\cal N} \big| \,T_{ K \bar{K}\to \pi\eta}(s) + c\, T_{\pi\eta \to \pi\eta}(s)  \, \big|^2\,.
\label{a0.inv}
\end{eqnarray}
Here, $q_{\pi\eta}$ is the three momentum of $\pi\eta$ system in CM and 
$E_{\pi\eta} = \sqrt{s}$ is the energy in the same frame. The parameter ${\cal N}$ accounts for the fact that 
the event distribution is not normalized. The linear combination of 
amplitudes in the previous equation originates because  the $a_0(980)$ shows up differently in them. 
See also Ref.~\cite{plbkn} for a more detailed explanation of the derivation of Eq.~\eqref{a0.inv} in connection 
with  the invariant mass distribution of  $\bar{K}N$ around the $\Lambda(1405)$ resonance.

In the end,  we have 16 free parameters  and the fitted results are 
\begin{eqnarray}\label{fitresult}
&& c_d= (15.6^{ +4.2}_{-3.4} )\,{\rm MeV}\,, \qquad \quad c_m= (31.5^{ +19.5}_{-22.5} )\,{\rm MeV}\,, 
\nonumber \\&&
\widetilde{c}_d = (8.7^{ +2.5}_{-1.7} )\,{\rm MeV}\,, \qquad \quad\quad
\widetilde{c}_m = (15.8^{ +3.3}_{-3.0} )\,{\rm MeV}\,, \nonumber \\&&  
M_{S_8}= (1370^{ +132}_{-57} )\,{\rm MeV}\,, \qquad
M_{S_1}= (1063^{ +53}_{-31} )\,{\rm MeV}\,, \nonumber \\&&
M_{\rho}= (801.0^{ +7.0}_{-7.5} )\,{\rm MeV}\,, \qquad
M_{K^*}= (909.0^{ +7.5}_{-6.9} )\,{\rm MeV}\,, \nonumber \\&&
G_V= (61.9^{ +1.9}_{-1.9} )\,{\rm MeV}\,, 
\qquad \quad 
a_{SL}^{1 \,0\,,\pi\eta}= 2.0^{ +3.1}_{-3.4} \,, \nonumber \\ &&
a_{SL}^{00}= (-1.15^{ +0.07}_{-0.09} )\,, \qquad \quad \,\,\,\,
a_{SL}^{\frac{1}{2} \,0}= (-0.96^{ +0.10}_{-0.16} )\,, \nonumber \\ &&
{\cal N} = (0.6^{ +0.3}_{-0.3} ) \,{\rm MeV^{-2}}\,, \qquad \,\,\,
c = (1.0^{ +0.6}_{-0.4} )\,, \nonumber \\ &&
M_0 = (954^{ +102}_{-95} ) \,{\rm MeV}\,, \qquad \quad
\Lambda_2 = ( -0.6^{ +0.5}_{-0.4} )\,,
\end{eqnarray}
with $\chi^2/{\rm d.o.f} = 714/(348-16) \simeq 2.15 $. 

The corresponding figures from the fit are displayed in 
Figs.~\ref{fitfig1}, \ref{fitfig2} and \ref{fitfig3}. The width of the bands shown and  the errors given in the fitted parameters 
in Eq.~\eqref{fitresult} represent our statistical uncertainties at the level of two standard
deviations \cite{etkin}
\begin{equation}\label{errorbase}
n_\sigma=\Delta \chi^2/(2 \chi_0^2)^{1/2}\,,
\end{equation} 
being $\chi_0^2$ the minimum of the $\chi^2$ obtained corresponding to the best fit, $n_\sigma$ the number of 
standard deviations and $\Delta \chi^2=\chi^2-\chi^2_0$. 

As one can see in Figs.~\ref{fitfig1}, \ref{fitfig2} and \ref{fitfig3}, 
most of the observables are well reproduced through the fit.  
The phase shifts of $\pi\pi \to \pi\pi$ 
with $IJ=0 0$ in the low energy region from the recent measurement of $K_{e4}$ decay \cite{na48} are included in our 
fit and they can be perfectly reproduced. 
For the $\pi\eta$ invariant mass distribution, we explicitly subtract the background 
as analyzed in Ref.~\cite{armstrong84zpc}, which are mainly caused by the tail of  
higher resonances. Compared with observables in the scalar channels, 
the errors in the vector channels are rather small. 
Among all of the curves, the least satisfactory results from our fit correspond 
to the channels with exotic quantum numbers with $IJ = 2 0$ and $IJ = \frac{3}{2} 0$, 
which is also an important source for the $\chi^2$ resulting  from our fit. 
For example, with 42 data points, they contribute 281 to the total $\chi^2$, i.e. 
$12\%$ of degrees of freedom carry $39\%$ of the $\chi^2$. Nevertheless, one also has to 
remark that the different experimental data in these channels are not always compatible at the 
level of one $\sigma$, which also makes the $\chi^2$ to increase. 
  On the other hand, since no resonances appear in those exotic channels, 
it may indicate that unitarity is less important as compared with  the other channels. 
In contrast, the LHC contribution could play a more important role in this case \cite{oller99prd,jamin}. 
We recall that precisely our unitarization scheme only incorporates the LHC  effects perturbatively. 
Thus, an improvement on the treatment of the LHC   
should provide  a better fit for these two channels.  

  With the fitted parameters in Eq.~\eqref{fitresult} we can also calculate the values for the masses 
of the $\eta$ and $\eta'$ that are obtained by diagonalizing Eq.~\eqref{defmixingpara}. We then obtain from Eq.~\eqref{mixnlo} the values
\begin{eqnarray}
m_{\eta} = ( 526^{ + 33}_{-41} )\, {\rm MeV}\,,\qquad  m_{\eta'} = ( 951^{ + 47}_{-60} )\, {\rm MeV}\,.
\end{eqnarray}
The leading order mixing angle of $\eta-\eta'$ that results  from Eq.~(\ref{lolagrangian}) is 
\begin{eqnarray}
\label{mixingangle}
\theta = -{16.2^\text{{\tiny{o}}}}^{+2.8^\text{{\tiny{o}}}}_{-2.9^\text{{\tiny{o}}}}\,.
\end{eqnarray}

Concerning the resonance parameter $c_d$ in Eq.(\ref{fitresult}), 
the value from our fit is a bit smaller than  
the previous determinations, such as $c_d = 19.1^{+2.4}_{-2.1}$~MeV given in Ref.~\cite{oller99prd}, 
$c_d = 23.8$~MeV reported in Ref.~\cite{jamin} and $c_d = (26 \pm 7)$~MeV in Ref.~\cite{guo09prd}.   
Nevertheless, they are compatible within the error bands. 
For the coupling $c_m$, though different approaches predict a broad range for its central value, 
they always accompany a large error, for example $c_m= 31.5^{ +19.5}_{-22.5} $~MeV in the present case, 
$c_m = (15 \pm 30)$~MeV in Ref.~\cite{oller99prd} and $c_m = (80 \pm 20 )$~MeV given in Ref.~\cite{guo09prd}. 
Comparing the couplings related to the singlet scalar with Ref.~\cite{oller99prd}, 
our determination of $\widetilde{c}_m $ is compatible with $\widetilde{c}_m = 10.6^{ +4.5}_{-3.5}$~MeV in 
Ref.~\cite{oller99prd}, while for $\widetilde{c}_d$ the current fit leads to a smaller 
value, compared with $\widetilde{c}_d = 20.9^{+1.6}_{-1.0}$~MeV given in the same reference. 
 For the bare masses of resonances, our present results agree well with the previous determinations: 
$M_{S_8} = (1390 \pm 20)$~MeV, $M_{S_1} = (1020^{+40}_{-20})$~MeV in Ref.~\cite{oller99prd} and   
$M_{S_8} = 1400 $~MeV in Ref.~\cite{jamin}.  Our values in Eq.~\eqref{fitresult} are  compatible with the estimates $\widetilde{c}_d\simeq c_d/\sqrt{3}$ and 
$\widetilde{c}_m\simeq c_m/\sqrt{3}$ based on 
large $N_C$ \cite{ecker89npb}, however the bare masses for the singlet and octet scalar resonances are 
 different, at around a 30$\%$, as obtained in previous studies \cite{oller99prd,oller08prl}. 
The current determination for the subtraction constant $a_{SL}$ is consistent with the previous 
result in ~Ref.\cite{oller99prd} as well. 
For the vector resonance coupling $G_V$, our result is in good agreement with 
$G_V = ( 63.9 \pm 0.6 )$~MeV determined in Ref.~\cite{guo09prd}.  
The $U_A(1)$ anomaly mass $M_0$ carries a large error and agrees with the conclusion 
in Ref.~\cite{feldmann}. The $1/N_C$ suppressed parameter $\Lambda_2$ is 
poorly known in the literature and in Ref.~\cite{escribano10} $\Lambda_2 = 0.3$ 
is estimated, which is somewhat incompatible with ours. 
Nevertheless in Ref.~\cite{escribano10} the value of $\Lambda_2$ is not determined by 
any physical observables, but only by naive dimensional analysis. Moreover in their case 
the counterterm $\Lambda_2$ is always accompanied by the factor of $m_\pi^2$, indicating the 
insensitivity of this parameter there. 
  
\begin{table}[ht]
\renewcommand{\tabcolsep}{0.05cm}
\renewcommand{\arraystretch}{1.2}
\begin{center}
{\small 
\begin{tabular}{|l|l|l|l|ll|}
\hline
R & M (MeV) & $\Gamma/2$ (MeV)&   $|$Residues$|^{1/2}$ (GeV) & Ratios &  
\\
[0.1cm] 
\hline
\hline
$\sigma$& $440^{+3}_{-3}$ & $258^{+2}_{-3}$ & $3.02^{+0.03}_{-0.03}$ $(\pi\pi)$ 
&  $0.51^{+0.03}_{-0.02}$\,($KK/\pi\pi$) &  $0.06^{+0.03}_{-0.01}$\,($\eta\eta/\pi\pi$) 
\\[0.1cm] 
& & & & $0.16^{+0.03}_{-0.02}$($\eta\eta'/\pi\pi$)\, & $0.05^{+0.05}_{-0.03}$($\eta'\eta'/\pi\pi$) 
\\[0.1cm]   
\hline 
$f_0(980)$& $981^{+9}_{-7}$ & $22^{+5}_{-7}$ & $1.7^{+0.3}_{-0.3}$($\pi\pi$)\,
& $2.3^{+0.3}_{-0.2}$($KK/\pi\pi$)\,& $1.6^{+0.3}_{-0.3}$($\eta\eta/\pi\pi$)  
\\[0.1cm]
& & & & $1.2^{+0.1}_{-0.2}$($\eta\eta'/\pi\pi$)\,& $0.7^{+0.4}_{-0.5}$($\eta'\eta'/\pi\pi$)  
\\[0.1cm]  \hline 
$f_0(1370)$& $1401^{+58}_{-37}$ & $106^{+36}_{-23}$ & $2.4^{+0.2}_{-0.1}$($\pi\pi$)\, 
& $0.62^{+0.04}_{-0.05}$($KK/\pi\pi$)\,& $0.9^{+0.1}_{-0.1}$($\eta\eta/\pi\pi$) 
\\[0.1cm]
& & & & $1.7^{+0.4}_{-0.6}$($\eta\eta'/\pi\pi$)\,& $1.1^{+0.4}_{-0.6}$($\eta'\eta'/\pi\pi$) 
\\[0.1cm]  
\hline
$\kappa$& $665^{+9}_{-9}$ & $268^{+21}_{-6}$ & $4.2^{+0.2}_{-0.2}$($K\pi$)\,
& $0.7^{+0.1}_{-0.1}$($K\eta/K\pi$)\,  & $0.5^{+0.1}_{-0.1}$($K\eta'/K\pi$)  
\\[0.1cm] 
 \hline  
$K^*_0(1430)$&  $1428^{+56}_{-23}$ & $87^{+53}_{-28}$ & $3.3^{+0.5}_{-0.4}$($K\pi$)\,
 & $0.54^{+0.07}_{-0.02}$ ($K\eta/K\pi$)\, & $1.2^{+0.2}_{-0.3}$($K\eta'/K\pi$) \\
[0.1cm]
  \hline  
$a_0(980)$
& $1012^{+25}_{-7}$ & $16^{+50}_{-13}$ & $2.5^{+1.3}_{-0.8}$($\pi\eta$)\, 
& $1.9^{+0.2}_{-0.3}$ ($KK/\pi\eta$)\, & $0.01^{+0.03}_{-0.01}$($\pi\eta'/\pi\eta$)
 \\[0.1cm]  \hline 
$a_0(1450)$
& $1368^{+68}_{-68}$ & $71^{+48}_{-23}$ & $2.3^{+0.4}_{-0.5}$($\pi\eta$)\,
& $0.6^{+0.7}_{-0.2}$($KK/\pi\eta$)\, & $0.6^{+0.2}_{-0.1}$($\pi\eta'/\pi\eta$) 
 \\[0.1cm]  \hline  
$\rho(770)$& $762^{+4}_{-4}$ & $72^{+2}_{-2}$ & $2.48^{+0.03}_{-0.05}$($\pi\pi$)\, 
& $0.64^{+0.01}_{-0.01}$($KK/\pi\pi$) & 
 \\[0.1cm]  \hline 
$K^*(892)$& $891^{+3}_{-4}$ & $25^{+2}_{-1}$ & $1.86^{+0.05}_{-0.05}$($K\pi$)\,
& $0.91^{+0.03}_{-0.02}$($K\eta/K\pi$)\,& $0.45^{+0.08}_{-0.08}$($K\eta'/K\pi$) 
 \\[0.1cm] \hline 
$\phi(1020)$& $1019.5^{+0.3}_{-0.3}$ & $2.0^{+0.04}_{-0.08}$ & $0.85^{+0.01}_{-0.02}$($K\bar{K}$)\,
&   &  
 \\[0.1cm]  
\hline
\end{tabular}
}
 \caption{ {\small Pole positions for the different resonances in 
 $\sqrt{s} \equiv (\text{M},-i\frac{\Gamma}{2})$.
 The mass (M) and the half width ($\Gamma/2$)  are given in units of MeV. 
The modulus of the square root of a residue is given in units of GeV, 
which corresponds to  the coupling of the resonance with the first channel 
(specified inside the parentheses).
The last two columns are the ratios of the coupling strengths of
the same resonance to the remaining channels with respect to the first one. 
The corresponding Riemann sheets where the resonance poles are located are explained 
in detail in the text. Note that here the residues for $\pi\pi$, $ \eta\eta$ and $\eta'\eta'$ 
are given in the unitary normalization, due to the extra factors of $\sqrt{2}$ 
dividing Eq.~\eqref{defpwp}. Thus, one should multiply by $\sqrt{2}$ these couplings if one 
wishes to restore standard physical normalization to 1 for these states. 
The error bands of the resonance parameters appearing in the table only correspond to 
the statistical error from the fit in Eq.(\ref{fitresult}). }
 }  
\label{tab:pole}
\end{center}
\end{table}

We also show another more constrained fit by employing relations and values in the literature, already 
commented above, for the 
parameters shown in the fit of Eq.~\eqref{fitresult}.  In particular we impose from the beginning the large 
$N_C$ constraints \cite{ecker89npb}  $\widetilde{c}_d=c_d/\sqrt{3}$ and $\widetilde{c}_m=c_m/\sqrt{3}$. We also 
take  $M_{S_8}=1390~$MeV and $M_{S_1}=1020$~MeV, according to Ref.~\cite{oller99prd}. For $G_V$  we take 60~MeV from the averages of values 
 taken from Refs.~\cite{ecker89plb,guo2,guo09prd}. For $M_0$ we take the value 850~MeV from Ref.~\cite{feldmann}. The subtraction 
constant $a_{SL}^{10,\pi\eta}$, given with large errors in Eq.~\eqref{fitresult}, is now fixed at $+2$.
 The resulting fit is shown by the dashed lines in Figs.~\ref{fitfig1}-\ref{fitfig3}, having now only 9 free parameters. 
We see that the reproduction of  data is of similar quality as the one achieved by the fit in Eq.~\eqref{fitresult} and the 
 $\chi^2/_{d.o.f}=843/(348-9)=2.5$  is not much larger. The resulting values for the free parameters are now 
\begin{equation}
\label{newfit}
\begin{array}{ll}
             c_d =   17.4~\text{MeV} &  c_m  =   28.1~\text{MeV}\\
           M_\rho   =  800.4~\text{MeV} & M_{K^*} = 910.0~\text{MeV}\\
            a_{SL}^{00} =   -1.14 &   a_{SL}^{\frac{1}{2}0}  =  -0.89 \\
           \Lambda_2  =   -0.22 &  {\cal N}=  0.55~\text{MeV}^{-2}\\
              c  =    0.84 & \\
\end{array}
\end{equation}
Comparing with the values in Eq.~\eqref{fitresult} one observes rather similar values, compatible within the errors given in 
Eq.~\eqref{fitresult}. The biggest change 
occurs for the value of $\Lambda_2$, although the new value lies well within the error band given  in Eq.~\eqref{fitresult}. 
The resulting singlet couplings are $\widetilde{c}_d=c_d/\sqrt{3}=10.05$~MeV and $\widetilde{c}_m=c_m/\sqrt{3}=16.20$~MeV.


\subsection{Resonances generated from unitarized amplitudes}

We summarize the masses, widths and residues of the various resonances 
in Table~\ref{tab:pole}. Resonances are characterized by their pole positions in the partial wave amplitudes in unphysical Riemann sheets. Around a  resonance pole $s_R$, corresponding to a resonance $R$, 
 the partial wave amplitude $T_J^I(s)_{i\to j}$ tends to 
\begin{align}
T_J^I(s)_{i\to j}\to -\frac{g_{R\to i}\, g_{R\to j}}{s-s_R}~.
\end{align}
 By calculating the residue of the resonance pole we then obtain the 
 product of the couplings  to the corresponding decay modes. 
 At the practical level we calculate the residues by applying the Cauchy integral formula
\begin{equation}
\label{cauchycc}
 g_{R\to i}\, g_{R\to j} =- \frac{1}{2\pi i}\oint_{|s-s_R|\to 0 } T_J^I(s)_{i\to j} \,d s~. 
\end{equation}
  For every pole one has further to indicate in which  unphysical Riemann sheet it lies.  
 Each function $g(s)$  has 2 sheets and their relation is given  by \cite{oller97}
\begin{eqnarray}
g_{II}(s)=g_I(s)+2 i \rho(s)~,
\end{eqnarray}
with $g_{II}(s)$ the function analytically extrapolated to its second Riemann sheet and $g_I(s)$ the function in 
its first Riemann sheet, given in Eq.~\eqref{defgfuncionloop}. Different Riemann sheets are easily 
accessed by deciding on which sheet every  $g_i(s)$ function, associated to channel $i$, is calculated. 
In this way, for an $IJ$ channel there are $2^N$ possible sheets, with $N$ the number of coupled states with the same 
$IJ$ quantum numbers. Along the real $s$-axis above threshold  changing sheet implies to reverse the sign   
of the imaginary part of the $g_i(s)$ function. In the following we conventionally 
label the physical or first Riemann sheet as 
$(+,+,+,\ldots)$.  The second Riemann sheet can be reached 
by  changing the sign of the first momentum, which is labeled as $(-,+,+,+,\ldots)$. 
The third, fourth and fifth sheets, also considered in this work, correspond to $(-,-,+,+,\ldots)$, $(+,-,+,+,\ldots)$ 
and $(-,-,-,+,\ldots)$, in order. More sheets can be obtained by taking more combinations of plus and minus signs between the brackets.  At a given energy value $\sqrt{s}$, there is one unphysical sheet 
to which  one can directly access  from the physical sheet by  crossing from $s+i0^+$ to $s-i0^+$ the branch cut 
between the two thresholds $T_{n}$ and $T_{n+1}$, with $T_{n} < \sqrt{s}  <  T_{n+1} $. 
In our current notation, this specific unphysical sheet corresponds to changing the signs 
of all the three-momenta below the considered energy point $\sqrt{s}$, i.e 
\begin{align}
(\underbrace{-,-,..-}_n,+,+,\ldots)
\end{align} 
In fact this is also the most relevant Riemann sheet where one should search the pole for 
a given resonance affecting that energy region, although other shadow poles may also appear in other 
unphysical sheets~\cite{eden64pr}. Thus in the following, we mainly present 
our findings for a given resonance on the complex plane in this specific sheet  and 
simply comment the results appearing in other unphysical sheets.

Through the unitarization procedure described previously, 
we can simultaneously get the relevant resonances in the considered energy region, 
such as $\sigma$, $f_0(980)$ and $f_0(1370)$ in the $IJ = 00 $ channel, 
$\kappa$ and $K^*_0(1430)$ in $IJ = \frac{1}{2}\,0 $, 
$a_0(980)$ and $a_0(1450)$ in $IJ = 1\,0 $, $\phi(1020)$ in $IJ=01$, 
$\rho(770)$ in $IJ=1 1$ and $K^*(892)$ with $IJ = \frac{1}{2}\,1 $. The pole positions and residues of 
these resonances are collected in Table~\ref{tab:pole}. 
We comment on them channel by channel next. 

\subsubsection*{ $IJ = 0 0$} 
Three kinds of resonance poles in the complex plane have been found,  
which we identify with the $\sigma$, $f_0(980)$ and $f_0(1370)$ resonances. 
According to our previous discussion, the most relevant Riemann sheets 
for $\sigma$, $f_0(980)$ and $f_0(1370)$ correspond to those with 
$(-,+,+,+,+)$, $(-,+,+,+,+)$ and $(-,-,-,+,+)$, respectively. 
 Their pole positions are compatible within errors with those given in the PDG \cite{pdg}. The pole 
for the $\sigma$ resonance in Table~\ref{tab:pole} is in close agreement with the pole position 
$(484\pm 14-255\pm 10\,i)$~MeV of Ref.~\cite{ruben76}, and slightly narrower, considering errors, than that of  
Ref.~\cite{leutwyler06}  $(441^{+16}_{-8}-272^{+9}_{12}\,i)$~MeV. 


From the residues given in Table~\ref{tab:pole}, one can see that the biggest coupling of the 
$\sigma$ resonance is to the $\pi\pi$ channel. Nonetheless, it also has a large 
coupling to the $K\bar{K}$ channel. From the value in 
Table~\ref{tab:pole} one has 
\begin{align}
\left|\frac{g_{\sigma\to K^+K^-}}{g_{\sigma\to \pi^+\pi^-}}\right|=\frac{\sqrt{3}}{2}\left| 
\frac{g_{\sigma\to K\bar{K}}}{g_{\sigma\to\pi\pi}} \right|=0.44^{+0.03}_{-0.02}~,
\label{sig.coup.kpkm}
\end{align}
compatible with  $0.37\pm 0.06$ obtained in Ref.~\cite{narison22}. In this reference this value  is interpreted as an indication that the $\sigma$ resonance 
has a strong glueball component.

One observes from Table~\ref{tab:pole} that the $\sigma$ couples very weakly to $\eta\eta$. This is typical of unitarized 
 $\chi$PT studies \cite{mixing}. However, from QCD spectral sum rules \cite{SNVeneziano89,SN98} the presence   around 1~GeV of a broad glueball, $\sigma_B$, together with a $S_2=(\bar{u}u+\bar{d}d)/\sqrt{2}$ quarkonium, is required in order to fulfill the corresponding scalar sum rules. If the $\sigma$ resonance were purely gluonium then it would be an $SU(3)$ singlet and its coupling to pseudoscalar pairs of $\pi$, $K$ and $\eta_8$ would be universal. In this case the coupling of the $\sigma$ to $\eta\eta$ would be the same as that to 
  $\pi^+\pi^-$. On the other hand, in Ref.~\cite{SN98} a maximal mixing  between $\sigma_B$ and $S_2$ is proposed, with a mixing angle $\theta_S\simeq (40-45)^{\text{o}}$. In this case the size of the coupling of the $\sigma$ to $ \eta\eta$ depends strongly 
on the value taken for the coupling of the $\sigma$ to a kaon pair. E.g. if one takes as in Ref.~\cite{SN98} that 
$|g_{\sigma\to \pi^+\pi^-}/g_{\sigma\to K^+ K^-}|\gtrsim 1$, for definiteness we employ 1.2, then  
\begin{align}
\left|\frac{g_{\sigma\to\eta_8\eta_8}}{g_{\sigma\to\pi^+\pi^-}} \right|\simeq 0.8~.
\end{align} 
However, if the latest value 0.37 \cite{narison22} for $|g_{\sigma\to K^+ K^-}/g_{\sigma\to \pi^+\pi^-}|$ is considered 
then
\begin{align}
\left|\frac{g_{\sigma\to\eta_8\eta_8}}{g_{\sigma\to\pi^+\pi^-}} \right|\simeq 0.15~.
\end{align}
 This result  is remarkably  close to our prediction from Table~\ref{tab:pole} 
\begin{align}
\left|\frac{\tilde{g}_{\sigma\to\eta\eta}}{g_{\sigma\to\pi^+\pi^-}} \right|
=\sqrt{3}\left|\frac{g_{\sigma\to\eta\eta}}{g_{\sigma\to\pi\pi}}\right|
=0.10^{+0.05}_{-0.02}~.
\label{sig.coup.e8e8}
\end{align}
In the previous equation, as well as in the following, a tilde over a coupling of the $\sigma$ to a state made of identical mesons means that it is considered 
with physical normalization and not with the unitary one, used in Table~\ref{tab:pole}, see Eq.~\eqref{defpwp}.

 Up to the authors' knowledge this is the first time  in the literature that the couplings of the 
$\sigma$ to states involving the pseudoscalar singlet $\eta_1$ are calculated. 
In Ref.~\cite{SNVeneziano89} one has the upper limit 
\begin{align}
\left|\frac{g_{\sigma\to\eta\eta'}}{g_{\sigma\to \pi^+\pi^-}}\right|&\leq 0.23~.
\label{sneep}
 \end{align}   
From this value one can obtain the corresponding upper limit for the coupling of the $\sigma$ to 
$\eta'\eta'$ dividing by $\tan\theta$ the one to $\eta\eta'$.\footnote{We thank S. Narison for suggesting us 
this procedure.} Taking  for $\theta$ our value given in Eq.~\eqref{mixingangle} we then obtain the upper bound
\begin{align}
\left|\frac{g_{\sigma\to\eta'\eta'}}{g_{\sigma\to \pi^+\pi^-}}\right|&\leq 0.07~.
\label{snepep}
\end{align} 
 From the values obtained in Table~\ref{tab:pole} we have 
\begin{align}
\left|\frac{g_{\sigma\to\eta\eta'}}{g_{\sigma\to \pi^+\pi^-}}\right|&= \sqrt{\frac{3}{2}}\left|\frac{g_{\sigma\to\eta\eta'}}{g_{\sigma\to \pi\pi}}\right|=0.20^{+0.04}_{-0.02}~,\nn\\
\left|\frac{\tilde{g}_{\sigma\to\eta'\eta'}}{g_{\sigma\to \pi^+\pi^-}}\right|&= \sqrt{3}\left|\frac{g_{\sigma\to\eta'\eta'}}{g_{\sigma\to \pi\pi}}\right|= 0.09^{+0.09}_{-0.05}~,
\end{align}
which are perfectly compatible with the bounds given in Eqs.~\eqref{sneep} and \eqref{snepep} calculated from ef.~\cite{SNVeneziano89}. 
 
We then obtain a remarkable compatibility between our results for the couplings of the $\sigma$ in table~\ref{tab:pole} and those 
 obtained from QCD spectral sum rules \cite{SNVeneziano89,SN98}, once the latest value for the ratio of couplings $|g_{\sigma\to \pi^+\pi^-}/g_{\sigma\to K^+ K^-}|$ is taken from Ref.\cite{narison22}. Nevertheless, in our approach there is  no a  corresponding bare state at the Lagrangian from which the $\sigma$ pole could stem,  it is mainly generated by the strong interactions between pseudo-Goldstone bosons.

 The $f_0(980)$ resonance has its strongest coupling to $K\bar{K}$, 
though it also couples almost as equally strong to $\eta\eta$. Notice that the ratios of the 
couplings to $K\bar{K}$ and $\pi\pi$ between the $\sigma$ and $f_0(980)$ are nearly inverse  each other. 
The importance of the $\eta\eta$ channel for understanding properly  the $f_0(980)$ 
was stressed in Ref.~\cite{oller99prd}, because once this channel is considered the inclusion of a bare 
single state around 1~GeV is required \cite{oller99prd,oller08prl,mixing}. 
 The $f_0(980)$ pole in fact moves continuously to the bare state around 1~GeV if the $g_0^0(s)$ matrix is removed, 
e.g. by multiplying it with a factor $\lambda\in [0,1]$ and taking $\lambda\to 0$.  But let us remark that 
 the strong $K\bar{K}$ interactions give rise to a bound state 
close to the actual mass of this resonance \cite{oller97,isgur,wein,speth,alar1}, around the $K\bar{K}$ threshold.

 The  $f_0(1370)$ resonance owns its origin  to the octet scalar 
$\sigma_8$ with the bare mass of 1370~MeV, as already noticed in Ref.~\cite{oller08prl}.
These bare poles gain their widths through the unitarization procedure.  
 The $f_0(1370)$ couples most strongly with the $\eta\eta'$ channel.
However,  a deeper study of the $f_0(1370)$ resonance requires to include explicitly the $4\pi$ channel \cite{oller08prl}. 
     
These conclusions are further supported by the $N_C$ trajectories of those poles, 
which we will discuss later. 

\subsubsection*{ $IJ = \frac{1}{2}\, 0$}
For these quantum numbers we obtain two poles corresponding to the $\kappa$ and $K^{*}_0(1430)$ resonances. 
 The most relevant Riemann sheets correspond to $(-,+,+)$ and $(-,-,+)$ respectively, with poles 
at $\sqrt{s} = (665^{+9}_{-9} - 268^{+21}_{-6}\,i)$~MeV and $\sqrt{s} = (1428^{+56}_{-23} - 87^{+53}_{-28}\,i )$~MeV, in order. 
Their pole positions are compatible with those in the PDG \cite{pdg}. For the $\kappa$ resonance the pole position $(658\pm 13 -279\pm 12\,i )$~MeV is found in Ref.~\cite{descotes06} from a  Roy-Steiner representation of $K\pi$ scattering, also in good agreement with our result.
 In the fourth Riemann sheet, i.e with $(+,-,+)$, a shadow pole at  
$\sqrt{s} = (717^{+8}_{-5} - 280^{+30}_{-24}\,i)$~MeV is also found for the $\kappa$ resonance. This resonance couples most strongly with the $K\pi$ channel, although it has also a large coupling to the other channels. In turn, the $K^*_0(1430)$ has its largest 
coupling to the $K\eta'$ channel, which has its threshold quite close to the resonance mass.

We also find a good agreement with the previous 
study of Ref.~\cite{jamin} that shares for this channel many facts in common with ours. 
The $\kappa$ resonance, similar to the case of $\sigma$, is generated mainly from 
the pseudo-Goldstone interactions. The $K^{*}_0(1430)$ pole originates from the 
bare octet of scalar resonances in the Lagrangian Eq.~\eqref{lagscalar}. Again, these 
conclusions are further supported below when discussing the $N_C$ dependence of the pole positions.

\subsubsection*{ $IJ = 1\, 0$} 
This is the most problematic channel in our analysis.  
By imposing the $U(3)$ symmetry on the subtraction constants, 
i.e. with the constraint of 
$a_{ SL}^{1 0,\pi\eta} = a_{ SL}^{1 0,K\bar{K}}=a_{ SL}^{1 0,\pi\eta'}$, 
no good fit for this channel could be obtained simultaneously with the other data. 
Within a reasonable range for the $\pi\eta$ subtraction 
constant the best fit prefers positive values, in contrast with the negative ones 
for the other channels, though the error for this parameter is large. However, we notice that, similarly to previous studies \cite{oller97,oller99prd}, 
if only the tree level amplitudes are taken for the interacting kernel one can fit well the same 
data and get  a reasonable pole for the $a_0(980)$ resonance on the second sheet. 
Also a similar negative value for the $\pi\eta$ subtraction constant results then as in the other channels,  
with the constraint $a_{ SL}^{1 0,\pi\eta} = a_{ SL}^{1 0,K\bar{K}}=a_{ SL}^{1 0,\pi\eta'}$ \cite{oller99prd}. 
We have checked too that the influence of the $\pi\eta'$ channel to the mass distribution 
of $\pi\eta$ Eq.~\eqref{a0.inv} is actually  tiny. In addition, no two-body scattering data are available for this channel which also 
reduces its statistical weight in the fits performed.

We show in Table~\ref{tab:pole} the $a_0(980)$ and $a_0(1450)$ pole positions and residues.  
For $a_0(980)$, the most relevant sheet should correspond to the one with $(-,+,+)$ or $(-,-,+)$. 
However with our current best fit, no reasonable poles are found in those Riemann sheets. 
Nevertheless, a pole located in the fourth sheet, i.e. with $(+,-,+)$,  at $\sqrt{s}=(1012^{+25}_{-7}-16^{+50}_{-13}\,i)$~MeV appears. This rather indirect location of the pole associated with  the $a_0(980)$ resonance indicates that this resonance 
is to a large extent a dynamical cusp effect within our approach. It couples mostly with the $K\bar{K}$ channel as shown in 
Table~\ref{tab:pole}.

The most relevant Riemann sheet for the $a_0(1450)$ corresponds to the one with $(-,-,-)$, since 
 its mass is larger than the $\pi\eta'$ threshold, where one has the pole $\sqrt{s}=(1368^{+68}_{-68}-71^{+48}_{-23}\,i)$. 
Other channels are lacking at the typical energies for the $a_0(1450)$, 
as listed in the PDG \cite{pdg} for its decay widths. 
This lack of decay channels (e.g. $\omega\pi \pi$ and $a_0(980)\pi\pi$) 
explains why our pole position for this resonance  gives a smaller width than in \cite{pdg}, though the mass 
is compatible within errors. In our approach this resonance dues its origin to the bare octet of scalar 
resonances in the Lagrangian of Eq.~\eqref{lagscalar}.

\subsubsection*{ $IJ = 1\, 1$, $\frac{1}{2}\, 1$ and $01$ } 

For the vector channels we find poles corresponding to the $\rho(770)$ and $K^*(892)$ in their respective 2nd Riemann sheets at 
$\sqrt{s}=(762^{+4}_{-4}-72^{+2}_{-2}\,i)$~MeV and $\sqrt{s}=(891^{+3}_{-4}-25^{+2}_{-1}\,i)$~MeV, respectively. 
 Data are reproduced in a straightforward way, though we have to distinguish between the bare $\rho$ and $K^*$ masses 
with fitted values given in Eq.~\eqref{fitresult}. The quality of the fit is the same regardless we free their 
subtraction constants or fix them to the scalar channels. This indicates that unitarity effects 
are not so important, though they provide the right widths to the bare poles \cite{oller99prd}.  
The masses and widths from the resonance poles agree well with the PDG values~\cite{pdg}.

In addition, we have another pole corresponding to the $\phi(1020)$ resonance in the vector isoscalar channel. We obtain a width for this resonance to two kaons of around 4~MeV, which is close to the experimental partial decay width of the $\phi(1020)$ to this decay channel of 3.5~MeV \cite{pdg}. Nonetheless, another important decay channel for the $\phi(1020)$ is the $ \pi \pi \pi$ state which is not considered in our approach.


\subsection{$N_C$ trajectories of the resonance poles }

As discussed above in the Introduction, one of the important improvements of our current work is to take 
into account the $N_C$ dependence of the pseudo-Goldstone boson  masses  when discussing the 
$N_C$ trajectories for the resonances. This may potentially cause some significant effects
because in the large $N_C$ limit the $U_A(1)$ anomaly disappears and the $\eta'$ becomes also a 
pseudo-Goldstone boson \cite{ua1innc}. 

Again we clarify that when fitting our theoretical 
formulas with the experimental data, all the masses of the  pseudo-Goldstone bosons are taken from 
PDG~\cite{pdg}, which we summarize in the Appendix~\ref{appendix-inputs} for completeness. 
Only when discussing the $N_C$ trajectories we start to use our prediction for the pseudo-Goldstone masses. 
The leading order mass parameters $\overline{m}_\pi = 2 B m_q $ and $\overline{m}_K = B(m_q + m_s)$ 
in Eq.(\ref{lolagrangian}),  do not vary with $N_C$ because $B\sim {\cal O}(N_C^0)$. This follows from the 
 expression for the quark condensate in the chiral limit from Eq.~\eqref{lolagrangian}, 
$\langle 0|\bar{q}^iq^j|0\rangle =- F^2 B\delta^{ij}$, taking into account that both the 
quark condensate and $F^2$ are proportional to $N_C$.  
 The bare masses are fixed in terms of the physical masses of the pion and kaon by employing the expressions 
given in Eqs.~\eqref{mpi2.sig} and \eqref{mk2.sig}, with the resulting values
\begin{eqnarray} 
\label{pikbmass}
\overline{m}_\pi= 139.5^{+4.4}_{-4.6}~{\rm MeV}\,, \qquad 
\overline{m}_K  = 519.6^{+12.0}_{-7.5}~{\rm MeV}\,.
\end{eqnarray}
We summarize here the leading $N_C$ scalings of the remaining parameters entering in our equations \cite{kaiser00,ecker89npb} 
\begin{eqnarray}\label{leading_nc_run}
&&\Lambda_2 \sim \frac{1}{N_C}\,, \quad  M_0 \sim \frac{1}{\sqrt{N_C}}\,, \quad
\big\{ c_d\,, c_m\,, \tilde{c}_d\,, \tilde{c}_m\,, G_V \big\} \sim \sqrt{N_C}\,,
\nonumber \\  &&
\big\{ M_{S_1}\,, M_{S_8}\,, M_{\rho}\,, M_{K^*}\,, M_{\omega}\,, M_{\phi}\,, a_{ SL} \big\} \sim \cO(1)\,.
\end{eqnarray}
As commented above  the $U(3)$ large $N_C$ relations between the singlet and octet scalar couplings \cite{ecker89npb}
\begin{equation}
\tilde{c}_d = \frac{c_d}{\sqrt3}\,,\quad \tilde{c}_m = \frac{c_m}{\sqrt3}\,,
\end{equation}
are well fulfilled by our fitted values in Eq.~\eqref{fitresult}, while the mass relation $M_{S_1} = M_{S_8}$ 
is affected by  a $30\%$ variation, still within expectations for a leading large $N_C$ prediction. 

The weak pion decay constant $F_\pi$ deserves a special attention because we can also calculate the subleading term in  $1/N_C$. The piont is that we have a $\delta$-expansion for this important parameter in terms of $F$, the pseudoscalar weak decay constant in the chiral and large $N_C$ limit appearing in the chiral Lagrangians, and that scales as $\sqrt{N_C}$. 
The relation between $F_\pi$ and $F$ is given  in Eq.~\eqref{ffpi}. From this expression the subleading $1/\sqrt{N_C}$ contribution to the running with $N_C$ of  $F_\pi$, is determined. This is, of course, a specific feature of employing $U(3)$ $\chi$PT with its associated $\delta$-expansion.

\begin{figure}[ht]
\begin{center}
\includegraphics[angle=0, width=0.7\textwidth]{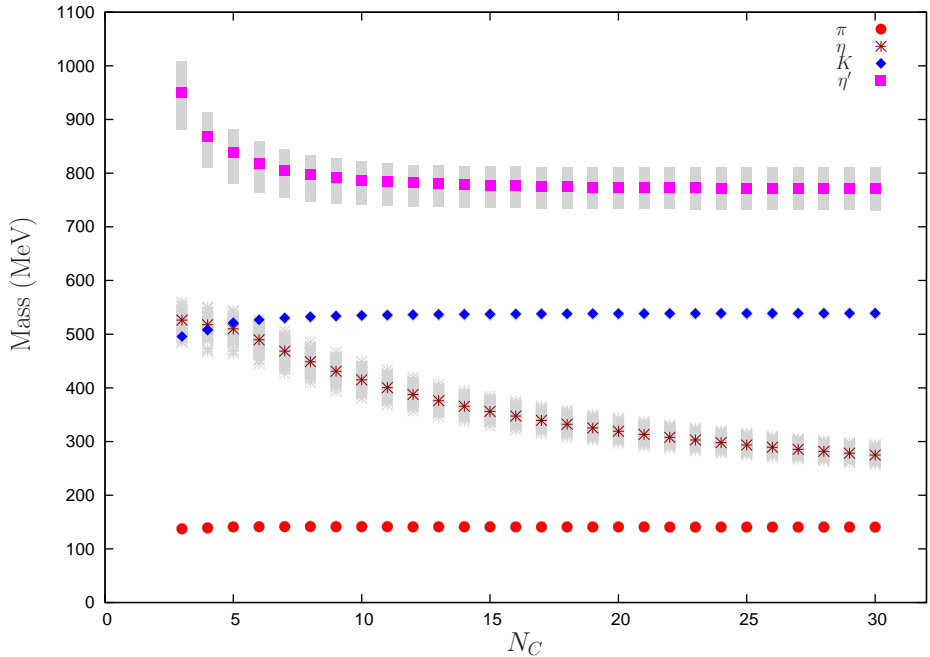}
\caption{(Color online.) Masses for $\pi$, $K$, $\eta$ and $\eta'$ as a function of $N_C$  from 3 to 30 with  one unit step. 
The  different points are obtained by using the best fit given in Eq.(\ref{fitresult}) 
and the shadowed regions correspond to the error bands. Squares (magenta) are for $\eta'$, diamonds (blue) for $K$, 
bursts (brown) for $\eta$ and circles (red) for $\pi$.  }
 \label{fig-nc-mass-err}
 \end{center}
 \end{figure}

\begin{figure}[ht]
\begin{center}
 \includegraphics[angle=0, width=.65\textwidth]{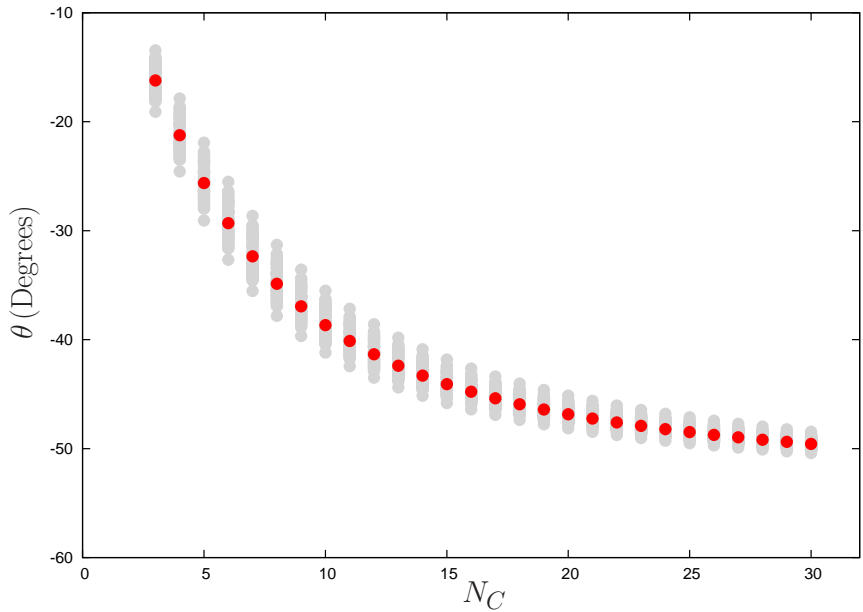}
 \caption{(Color online.) Leading order $\eta-\eta'$ mixing angle $\theta$ Eq.~\eqref{deftheta0} as a function of $N_C$  from 3 to 30 with  one unit step. 
Similar to Fig.~\ref{fig-nc-mass-err}, the circle (red) points are from  
the best fit given in Eq.(\ref{fitresult}) 
and the shadowed area corresponds to the error bands. }
 \label{fig-nc-theta-err}
 \end{center}
 \end{figure}

 All the parameters in Eq.~\eqref{leading_nc_run}, except  $a_{SL}$, are present in chiral Lagrangians, namely 
in Eqs.~\eqref{lolagrangian},  \eqref{lagscalar}, \eqref{lagvector}, \eqref{kinerv}, \eqref{kiners} and \eqref{deflam2}. It is important to notice 
 that by construction, since $U(3)$ $\chi PT$ \cite{oriua,herrera97,kaiser00} is a combined $1/N_C$ and chiral expansion, 
every coefficient multiplying a monomial of the fields does not contain 
any extra subleading piece in the $1/N_C$ and chiral quark mass expansion. These extra terms would be part of higher order monomials in   the $\delta$-expansion. However, when fitting data, the actual numerical values in Eq.~\eqref{fitresult} certainly reabsorb {\it de facto} higher order contributions in this expansion. This is also the case for the subtraction constant $a_{SL}$.

We first consider the dependence on $N_C$ of our results by taking the leading scaling with $N_C$ of the parameters in Eq.~\eqref{leading_nc_run} (but keeping the full $N_C$ dependence of $F_\pi$ from Eq.~\eqref{ffpi}, as follows from our calculation.) Later, when considering the properties of the $\sigma$ meson, we will also discuss variations in our results by considering subleading terms in the running with $N_C$ for many of the parameters in Eq.~\eqref{leading_nc_run}.

Our full result for the $N_C$ dependence of the masses of $\pi$ Eq.~\eqref{mpi2.sig}, $K$ Eq.~\eqref{mk2.sig}, $\eta$ and $\eta'$ Eq.~\eqref{mixnlo}, as well as  
the leading order mixing angle $\theta$ Eq.~\eqref{deftheta0}, are shown in Figs.~\ref{fig-nc-mass-err} 
and \ref{fig-nc-theta-err}, respectively. 
 The most streaking fact is the reduction by more than a factor of 2 of the mass of the $\eta$ with $N_C$. The mass of the 
$\eta'$ also diminishes significantly. The $K$ and $\pi$ masses vary little, specially the latter. 
 For large $N_C$  all the masses of the  
pseudo-Goldstone bosons $\pi\,, K\,,\eta$ and $\eta'$ go to zero in the chiral limit. 
 However, for non-vanishing quark masses the 
$\eta'$ meson still gains a relatively large mass in the large $N_C$ limit while the $\eta$ becomes 
similarly light as the pion  \cite{weinberg75}. This can be easily 
understood by looking at the leading order prediction in the large $N_C$ limit for the pseudoscalar masses,
\begin{eqnarray}
\overline{m}^2_\eta &=& \overline{m}_\pi^2\,, \qquad \nn\\
\overline{m}^2_{\eta'} &=& 2 \overline{m}_K^2 - \overline{m}_\pi^2\,,
\end{eqnarray}
where $\overline{m}_\eta$ and $\overline{m}_{\eta'}$ stand for the
masses of the $\eta$ and $\eta'$ mesons, corresponding to the notations of 
$m_{\overline{\eta}}$ and $m_{\overline{\eta}'}$ in 
Eqs.(\ref{defmetab2}) and (\ref{defmetaPb2}) with $M_0 \to 0$. 
 Now, taking into account the values  given in Eq.~\eqref{pikbmass} for $\overline{m}_\pi$ and $\overline{m}_K$ 
we then end with the following prediction for the  leading order masses of 
$\eta$ and $\eta'$ in the large $N_C$ limit  
\begin{eqnarray} 
\overline{m}_\eta     = 139.5^{+4.4}_{-4.6} ~{\rm MeV}\,, \qquad 
\overline{m}_{\eta'}   = 721.5^{+17.4}_{-11.1} ~{\rm MeV}\,. 
\end{eqnarray} 
This leading order result already explains  qualitatively  the $N_C$ behaviors 
for the masses of $\eta$ and $\eta'$ mesons shown in Fig.~\ref{fig-nc-mass-err}.

For the leading order mixing angle $\theta$, its $N_C$ dependence is dominantly 
governed by the $U_A(1)$ anomaly mass  $M_0$, as one can see from Eq.(\ref{deftheta0}). 
It is also easy to demonstrate that in the large $N_C$ limit, i.e. $M_0 = 0$, 
the leading order mixing of $ \eta-\eta'$ turns out to be the 
ideal mixing: $\theta = -54.7^\text{{\tiny{o}}}$, as it should.

It is worth stressing   that  previous works discussing the $N_C$ behaviors of 
resonances, such as Refs.\cite{pelaez04,sun07}, 
directly identify the $\eta$ meson with $\eta_8$, since it is based on 
$SU(3)$ $\chi$PT. The $\eta_1$ is considered implicitly through higher order counterterms,  as any other heavy field in $\chi$PT. In addition, all of the 
pseudo-Goldstone masses are not 
changed in the variation of $N_C$ \cite{pelaez04,pelaez06prl,sun07,zhou11prd}. However, as we have just shown, the change  with $N_C$ for the $\eta$ mass 
 is very pronounced. 

Next we discuss our findings for the $N_C$ dependence of the resonance 
properties channel by channel. As a clarifying remark let us mention that the following discussions 
are based mainly on the resonance poles appearing in the most relevant Riemann sheets for the 
energy region under discussion, as 
 previously elaborated, unless a specific statement is given. 

\subsubsection*{Poles in the $IJ = 0 0$ channel}

\begin{figure}[ht]
 \begin{center}
 \includegraphics[angle=0, width=.95\textwidth]{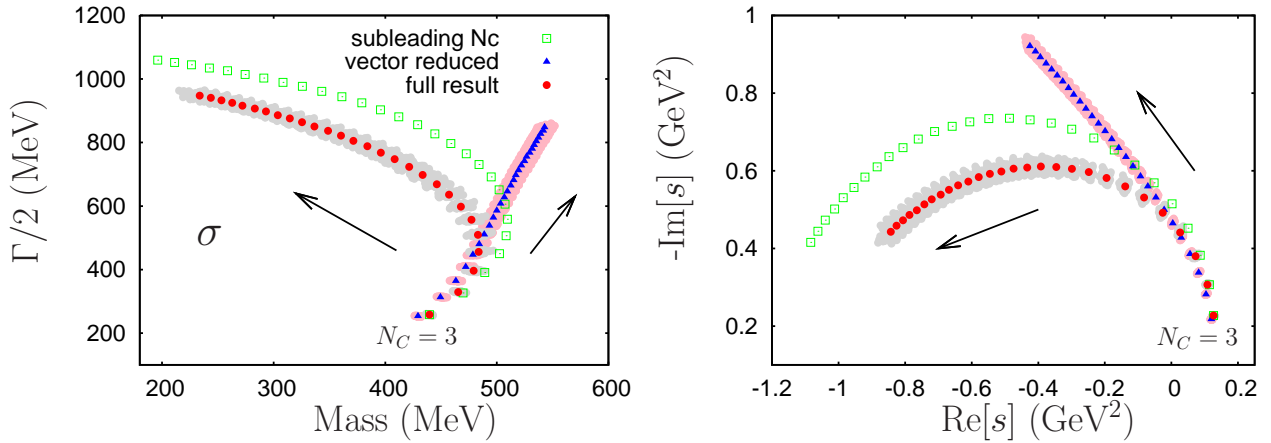}
 \caption{ \small (Color online.) Dependence of the $\sigma$ pole position, $s_\sigma$ as a function of $N_C$ from 3 to 30 with one unit step. In the left panel we show  $\sqrt{s_\sigma}$ and in the right one  $s_\sigma$. 
The {\it full result} (circle in red) corresponds to our calculation without any approximation. 
The points labeled as  {\it vector  reduced} (triangle in blue) are obtained by keeping only the leading local 
terms from the vector resonance propagators. The empty squares (in green) correspond also to our full result but taking 
into account subleading terms in the running of $N_C$ for the resonance couplings, according to the model of Eq.~\eqref{run1}.   
The shadowed areas  enveloping the circles and the triangles show the error bands of our full result and {\it vector reduced} approximation.}
 \label{fig-nc-sig}
 \end{center}
 \end{figure}

\begin{figure}[H]
 \begin{center}
 \includegraphics[angle=0, width=0.9\textwidth]{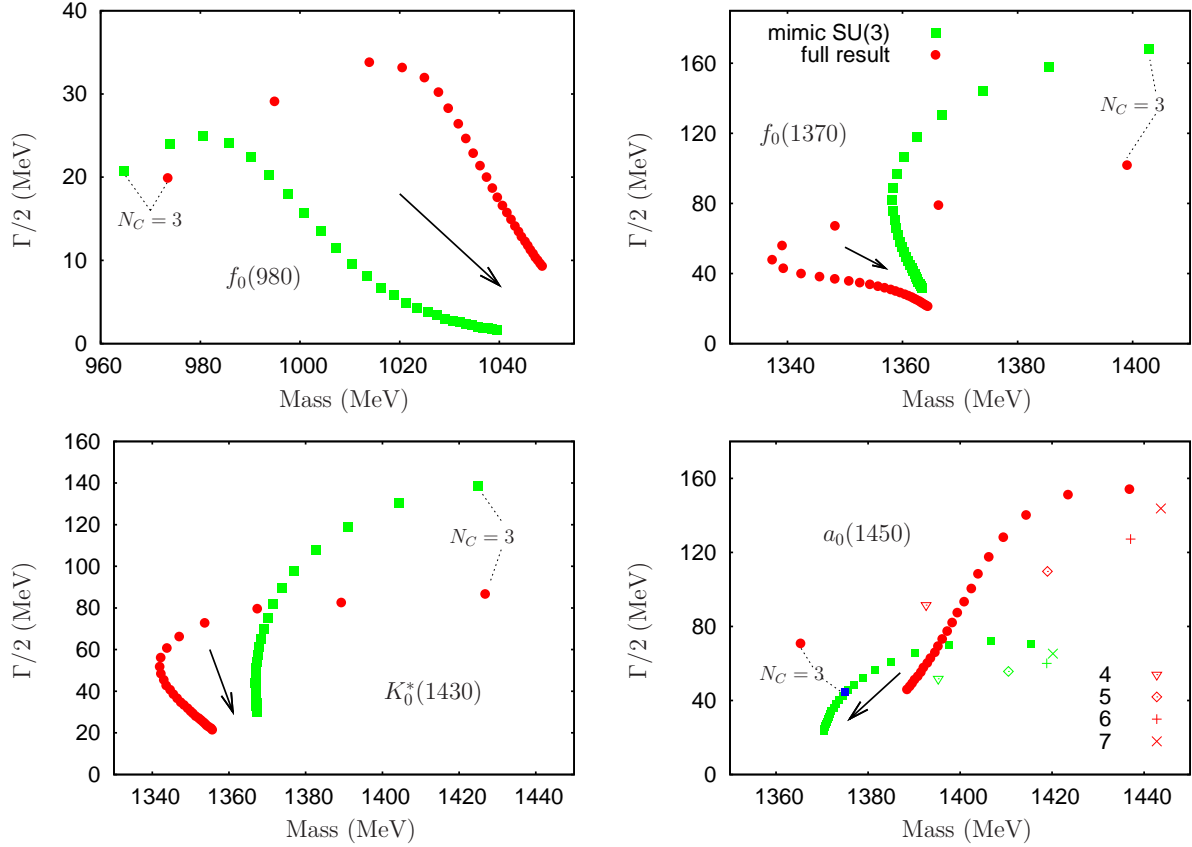}
 \caption{(Color online.) From top to bottom and left to right: Dependence on $N_C$ from 3 to 30 with one unit step for the pole positions of 
the $f_0(908)$, $f_0(1370)$, $K^*_0(1430)$ and $a_0(1450)$, respectively. The notation is the same as that in Fig.~\ref{fig-nc-sig}. 
For the last panel dedicated to the $a_0(1450)$ the $N_C$ points for $N_C\leq 7$ are indicated explicitly for clarifying purposes. 
All the points with the same color correspond to the same framework (except the blue point in the $a_0(1450)$-panel that corresponds 
to the {\it mimic SU(3)} case with $N_C=3$.)}
 \label{fig-nc-s1}
 \end{center}
 \end{figure}

We can easily track the pole trajectories for the $\sigma$, $f_0(980)$ and $f_0(1370)$ resonances while varying 
 $N_C$. We first discuss the case of the $\sigma$ in  Fig.~\ref{fig-nc-sig} and latter we will turn our 
attention to the more massive $f_0(980)$ and $f_0(1370)$ resonances in Fig.~\ref{fig-nc-s1}.

Fig.~\ref{fig-nc-sig} has two panels. In the left one we show the running with $N_C$, starting with $N_C=3$ in one 
unit steps, of the $\sigma$ pole position in the variable $\sqrt{s}$, similarly as done in Refs.~\cite{pelaez04,pelaez06prl,pelaezconf,sun07}, while in the right panel the same is shown employing 
the variable $s$, as in Refs.~\cite{sun05,sun07}.  Our full results correspond to the filled circles. 
 The shadowed areas around every curve are generated by employing  
 the same configurations of  parameters that we exploit before in calculating the 
error bands from our fits in Eq.~\eqref{fitresult} and in Figs.~\ref{fitfig1}-\ref{fitfig3}.
 In this way we obtain the error bands for  the shown $N_C$ trajectories of the $\sigma$ in Fig.~\ref{fig-nc-sig}. 
As one can see from this figure 
the resulting curves 
are quite stable even after taking into account the uncertainties of the inputs. 
 From the left panel we observe that the width increases very fast with $N_C$,  so that already for $N_C=7$ it doubles as 
compared with its value at $N_C=3$. In regards to its mass it first increases with $N_C$  but for $N_C$ above 7
 it decreases. In the variable $\sqrt{s}$ the pole stays deep in the complex plane, as observed already 
in Refs.~\cite{pelaez04,sun05,sun07,arriola1}. However,  there is no point to keep interpreting the imaginary part 
of $\sqrt{s_\sigma}$ at the pole position as one half of the width when the mass is by far much smaller. The correct interpretation, as pointed out in Refs.~\cite{sun05,sun07},  is obtained by considering the pole 
trajectory in the $s$ complex plane: the mass square of the $\sigma$ resonance becomes real negative.

In Refs.~\cite{sun07,arriola1}  the large $N_C$ limit of the one-loop IAM  is studied in the chiral limit and the following condition in terms of the ${\cal O}(p^4)$ $SU(3)$ chiral counterterms $L_2$ and $L_3$ \cite{gasserleu} is obtained
\begin{align}\label{l2l3comb}
25L_2+11 L_3
\left\{
\begin{array}{ll}
> 0~, & s_\sigma ~\text{approaches positive real}~s~\text{axis}~,\\
=0~,  & s_\sigma ~\text{moves to }\infty~,\\
< 0~, & s_\sigma ~\text{approaches negative}~s~\text{axis}~. 
\end{array}
\right.
\end{align}
Making use of resonance saturation \cite{ecker89npb} 
\begin{align}
L_2&=\frac{G_V^2}{4M_V^2}~,\nn \\
L_3&=-\frac{3G_V^2}{4M_V^2}+\frac{c_d^2}{2 M_{S_8}^2}~,
\label{satura}
\end{align}
from the fitted values of Eq.~\eqref{fitresult}, and taking $M_V$ as $M_\rho$, we have 
 \begin{align}
25L_2+11 L_3=-11.2^{+0.8}_{-0.6}\times 10^{-3}<0~.
\end{align}
Thus, our curve with circles for $s_\sigma$ in the right panel of Fig.~\ref{fig-nc-sig}, bending towards  
negative large real values of $s$, is in agreement with the condition Eq.~\eqref{l2l3comb} obtained in 
Ref.~\cite{sun07}.  

Compared with the one-loop IAM result of Ref.~\cite{sun07} our curve bends faster to the left in the $s$-complex plane.  
It is interesting to point out that we can give rise to closer curves to those of Refs.~\cite{pelaez04,sun07}  if we freeze out the full propagators of 
the vector resonances exchanged in the crossed channels and only 
take the leading local terms generated by them.  This in fact corresponds to integrating out the vector resonance states in our theory 
and keeping only the leading contributions in the low energy sector, which are $\cO{(p^4)}$. These 
terms are the ones incorporated in the one-loop $\chi$PT calculation used in NLO IAM Refs.\cite{pelaez04,sun07}, but not the higher order ones that arise by keeping the full vector resonance 
propagators that enter in our full calculation.
The corresponding trajectory within this approximation is shown in Fig.~\ref{fig-nc-sig} by the triangles, 
labeled as {\it vector reduced} throughout. Notice that now the mass of the $\sigma$ keeps increasing with $N_C$, and the 
resulting trajectory is quite similar  to that obtained in Ref.~\cite{pelaez04}, particularly for 
 $N_C\lesssim 10$.  Let us remark also that Refs.~\cite{pelaez04,pelaezconf}, within their estimated uncertainties, also have $\sigma$ pole trajectories in the $\sqrt{s}$ variable with decreasing mass (all of them has increasing width as our case.) 

One could also interpret the discrepancy between the circles and triangles in Fig.~\ref{fig-nc-sig} arguing  that higher order effects in the expansion employed  could potentially have a large impact on the results. Let us first note that this criticism,  from what we explicitly showed (the comparison between the triangles and circles in  Fig.~\ref{fig-nc-sig}), applies strictly to the IAM results.  The higher orders included in our approach give rise to a large  variation  with respect to the results obtained when only their local reduction is kept, as done in the IAM. 
Second, from this one could of course infer the  possibility that higher orders not considered in our approach, as well in any of other unitarization methods, could give rise to quite different $\sigma$ pole trajectories for large enough $N_C$. This is further discussed below, see also Ref.~\cite{pelaezconf}.  

In contrast, to integrate out or  keep the full contributions of the bare scalar resonance exchanges from ${\cal L}_S$ 
Eq.~\eqref{lagscalar}  does not change the $N_C$ trajectory of $\sigma$ in a significant way. This also indicates 
that the bare scalar nonet has little influence on the $\sigma$ pole, as already pointed out in Ref.~\cite{oller99prd}. There it was 
stressed that the $\sigma$, $\kappa$ and $a_0(980)$  originate independently of whether the bare scalar singlet and octet resonances are included in the formalism. 
 This is also  understandable by looking at how the resonances contribute 
to the LECs in $\chi$PT. For example, $L_1\,,L_2$ in $SU(3)$ $\chi$PT are purely 
contributed by the vector resonances, which also  dominate 
$L_3$ \cite{ecker89npb}. For $L_4\,,L_5\,,L_6\,,L_8$, they are 
dominated by the scalar resonances~\cite{ecker89npb}. Precisely in the $\pi\pi \to \pi\pi$ process, 
the factors accompanying $L_4\,,L_5\,,L_6\,,L_8$ are always proportional 
to $s\,m_\pi^2$, $t\,m_\pi^2$ or $m_\pi^4$, which are less important in the resonance region 
than the terms $L_1\,,L_2\,,L_3$ that are accompanied by $s^2$, $s t$ or $t^2$.  Indeed, in the chiral limit 
only the contributions to $\pi\pi$ scattering proportional to $L_1$, $L_2$ and $L_3$ survive and with those 
LECs $\sigma$ and $\rho$ were perfectly generated in Ref.~\cite{sun07,arriola1}. 

To conclude we can say that our results for the trajectory of the $\sigma$ resonance pole 
are in qualitative agreement with those from the one-loop IAM \cite{pelaez04,sun07,arriola1} and differences 
can be understood in terms of the higher orders included by using full resonance propagators for the 
crossed vector exchanges. 
 
Indeed, one-loop IAM is a particular case of our approach that results by expanding\footnote{This is also derived in Ref.~\cite{higgs}.} $N^{-1}={N^{(2)}}^{-1}(1-N^{(4)} {N^{(2)}}^{-1}+\ldots)$. Then 
from Eq.~\eqref{defnd} one has
\begin{align}
T=N^{(2)} \left(N^{(2)}-N^{(4)}+N^{(2)} g N^{(2)}\right)^{-1}N^{(2)}~.
\end{align}
In the previous expression the chiral order  is indicated by the superscripts. For simplicity in  the notation we have removed the subscript $J$ and superscript $I$ present in Eq.~\eqref{defnd}. The derivations apply to any specific partial wave. Now, reducing Eq.~\eqref{defnd} to ${\cal O}(p^4)$ and applying resonance saturation one has that 
\begin{align}
N^{(2)}&=T^{(2)}~,\nn\\
N^{(4)}&=T^{(4)}+N^{(2)} g N^{(2)} ~,\nn\\
T&=T^{(2)}\left[T^{(2)}-T^{(4)}\right]^{-1}T^{(2)}~,
\end{align}
which is the one-loop IAM result.

However, our  results are quite different from those obtained in Refs.~\cite{sun07,pelaez06prl} by employing the  
IAM  with a  two loop  $SU(2)$ $\chi$PT calculation for $\pi\pi$ scattering,  where for large $N_C$ the $\sigma$ pole falls 
down to the positive real $s$-axis at around 1~GeV$^2$.  
 Indeed, as discussed in more detail below, we also obtain a pole in the large $N_C$ limit at $\sqrt{s}=M_{S_1}\simeq 1$~GeV, 
but it comes from the bare singlet state which, for $N_C=3$, is part of the $f_0(980)$ resonance. However, 
the $ \sigma$ pole is not affected by its presence in a first approximation and this is not the reason for the different
 $\sigma$-pole trajectory obtained by us in comparison with Refs.~\cite{sun07,pelaez06prl}. 

In order to understand better this discrepancy with the two-loop IAM calculations \cite{pelaez06prl,sun07} let us  discuss the effects of several contributions particular 
to our approach. The first one corresponds to the influence of having  $U(3)$ $\chi$PT.  Compared with  $SU(3)$ $\chi$PT, where $\eta$ is identified with the octet $\eta_8$ 
and $\eta_1$ is only implicitly included at ${\cal O}(p^4)$ through the LEC $L_7$ , the present discussion 
incorporates the explicit contributions from the physical states of $\eta$ and $\eta'$ within $U(3)$ 
$\chi$PT with the  $\delta$ counting.  As a result we take into account the drastic reduction 
of the $\eta$ mass with $N_C$ as discussed above, see Fig.~\ref{fig-nc-mass-err}. However, since the couplings of the $\sigma$  to  
$\eta\eta$, $\eta\eta'$ and $\eta'\eta'$ are weak in our calculation (but not necessarily in others \cite{SNVeneziano89,SN98}), Table~\ref{tab:pole}, one should then expect small effects on the 
$\sigma$-pole trajectory from this more exhaustive treatment of the $\eta$ and $\eta'$ pseudoscalars in $U(3)$ $\chi$PT.\footnote{The weak couplings of the $\sigma$ to states with $\eta$ and $\eta'$  is kept when varying $N_C$ as we have 
explicitly checked, even though  the $\eta$ mass decreases significantly with $N_C$, as shown in Fig.~\ref{fig-nc-theta-err}.}
To make this statement more quantitative it is interesting  to find a way to make closer 
our approach to the  $SU(3)$ one and then compare. This can be achieved by imposing the following conditions  when varying $N_C$:  (i) 
Freezing  the mixing angle $\theta$ at leading order in Eq.(\ref{deflomixing}) to zero, together with the higher order mixing 
parameters in Eq.(\ref{defmixingpara}). (ii) The masses  of the $\eta$ (then  $\eta_8$), $\pi\,, K$ are fixed to their 
physical values and do not vary with $N_C$. (iii) The $\eta_1$ mass is also fixed by using the leading order result from 
Eq.(\ref{lolagrangian}) without mixing, which gives the value 1040 MeV. 
 In this way, the $\eta$ meson in $U(3)$ $\chi$PT 
resembles the $\eta_8$  in  $SU(3)$ $\chi$PT  and the $\eta_1$ only appears in the scattering amplitudes 
involving $\pi\,, K\,, \eta$ mesons through the loops, which is suppressed by $1/N_C$. 
This mimics the role of $L_7$ in $SU(3)$ $\chi$PT~\cite{pelaez04}, although one 
needs to be careful about the $N_C$ counting for $L_7$~\cite{peris95plb}. The corresponding $s_\sigma$ $N_C$ trajectory within this approximation, named as {\it mimic SU(3)}, is so similar to our full results in Fig.~\ref{fig-nc-sig} that we do not show it 
explicitly. Thus, for the $\sigma$ case differences that arise by having used the $U(3)$ $\chi$PT in our approach are not significant and are not certainly  responsible for the differences with respect to the two-loop IAM results \cite{sun07,pelaez06prl}. However, 
the situation could change dramatically for other resonances, e.g. for the $f_0(980)$ and $f_0(1370)$, and in general for any other resonance that had  large  couplings to states including the $\eta$ and $\eta'$ mesons.   We also want to point out that, by the same token, our study clarifies that previous results, e.g. those from Refs.~\cite{pelaez04,sun07,pelaez06prl}, are stable under the explicit inclusion of the $\eta_1$ singlet when increasing $N_C$.

Another issue to be considered is the fact, pointed out repeatedly in Refs.~\cite{pelaezconf} regarding the IAM, that 
the large $N_C$ limit is a weakly interacting limit while the phenomenological success of unitarization methods, e.g. ours, typically rests on resumming infinite string of diagrams which are particularly strong.  In our approach two types of resummations are 
considered simultaneously: (i) At the tree-level, by employing explicit resonance fields, so that infinite 
local terms are resummed by taking vector and scalar bare resonance exchanges. (ii) At the loop level, by resumming the 
RHC. The former resummation is leading in large $N_C$ while the latter is subleading. Point (i) is the most important for studying 
the vector resonances, while point (ii) dominates for the scalar ones in real life \cite{oller99prd,pelaezconf,arriola1}.  

Concerning the point (i), let us illustrate it by considering tree-level $IJ=11$ $\pi\pi$ scattering from  the leading order Lagrangian in 
Eq.(\ref{lolagrangian}), required 
by current algebra, and the vector resonance Lagrangians in Eqs.(\ref{lagvector}) and (\ref{kinerv}). It is well known the dominant role of 
the $\rho$ resonance in  $IJ=11$ $\pi\pi$ scattering at low and intermediate energies. These tree-level contributions are the ones 
that survive in the large $N_C$ limit. The corresponding amplitude is 
\begin{align}\label{input.ex}
T_{1 }^{1 \,\pi\pi \infty}(s) =& \, \frac{s-4m_\pi^2}{6 F_\pi^2 }  + \frac{G_V^2 s (s-4m_\pi^2)}{ 3 F_\pi^4(M_\rho^2 - s ) } 
+ \frac{G_V^2}{6F_\pi^4 (s-4m_\pi^2)^2} \bigg\{ 
(4m_\pi^2 - s )\big[ 16 m_\pi^4 + s^2 - 8 m_\pi^2(6M_\rho^2 + s ) 
\nonumber \\ &
+ 12 M_\rho^4 + 24 M_\rho^2 s  \big] + 6 M_\rho^2(4m_\pi^2 -s -2M_\rho^2) (2s + M_\rho^2-4m_\pi^2)\log{\frac{M_\rho^2}{s + M_\rho^2 - 4m_\pi^2}}
\bigg\} \,, 
\end{align}
where the first term in the right hand side of the above equation is from the current algebra in Eq.(\ref{lolagrangian}), 
the second term corresponds to the $s$-channel exchange of a $\rho$ resonance and the last term is contributed by 
the $\rho$-exchanges in crossed channels, $t$ and $u$.

Let us now calculate the local terms at ${\cal O}(p^2)$,  $T_{1 }^{1 \,\pi\pi \infty}(s)_2$, and ${\cal O}(p^4)$, $T_{1 }^{1 \,\pi\pi \infty}(s)_4$ 
corresponding to the chiral expansion of Eq.(\ref{input.ex})   at low energy. We  have for these terms: 
\begin{align}
T_{1 }^{1 \,\pi\pi \infty}(s)_2&= \frac{s - 4m_\pi^2}{6F_\pi^2}~,\nn\\
T_{1 }^{1 \,\pi\pi \infty}(s)_4&=\frac{  s (s - 4m_\pi^2) G_V^2 }{2 F_\pi^4 M_\rho^2}~.
\label{ord2.4}
\end{align} 
An interesting task is to apply the IAM at NLO \cite{truong,dobado,dobado97,sun07,arriola1} in order to 
reconstruct the full $T_1^{1\,\pi\pi\infty}(s)$ amplitude from the contributions in Eq.~\eqref{ord2.4}. It results 
\begin{align}
\frac{T_1^{1\,\pi\pi\infty}(s)_2^2}{T_1^{1\,\pi\pi\infty}(s)_2-T_1^{1\,\pi\pi\infty}(s)_4}&= 
-\frac{s- 4m_\pi^2}{18 G_V^2 }\frac{M_\rho^2}{ s-\frac{F_\pi^2 M_\rho^2}{3 G_V^2} }~.
\label{iam.tree}
\end{align} 
Then, we see that only when 
\begin{equation}\label{ksrfnew}
G_V = \frac{F_\pi}{\sqrt{3}}
\end{equation}
is fulfilled, one can reproduce the exact input $s = M_\rho^2$.  
The relation in Eq.(\ref{ksrfnew}) is nothing but the so-called KSRF relation~\cite{ksfr}. 
 Within chiral resonance Lagrangians it was derived originally  in Ref.~\cite{guo2}  by studying $\pi\pi$ scattering and later in 
Ref.~\cite{pich2011} considering the pion vector form factor. We confirm this relation here from a different point of view.  
 Due to the inclusion of the contributions from the crossed channels, the current version differs from the original one 
with $G_V = \frac{F_\pi}{\sqrt{2}}$. Nevertheless if the contributions from the crossed channes are neglected, by dropping the last 
term in the right hand side of Eq.(\ref{input.ex}), we can then recover the original KSRF relation $G_V = \frac{F_\pi}{\sqrt{2}}$ \cite{ecker89plb}

Another look at Eq.~\eqref{iam.tree} can be obtained by employing the results of Ref.~\cite{arriola1} which gives the $\rho$ 
mass square in the chiral and large $N_C$ limit from one-loop IAM at the position 
\begin{align}
s_\rho=-\frac{F_\pi^2}{4L_3}~.
\end{align}
Taking into account Eq.~\eqref{satura} (and keeping the vector contribution proportional to $G_V^2$ which is much larger than the scalar one) 
one then has
\begin{align}
s_\rho=\frac{M_V^2 F^2}{3 G_V^2}~.
\end{align}
Then only when $G_V=F_\pi/\sqrt{3}$ one can recover the exact position of $s = M_V^2 = M_\rho^2$, which is the same as Eq.(\ref{ksrfnew}).

This simple exercise shows that the success of the IAM in the vector channels \cite{arriola1} is intrinsically related with the KSRF relation and 
Vector-Meson-Dominance \cite{sakurai}. This success is not related with having resummed the RHC, as stated confusingly in Refs.~\cite{pelaezconf},
it occurs at the tree-level, and it is then expected to occur at any order in large $N_C$ because loop corrections are even further suppressed with increasing $N_C$. 
 Notice that the benefits of the tree-level IAM resummation of 
Eq.~\eqref{iam.tree} is accomplished in our approach by the explicit exchange of bare vector and scalar resonances. Indeed, 
if $G_V \neq  F_\pi/\sqrt{3} $ the IAM resummation would not be so accurate, while including explicit resonance exchanges is not affected by
 tuning coupling or mass parameters. For an explicit example on this, corresponding to the exchange of a Higgs particle 
in the spontaneous breaking of electroweak symmetry,  see Ref.~\cite{higgs}.

Regarding the RHC resummation, point (ii) above, the situation is the opposite since loops in general, and unitarity loops in particular, become less relevant when $N_C$ increases. In this respect, the situation is more worrisome than for the vector channels since when considering large values of $N_C$ one is testing contributions not relevant in the physical $N_C=3$ case. Within one-loop IAM it was shown in Ref.~\cite{arriola1} that this could imply inconsistencies. In Ref.~\cite{pelaezconf} it is argued that if IAM describes data and resonances within a 10 to $20\%$ errors, this means that the other contributions at $N_C=3$ are not badly approximated. But since meson loops scale as $3/N_C$, while tree-level inadequacies scale as ${\cal O}(1)$, those 10 to $20\%$ errors at $N_C=3$  become 100$\%$ for  values $N_C \sim 30$ and $\sim 15$, respectively. This criticism also applies to our approach because we rely in an expansion for the interacting kernel $N_J^I(s)$,  Eqs.~\eqref{defnd} and \eqref{defnfunction}. In this sense, at least, part of the differences between our results and two-loop IAM should correspond to different treatment of higher order terms in the chiral expansion, that at $N_C=3$ are not so relevant but become more important for large $N_C$. Nevertheless, we think that the final reason why our results differ with respect to the two-loop IAM ones should be deeper, since already from the very start our curve for the full results on the right panel of Fig.~\ref{fig-nc-sig} bends to the left while the analogous one for the two-loop IAM results of Ref.~\cite{sun07}  bends to the right for $N_C\geq 3$. There is also a big qualitative change by passing from one- to two-loop  IAM results (which is not either properly understood yet in the literature and that is related to the issue we are discussing.) Our calculation certainly resembles much more to the one-loop case. 
On the one hand, we employ explicit resonance fields, with its associated tree-level resummation as just discussed. Note that the tree-level resummation for the IAM only generates the full resonance exchanges in the $s$-channel. E.g. taking again the toy model described above the IAM result, Eq.~\eqref{iam.tree}, generates the $\rho$ pole in the $s$-channel but no the cut due to $\rho$-exchanges in the crossed channels. The latter correspond to the 
$\log$ in Eq.~\eqref{input.ex}.  However, on the other hand, standard $\chi$PT two-loop calculations \cite{bij97}, as those used in two-loop IAM \cite{sun07,pelaez06prl} include  more ${\cal O}(p^6)$ operators than those that can be reproduced by expanding the resonance propagators up to this order included in our study \cite{cirigliano06npb} (but let us recall that up to ${\cal O}(p^4)$, and also up to ${\cal O}(\delta)$, we have all of them.) The situation in regards to terms that become more important for the study of  scalar resonances for large $N_C$, but that are not so important at $N_C=3$, is then still  far from being settled.

 Interestingly, in our approach one knows beforehand which is the resonance spectrum for $N_C\to \infty$. The reason is just because in the equation that we use for calculating the $T_J^I$-matrix, Eq.~\eqref{defnd}, $N_J^I(s)^{-1}$ scales as ${\cal O}(N_C)$ while $g(s)$ scales as ${\cal O}(1)$. Then in the $N_C\to \infty$ limit $T_J^I(s)\to N_J^I(s)$ and the latter only contains the poles included explicitly as bare resonances (vector and scalars) from the Lagrangians Eq.~\eqref{lagscalar} and Eq.~\eqref{lagvector}. In this sense, when reproducing data  we are at the same time directly testing an $N_C\to \infty$ spectrum which is known beforehand. However, in the IAM due to the proliferation of higher order counterterms when passing from one- to two-loop order is not so clear which are the resonance poles that  come up from the model for $N_C\to \infty$, since they depend on the specific values of the chiral 
 counterterms.

  Another issue of interest on which we want to elaborate further about its possible implications is the 
assumption of using the leading running with $N_C$ for the parameters in Eq.~\eqref{leading_nc_run}. 
The possible impact of subleading terms in the running of $N_C$ for chiral counterterms was already pointed out in Refs.~\cite{pelaez04,pelaez06prl,pelaezconf,arriola1}.  Refs.~\cite{pelaez04,pelaez06prl,pelaezconf} estimated this uncertainty 
by varying the $\chi$PT renormalization scale in which the counterterms are calculated. These values are then taken 
as the initial ones to engage the simple leading large $N_C$ running. In our case,  
 we include a subleading dependence on $N_C$ on the coupling parameters. We think that for the bare masses this correction should be quite small because, as discussed below, the displacement of the bare resonance mass to its final pole position (at $N_C=3$) is quite short. Since both limits are so close it is then a good approximation to keep them as ${\cal O}(1)$ and not to elaborate further on subleading contributions for the bare masses. For the subtraction constant $a_{SL}^{00}$,  we do not consider any subleading term because at $N_C=3$ it is already a common number, being the same for all the five channels with $IJ=00$ as required by the large $N_C$ $U(3)$ symmetry. The parameters $\Lambda_2\sim 1/N_C$ and $M_0\sim 1/\sqrt{N_C}$ already vanish in the large $N_C$ limit so that subleading terms are not so interesting because they are even further suppressed. Then, their contributions in the large $N_C$ should be marginal compared with those that we keep. Coming back to the resonance couplings in Eq.~\eqref{fitresult} we take into account additional QCD-inspired assumptions of high-energy behavior, such as unsubtracted dispersion relations for the pion electromagnetic form factor \cite{guo2,pich2011} and for the scalar strangeness changing scalar form factor \cite{jamin,jaminff} one has the relations,
\begin{align}
G_V&=\frac{F_\pi}{\sqrt{3}}~,\nn\\
\sum_{i=1}^N c_{d,i}c_{d,i}&=\frac{F_\pi^2}{4}~.
\label{coup.fpi}
\end{align}
In the last equation the sum extends over the set of bare scalar resonances considered. In our approach we only include one scalar octet of resonances, $N=1$. This is the so-called single resonance approximation which is based on the $1/M^2$ suppression for more massive resonances. The single resonance approximation was also employed in Ref.~\cite{arriola1} for studying the $\rho$ and $\sigma$ resonance pole trajectories within one-loop IAM in the chiral limit.
Next we apply the large $N_C$ relations of Eq.~\eqref{coup.fpi} to  the physical case.  
 It follows then 
the scaling with $N_C$:
\begin{align}
\label{run1}
G_V(N_C)&=G_V(N_C=3)\frac{F_\pi(N_C)}{F_\pi(N_C=3)}~,\nn\\
c_d(N_C)&=c_d(N_C=3)\frac{F_\pi(N_C)}{F_\pi(N_C=3)}~,\nn\\
c_m(N_C)&=c_m(N_C=3)\frac{F_\pi(N_C)}{F_\pi(N_C=3)}~.
\end{align}
In the last equation we have assumed the same subleading dependence on $N_C$ for $c_m$ as for $c_d$. Note that $c_m$ is poorly determined by the fit in Eq.~\eqref{fitresult} and it is not excluded that $c_m\simeq c_d$, as used e.g. in Ref.~\cite{jamin}. We also make use of the already discussed $U(3)$ large $N_C$ relations $\widetilde{c}_d=c_d/\sqrt{3}$ and $\widetilde{c}_m=c_m/\sqrt{3}$ so that the scaling with $N_C$ of all the couplings in Eq.~\eqref{fitresult} is driven by $F_\pi$. Similar relationships were employed in Ref.~\cite{arriola1} to deduce constraints between resonance saturation and unitarization by one-loop IAM. 
 But note that here  we now take advantage of the fact that we can calculate both the leading and subleading terms in the $1/N_C$ expansion 
of $F_\pi$, Eq.~\eqref{ffpi}, within the $\delta$-expansion:\footnote{However, note that in Eq.~\eqref{ffpi} we do not have all the counterterms 
of ${\cal L}_{\delta^2}$ because we use resonance saturation from the Lagrangians \eqref{lagscalar}. Extra resonance operators would be needed which are 
beyond of our study \cite{cirigliano06npb}.}
\begin{align}
\label{run2}
\frac{F_\pi(N_C)}{F_\pi(N_C=3)}&=\sqrt{\frac{N_C}{3}}\Bigg\{1+\frac{1}{16\pi^2   F_\pi^2(N_C=3)}\left[A_0(\overline{m}^2_\pi)+\frac{1}{2}A_0(\overline{m}_K^2)\right]
 \left( \frac{3}{N_C}-1 \right) \Bigg\}~.
\end{align} 

 We show by the empty squares in Fig.~\ref{fig-nc-sig} the resulting $\sigma$ trajectories taking into account the subleading terms 
in the running with $N_C$ for the resonance couplings according to Eqs.~\eqref{run1} and \eqref{run2}.  We observe that the curve still bends to the left, and qualitatively speaking is not very different to the circles obtained by considering the leading 
scaling with $N_C$ shown in Eq.~\eqref{leading_nc_run} (we recall that we always include the subleading term in $1/N_C$ for $F_\pi$, Eq.~\eqref{run2}.) It seems plausible then that subleading terms in the large $N_C$ expansion are not either responsible for the difference between our results and two-loop IAM \cite{sun07,pelaez06prl}. 


The $N_C$ trajectory for $f_0(980)$ is not reported in the IAM method due to the difficulty to 
track this pole \cite{pelaez04}. In contrast, in our case its variation when moving $N_C$  can be easily followed. 
The trajectories of the $f_0(980)$ and $f_0(1370)$ also pose a 
strong evidence that they are originated from the singlet and octet scalar 
states in the Lagrangian, respectively. However, one should also take into account that the contribution 
to the physical $f_0(980)$ due to a $K\bar{K}$ bound state \cite{oller97,oller99prd,isgur,wein,speth}  
disappears in the large $N_C$ limit.  We see that the difference in the trajectories obtained in   
the full result and  {\it mimic SU(3)} cases differ significantly for the $f_0(980)$ and $f_0(1370)$ resonances. This is expected  because these resonances couple strongly to states with the $\eta$ and $\eta'$.  
For the $f_0(980)$ and $f_0(1370)$ resonances, since in our case  their poles always fall down to 
the real axis at the masses of the singlet and octet bare resonances in the Lagrangian, in order, we only show  the pole trajectories from the best fit in Fig.~\ref{fig-nc-s1}.  
For $N_C\to \infty$ the uncertainty is already given because their  pole positions correspond  to the bare masses of the singlet 
and octet scalar resonances, in order, given in Eq.~\eqref{fitresult} together with their errors.

 \begin{figure}[ht]
 \begin{center}
 \includegraphics[angle=0, width=.9\textwidth]{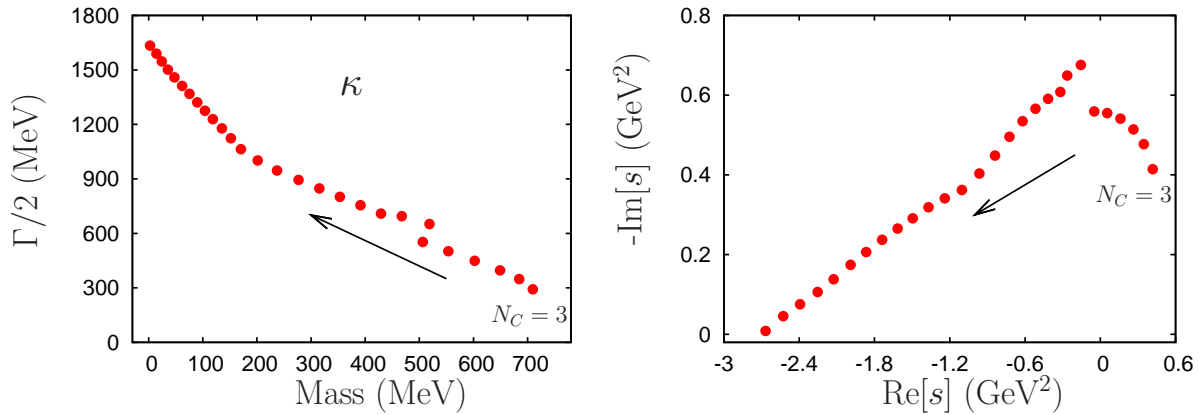}
 \caption{(Color online.) $N_C$ dependence of the $\kappa$ pole position in the fourth Riemann sheet $(+,-,+)$ from $N_C=3$ to 30 
with one unit step. The notation is the same as in Fig.~\ref{fig-nc-sig}.  }
 \label{fig-nc-kappa}
 \end{center}
 \end{figure}

\subsubsection*{Poles in the $IJ = \frac{1}{2} 0$ channel}

The $K^*_0(1430)$ pole can be easily tracked in the complex plane,  
which we show in the third panel of Fig.~\ref{fig-nc-s1}.  Like the $f_0(1370)$, this pole falls 
down  on the real axis for $N_C\to \infty$, at $s=M_{S_8}^2$, indicating that it originates from the bare octet scalar 
states with $I=1/2$. At $N_C = 3$, its pole is somewhat sensitive to the different modifications of our results already discussed, while for 
large $N_C$ they run towards the same point, as expected. 
From Table~\ref{tab:pole} we can see that the $K^*_0(1430)$ resonance couples to $K\eta'$  
as strongly as to  $K\pi$  for $N_C=3$, while its 
coupling to $K\eta$ is suppressed. This behavior is also kept when varying  $N_C$.

For the $\kappa$ pole trajectories, depicted in Fig.~\ref{fig-nc-kappa}, the situation is more involved. 
After $N_C = 4$, we could not find its pole in the second Riemann sheet. However, 
switching to the fourth sheet, i.e. with $(+,-,+)$, we can then track the $\kappa$ pole 
for any value of $N_C$. This is due to the more complicated cut structures in $IJ = \frac{1}{2} 0$ case so that  
the trajectories of $\kappa$ are not as smooth as for the $\sigma$ in the $IJ = 00$ case. This is the reason why the 
$\kappa$ pole in the second Riemann sheet cannot be tracked because at $N_C=5$ the pole 
moves to another Riemann sheet obtained by crossing the new cut in the complex plane, so that only a bump remains in the second Riemann sheet.  
 Indeed, for $N_C=3$ the $\kappa$ pole position in the fourth sheet is very similar to that 
in the second sheet for $N_C=3$ (as discussed above.) The resulting trajectory in the fourth Riemann sheet is then shown by the circle points 
in Fig.~\ref{fig-nc-kappa}. We  show the $\kappa$ pole trajectory with $N_C$ both 
in the $\sqrt{s}$, left panel, and $s$ variables, right panel. 
 The same remark as for the $\sigma$ is 
in order also for the $\kappa$. The correct interpretation is obtained by considering the pole 
trajectory in the $s$ complex plane: the mass square of the $\kappa$ resonance becomes negative, as stressed in 
Ref.~\cite{sun05} for the similar $\sigma$ case. From Fig.~\ref{fig-nc-kappa} one observes that the pole trajectory bends
 to the left for the final as well as for the {\it mimic SU(3)} approximation. However, for the {\it vector reduced} 
approximation it tends to the right for $N_C\gtrsim 10$ in the $\sqrt{s}$ complex plane. In Fig.~\ref{fig-nc-kappa} the {\it mimic SU(3)} and 
{\it vector reduced} approximations are not shown, they were already shown in Fig.~\ref{fig-nc-sig} for the similar $\sigma$ case.

\subsubsection*{Poles in the $IJ = 1 0$ channel}

The trajectories for the $a_0(1450)$ pole position are shown in the last panel of 
Fig.~\ref{fig-nc-s1}. One can see that for this resonance the trajectories bend and make a knot for  $N_C < 7$. 
In order to visualize this behavior different symbols for the pole positions up to $N_C=7$ have been used in the 
figure. Independently  from this peculiar behavior the $a_0(1450)$ pole position trajectories can be 
 followed easily in a smooth way as $N_C$ varies. As expected, for large $N_C$ this pole moves to the real axis with zero width at $s=M_{S_8}^2$, 
the same position as for the $f_0(1370)$ and $K^*_0(1430)$. It corresponds to the bare  isovector 
members of the scalar octet $S_8$, Eq.~\eqref{s1s8}. All $SU(3)$ breaking effects in the masses of these heavier scalar 
resonances originate through pseudoscalar loops and disappear in the large $N_C$ limit. 
\begin{figure}[ht]
 \begin{center}
 \includegraphics[angle=0, width=0.5\textwidth]{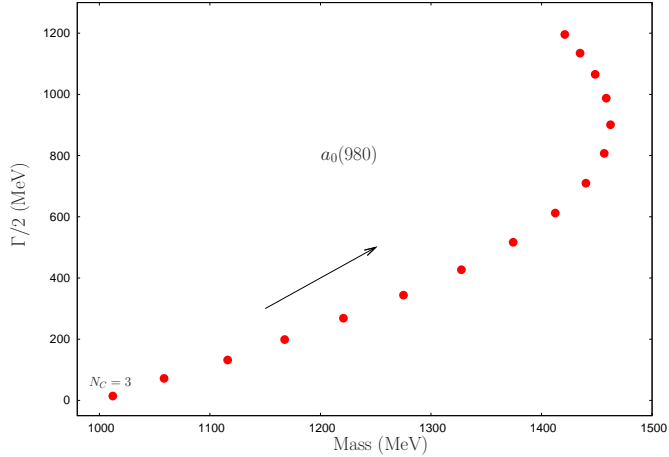}
 \caption{(Color online.) Dependence on $N_C$ for the $a_0(980)$ pole position from 3 to 18 with one  unit step. The notation 
is the same as in Fig.~\ref{fig-nc-sig}.}
 \label{fig-nc-a0}
 \end{center}
 \end{figure}

For the $a_0(980)$ pole in the fourth sheet, we can track its trajectory up 
to $N_C = 18$ without any difficulty, which we display in Fig.(\ref{fig-nc-a0}). 
After $N_C = 19$, only bumps appear in this sheet because there is a cut in the complex plane that connects 
the fourth Riemann sheet with another one, that would be  obtained by crossing continuously this cut in the complex 
plane (not the real axis on which the standard Riemann sheets based on unitarity are defined), where 
 the pole is finally located. Contrary to the $\kappa$ case we cannot 
find a unique sheet where we can track the pole position from $N_C=3$ onwards to high values of $N_C$. Although similar poles 
can be tracked if we switch to other sheets for $N_C>18$, there is some discontinuity when 
passing from one trajectory to another in different Riemann sheets. The trajectory shown in 
Fig.~\ref{fig-nc-a0} corresponding to our full results clearly indicates that this resonance is 
dynamically generated \cite{speth,oller97,oller99prd}, since it moves further and further in the complex plane with a huge width increasing very fast with $N_C$.

\subsubsection*{Poles in the $IJ = 1 1$, $\frac{1}{2}\, 1$ and $0 1$ channels}

\begin{figure}[H]
 \begin{center}
 \includegraphics[angle=0, width=1.0\textwidth]{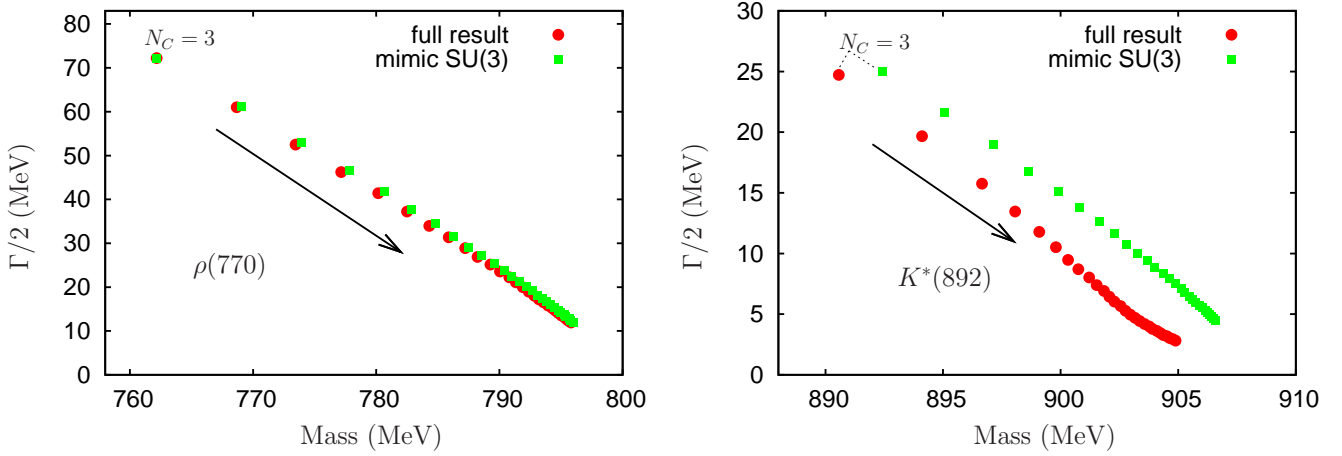}
 \caption{(Color online.) Dependence on $N_C$ of the $\rho(770)$ and $K^*(892)$ pole positions as a 
function of $N_C$  from 3 to 30 with one unit step. The notation is the same as in Fig.~\ref{fig-nc-sig}.}
 \label{fig-nc-v}
 \end{center}
 \end{figure}

The $N_C$ trajectories of the $\rho(770)$ and $K^*(892)$ vector resonances are shown 
in the left and right panels, respectively, of Fig.~\ref{fig-nc-v}. 
 Both the $\rho$ and $K^*$ move to the real axis with  increasing $N_C$, 
as one can see in Fig.(\ref{fig-nc-v}), indicating that their widths vanish in the 
large $N_C$ limit. Indeed they vanish exactly as $1/N_C$. 
 In addition their masses move very little. Both facts are in agreement with 
its standard interpretation as $\bar{q}q$ resonances which implies an $N_C^0$ scaling with $N_C$ for the mass and 
 $1/N_C$ for the width. 

\begin{figure}[H]
 \begin{center}
 \includegraphics[angle=0, width=1.0\textwidth]{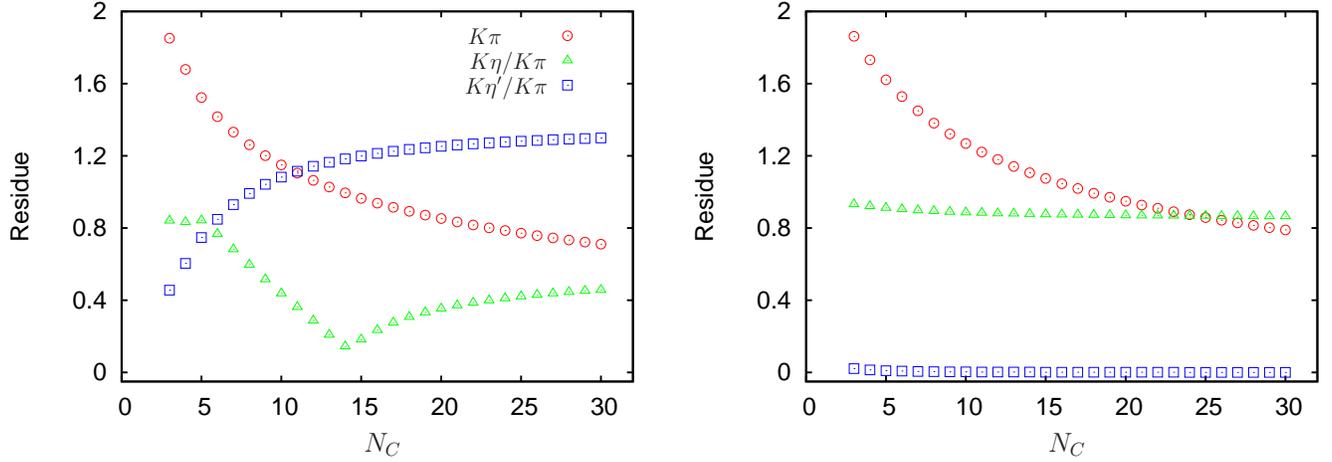}
 \caption{(Color online.) The $N_C$ running of the various residues of the $K^*(892)$ resonance. 
The absolute value of the coupling to $K\pi$ is given by the circular points, while for the other states the ratio to 
the $K\pi$ coupling is shown. Triangles apply to $K\eta$ ($K\eta_8$) and squares to $K\eta'$ ($K\eta_1$). 
The left (right) panel corresponds to our full ({\it mimic SU(3)}\,) results.} 
 \label{fig-nc-v-resi}
 \end{center}
 \end{figure}

Our results agree well with the previous conclusions on this respect~\cite{pelaez04,pelaez06prl,arriola1}. 
Nevertheless, the $N_C$ trajectories of the residues for the $K^*(892)$ in our full calculation show 
a clearly different feature, compared with the {\it mimic SU(3)} approximation. 
As one can see in Fig.~\ref{fig-nc-v-resi}, 
the residues of $K \pi$ are rather similar between the two frameworks,  
while the residues of $K\eta'$ in the full result are obviously larger than 
the $K\eta_1$ in the {\it mimic SU(3)} approximation, which can be attributed to the $\eta-\eta'$ mixing angle. 
In contrast to the almost flat residue of $K\eta$ (in fact $K\eta_8$) in 
{\it mimic SU(3)} approximation, a more complicated  structure appears in the full result. 
We verify this structure is caused by the $N_C$ variation of the $\eta$ mass. 
As one can see the kink in the residue of $K\eta$ happens around $N_C = 14$, where  
precisely the $K\eta$ threshold becomes lighter than the $K^*(892)$ mass.

The $N_C$ trajectory for the $\phi(1020)$ pole position is completely analogous to those discussed 
for the $\rho(770)$ and $K^*(892)$, but even simpler because it is an elastic channel and the width of the $\phi(1020)$ is 
very small, Table~\ref{tab:pole}. This is why we do not dedicate a separate figure for this case. 
This type of trajectory is then in agreement with the standard $\bar{q}q$ interpretation for 
this resonance.  Nevertheless, there is some movement in the $\phi(1020)$ mass due to the small variation 
of the nearby $K\bar{K}$ threshold with $N_C$, see Fig.~\ref{fig-nc-mass-err}.

\section{Conclusions}
\label{conclusion}

In this work, we complete the first one-loop calculation of meson-meson scattering in $U(3)$ 
$\chi$PT  in the literature, that comprises simultaneously a chiral and an $1/N_C$ expansion, the so-called 
$\delta$ counting. It is also the first work in which the  $N_C$ trajectories of 
resonance pole positions are studied taking into account the fact that the $\eta_1$ becomes the ninth Goldstone boson 
in the large $N_C$ limit. 
 This has a large impact in the hadron spectrum because the mass of the $\eta$ pseudoscalar in 
large $N_C$ decreases drastically and tend to be as light as the pion. 

In our one-loop $U(3)$ $\chi$PT calculation for scattering amplitudes, instead of the local terms at ${\cal O}(\delta^3)$, we include explicit resonance 
fields whose exchange generates these terms, as well as higher order ones. In order to compare with 
data, including the resonance region, we  have unitarized the previous one-loop amplitudes projected 
in partial waves. For the unitarization we employ  a non-perturbative scheme based on the N/D method, 
where the right-hand cut is resummed while the crossed-channel cuts 
 are perturbatively treated as given by the input partial waves from $U(3)$ $\chi$PT.   We achieve a good 
reproduction of meson-meson scattering data from $\pi\pi$ threshold up to energies between 1.2-1.6 GeV, 
depending on the particular partial wave. 

We have then studied the spectroscopy content of our solution by considering the poles and related residues
 of the  different partial waves.
 Various resonance poles in the complex energy plane are then found, namely, for the  $\sigma$, $f_0(980)$, $f_0(1370)$, $\kappa$, 
$K^*_0(1430)$, $a_0(980)$, $a_0(1450)$, $\rho(770)$,  $K^*(892)$ and $\phi(1020)$ resonances. 
 The pole positions agree  remarkably well with the PDG values~\cite{pdg}. 
The corresponding residues are calculated as well, which  give us the coupling strengths of 
every resonance to the different channels. 
  The couplings of the $\sigma$ resonances were studied in detail. We first discuss that the ratio of the couplings of the $\sigma$ 
to $K^+K^-$ and $\pi^+\pi^-$. Our result for this quotient, $0.44^{+0.03}_{-0.02}$, is in close agreement with the determination in Ref.~\cite{narison22}, where its not-small size is interpreted as an indication in favor of the glueball content of the $\sigma$. In our case, this resonance originates dynamically because of the $\pi\pi$ interactions \cite{oller97,oller99prd,arriola1}. We further compared our results for the couplings of the $\sigma$ to $\eta\eta$, $\eta\eta'$ and $\eta'\eta'$ with previous results from Narison and collaborators. These authors obtain a coupling of the $\sigma$ to $\eta\eta$ that depends markedly on the value taken for the ratio of the couplings of the $\sigma$ to $K^+K^-$ and $\pi^+\pi^-$, while ours is small. For the couplings to $\eta\eta'$ and $\eta'\eta'$  our calculation is in agreement with the upper bounds from Ref.~\cite{SNVeneziano89}. 

We pay special attention to the $N_C$ dependences of the poles for these resonances as well as 
for their residues. Our approach fills a gap in the literature  because previous studies in the 
literature did not take into account the Goldstone boson nature of the $\eta_1$ field in the large $N_C$ limit, 
due to the vanishing of the $U_A(1)$ anomaly with large $N_C$. The trajectories for the pole positions 
are obtained 
by taking the leading scaling  with $N_C$ of the $U_A(1)$ anomaly mass, $\Lambda_2$, the subtraction 
constants $a_{SL}$ and 
the bare resonance parameters. The scaling with large $N_C$ of $F_\pi$ also includes  subleading terms determined 
from our one-loop calculation in $U(3)$ $\chi$PT.
  The  $N_C$ dependences of the pseudo-Goldstone masses have been taken into account for determining the resulting trajectories. 
It is specially remarkable  the large reduction of the $\eta$ mass with $N_C$, that becomes similarly light as 
the pion for large $N_C$. We  discussed in depth the particular case of the $\sigma$ resonance. We show that our results  are in qualitative agreement with the ones of one-loop IAM. We obtain that $s_\sigma$, the pole position of the $\sigma$ resonance in 
the variable $s$, bends towards the real negative axis. This behavior is the expected one according to a relation obtained in Refs.~\cite{sun07,arriola1}, given the values of the large $N_C$ leading ${\cal O}(p^4)$ counterterms $L_2$ and $L_3$ that follow from our calculation making use of resonance saturation.  We also find that 
higher order terms arising within our approach by keeping the full resonance propagators in the crossed vector 
resonance exchanges give rise to a qualitative different behavior in the $\sqrt{s}$-complex plane for 
the $\sigma$  case for not too large $N_C$ values ($N_C\gtrsim 7$). This is easily understood 
because for increasing $N_C$ the loop contributions from crossed diagrams tend to vanish faster with $N_C$ than 
those from the resonance exchange diagrams. In this way, the 
cancellation between both contributions found in Ref.~\cite{oller99prd} at $N_C=3$ for the resonant scalar channels below 1~GeV,
is spoilt for higher values of $N_C$ and then there is more sensitivity to crossed channel dynamics.  As a result, we find 
that the pole position for the $\sigma$  resonances has a mass that typically decreases with $N_C$, while this is 
not the case if the vector resonance exchanges are reduced to the leading local term contribution. Similar results are also obtained for the $\kappa$ resonance. 

The last point is one of the most problematic issues in determining the large $N_C$ trajectory for the $\sigma$ resonance pole. As stressed in Refs.~\cite{pelaezconf} one is sensitive in this case to terms that are not so significant at $N_C=3$ so that a testing ground for the model is lacking. Particularly taking into account that  the tendency for $|s_\sigma|$ is to increase with $N_C$  at around or above 1~GeV$^2$, so that model dependence becomes more important. Indeed, our results are quite different from the two-loop IAM ones \cite{sun07,pelaez06prl} where $s_\sigma$ bends towards real positive values of $s$, falling down to the real axis at around 1~GeV$^2$. Surely the most important reason for such disagreement has to do with the different way higher orders are treated within two-loop IAM and our approach. In addition, we have argued about other reasons. We have excluded that the explicit consideration from our side of the $\eta_1$ as a degree of freedom, by extending  $SU(2)$ or $SU(3)$ $\chi$PT framework to  $U(3)$ $\chi$PT, is not responsible for such disagreement. The basic reason is that the $\sigma$ resonance within our approach couples weakly to states containing the $\eta$ and $\eta'$ and, then, is not so sensible to any improvement in the treatment related to those states.  This is an interesting result by itself because then our calculation establishes  that previous studies \cite{pelaez04,sun07,pelaez06prl}, are stable under the explicit inclusion of the $\eta_1$ singlet even when increasing $N_C$. However, this is not the case for all the other resonances studied that have large couplings to states with $\eta$ and/or $\eta'$ and that it is a necessary step forward to consider $U(3)$ $\chi$PT.

 We also considered the possible presence of subleading terms in the large $N_C$ running of the resonance parameters in the resonance Lagrangians by employing model results in the literature that are obtained by imposing QCD-inspired assumptions for the high energy behavior of some form factors. We then obtain that they scale as $F_\pi$, whose scaling we calculate within $U(3)$ $\chi$PT at the one-loop level. Subleading terms in the resonance masses are not considered because they move very little from $N_C=3$ (actual physical poles) to $N_C=\infty$ (bare masses). They are not either taken into account for parameters vanishing in the large $N_C$ limit. 
 We conclude tentatively that subleading terms are not  responsible for the discrepancy of our results with those 
of two-loop IAM \cite{sun07,pelaez06prl}. We also pointed out that within our approach we know beforehand the resonance spectrum in the  $N_C\to \infty$ limit, because it directly corresponds to the bare resonances introduced. In this way, when fitting data we are testing a known $N_C\to \infty$ set of resonances. In the IAM the knowledge of the $N_C\to\infty$ resonance spectrum is  more uncertain at the level of two-loops because of the proliferation of higher orders counterterms and the dependence of the precise spectrum with the explicit numerical values of the counterterms.

Regarding the $f_0(980)$ we obtain that in the large $N_C$ limit it tends to a zero 
width pole position corresponding to the bare  singlet scalar resonance around 1~GeV.  It is worth 
stressing that in other studies a zero width pole with mass of 1~GeV was already found for large 
$N_C$ from the evolution of the $\sigma$ pole trajectory \cite{sun07,pelaez06prl}. However, the pole at 1~GeV disappears in those studies 
for $N_C=3$ and only the $\sigma$ remains \cite{pelaez06prl}. In our case, both states remain.   On the one hand, we have the 
bare scalar pole around 1~GeV that contributes  to the $f_0(980)$ resonance pole in $ N_C=3$ while, on the other hand, 
the $\sigma$ resonance originates dynamically mainly from  pion interactions. In addition, the strong $K\bar{K}$ interactions 
near threshold in S-wave gives rise to another strong contribution to the $f_0(980)$ as a $K\bar{K}$ bound state. 
 This contribution disappears in the large $N_C$ limit and only the bare singlet scalar state pole contributes then.  It is an interesting exercise 
for future work to check that our bare singlet scalar pole at around 1~GeV is enough to guarantee  local-duality  \cite{localduality}. 

The $a_0(980)$ resonance  disappears deep in the complex plane for large $N_C$, by increasing its width and mass, which is always 
 positive and large. This behavior corresponds to a dynamically generated resonance. 
  We also discussed the $N_C$ trajectories for the $f_0(1370)$, $a_0(1450)$ and $K^*(1430)$. Asymptotically for large $N_C$, they 
tend to the zero width pole position of the bare octet of scalar resonances included at the tree-level around 1.4~GeV. 

  Finally, the vector resonances $\rho(770)$, $K^*(892)$ and $\phi(1020)$ are reproduced with properties in good agreement 
with the PDG \cite{pdg}. They have an $N_C$ pole position trajectory in good agreement with the expectations for a $\bar{q}q$ state, 
with a quenched mass running as $N_C^0$ and the width decreasing as $1/N_C$. The decreasing $\eta$ mass with $N_C$  
makes that the $K^*(892)$ resonance becomes heavier than the $K \eta$ channel for $N_C$ around 15. The crossover 
between the mass of the resonance and the $K\eta$ threshold manifests in a kink in the coupling to $K\eta$.

\section*{Acknowledgement}

  We would like to dedicate this work to Joaquim~Prades who passed away while this work was being done. He  
 encouraged us to get involved in this research and also collaborated in an earlier stage of the work.
We also thank Miguel~Albaladejo for technical assistance with computer software and Felipe~Llanes for discussions. 
  This work is partially funded by the grants  FPA2010-17806 and the Fundaci\'on S\'eneca one Ref.~11871/PI/09. 
 Financial support from  the EU-Research Infrastructure
Integrating Activity ``Study of Strongly Interacting Matter" (HadronPhysics2, grant n. 227431)
under the Seventh Framework Programme of EU and  the Consolider-Ingenio 2010 Programme CPAN (CSD2007-00042) is also 
acknowledged. 
  Z.H.G. acknowledges CPAN postdoc contract in the Universidad de Murcia and financial support from the grants 
Natural Science Foundation of Hebei Province with contract No. A2011205093 and Doctor Foundation of Hebei 
Normal University with contract No. L2010B04.

\vspace{1.0cm}
\appendix{\huge \bf Appendix}

\section{Convention of the loop functions}
\def\theequation{\Alph{section}.\arabic{equation}}
\setcounter{equation}{0}
\label{app.A}

The one-loop  functions that appear in our $U(3)$ $\chi$PT calculation are calculated 
in dimensional regularization within the $\overline{MS}-1$ renormalization scheme \cite{gasserleu}. 
They are defined as 
\begin{align}
A_0(m^2) &= \frac{(2\pi \mu)^{4-D}}{i \pi^2} \int d^D q \,\frac{1}{q^2-m^2} \nn\\
 &=\, - m^2 \ln{\frac{m^2}{\mu^2}}\,,\nn\\
B_0(s, m_a^2, m_b^2) &= 
\frac{(2\pi \mu)^{4-D}}{i \pi^2} \int d^D q \, \frac{1}{q^2-m_a^2} \frac{1}{(q-p)^2-m_b^2}
\nn \\ &= 1 -\log\frac{m_b^2}{\mu^2} 
+x_+\log\frac{x_+-1}{x_+}
+x_- \log\frac{x_--1}{x_-} \,,
\end{align}
where $\mu$ is the renormalization scale, $s=p^2$ and 
$x_{\pm}$ was defined in Eq.\eqref{defgfuncionloop}. In case of equal masses, 
the two-point function reduces to 
\begin{equation}
B_0(s,m^2) =  1 -\log{\frac{m^2}{\mu^2}} + \sigma(s) \log{\frac{\sigma(s)-1}{\sigma(s)+1}}\,,
\end{equation}
with $\sigma(s)$ defined in Eq.~\eqref{defsigfunc}.

\section{Renormalization of the wave functions and masses}
\def\theequation{\Alph{section}.\arabic{equation}}
\setcounter{equation}{0}
\label{app.B}
From the calculation of the $\pi$, $K$, $\overline{\eta}$ and $\overline{\eta}'$ self-energies Fig.~\ref{Fig.selfenergy}, with the latter 
fields defined in Eq.~\eqref{deflomixing}, we have:

\begin{align}
Z_\pi &= 1-\frac{1}{16 \pi^2 F_\pi^2}\bigg[ \frac{2}{3}A_0(m_\pi^2)+\frac{1}{3}A_0(m_K^2)  \bigg]
+\bigg[ -\frac{8\widetilde{c}_d \,\widetilde{c}_m (2m_K^2 + m_\pi^2)}{F_\pi^2 M_{S_1}^2}
+ \frac{16 c_d \,c_m (m_K^2 - m_\pi^2)}{3 F_\pi^2 M_{S_8}^2} \bigg] \,, \\
m_\pi^2&=  \overline{m}_\pi^2 -\frac{1}{16  \pi^2 F_\pi^2}\bigg[ \frac{m_\pi^2}{2}A_0(m_\pi^2)
-\frac{m_\pi^2(c_\theta-\sqrt2s_\theta)^2}{6}A_0(m_\eta^2)
-\frac{m_\pi^2(\sqrt2c_\theta+s_\theta)^2}{6}A_0(m_{\eta'}^2) \bigg]
\nonumber\\ &
\quad +\bigg[
\frac{8\widetilde{c}_m ( \widetilde{c}_m- \widetilde{c}_d) \, m_\pi^2 (2m_K^2 + m_\pi^2)}{F_\pi^2 M_{S_1}^2}
- \frac{16 c_m (c_m-c_d)\, m_\pi^2 (m_K^2 - m_\pi^2)}{3 F_\pi^2 M_{S_8}^2}  \bigg]
\,,
\label{mpi2.sig}\\
Z_K&= 1-\frac{1}{16 \pi^2 F_\pi^2}\bigg[ \frac{1}{4}A_0(m_\pi^2)+\frac{1}{2}A_0(m_K^2)
+\frac{c_\theta^2}{4}A_0(m_\eta^2) +\frac{s_\theta^2}{4}A_0(m_{\eta'}^2) \bigg]
\nonumber \\ &
\quad
+\bigg[ -\frac{8\widetilde{c}_d \,\widetilde{c}_m (2m_K^2 + m_\pi^2)}{F_\pi^2 M_{S_1}^2}
- \frac{8 c_d \,c_m (m_K^2 - m_\pi^2)}{3 F_\pi^2 M_{S_8}^2} \bigg] \,,  \\
m_K^2&=\overline{m}_K^2 -\frac{1}{16 \pi^2 F_\pi^2}\bigg[ \frac{c_\theta^2(3m_\eta^2+m_\pi^2)
+2\sqrt2 c_\theta s_\theta(m_\pi^2-2m_K^2)-4s_\theta^2 m_K^2}{12}A_0(m_\eta^2)
\nonumber\\ &
\qquad\qquad\qquad \quad+\frac{-4c_\theta^2 m_K^2+2\sqrt2 c_\theta s_\theta(2m_K^2-m_\pi^2)
+s_\theta^2(3m_{\eta'}^2+m_\pi^2) }{12}A_0(m_{\eta'}^2)
\bigg]
\nonumber\\ &
\quad +\bigg[
\frac{8\widetilde{c}_m ( \widetilde{c}_m- \widetilde{c}_d) \, m_K^2 (2m_K^2 + m_\pi^2)}{F_\pi^2 M_{S_1}^2}
+ \frac{8 c_m (c_m-c_d)\, m_K^2 (m_K^2 - m_\pi^2)}{3 F_\pi^2 M_{S_8}^2}  \bigg]
\,,
\label{mk2.sig}
\end{align}

The leading order masses of $\overline{\eta}\,, \overline{\eta}'$, 
i.e. $m_{\overline{\eta}}$ and $m_{\overline{\eta}'}$ defined in Eq.(\ref{defmixingpara}), 
and the $\overline{\eta}-\overline{\eta}'$ mixing angle from Eq.(\ref{lolagrangian}) 
are found to be 
\begin{eqnarray}
m_{\overline{\eta}}^2 &=& \frac{M_0^2}{2} + \overline{m}_K^2 
- \frac{\sqrt{M_0^4 - \frac{4 M_0^2 \Delta^2}{3}+ 4 \Delta^2 }}{2} \,, \label{defmetab2}  \\
m_{\overline{\eta}'}^2 &=& \frac{M_0^2}{2} + \overline{m}_K^2 
+ \frac{\sqrt{M_0^4 - \frac{4 M_0^2 \Delta^2}{3}+ 4 \Delta^2 }}{2} \,, \label{defmetaPb2}  \\
\sin{\theta} &=& -\left( \sqrt{1 + 
\frac{ \big(3M_0^2 - 2\Delta^2 +\sqrt{9M_0^4-12 M_0^2 \Delta^2 +36 \Delta^4 } \big)^2}{32 \Delta^4} } ~\right )^{-1}\,, 
\label{deftheta0}
\end{eqnarray}
with $\Delta^2 = \overline{m}_K^2 - \overline{m}_\pi^2$\,.

The higher order mixing parameters of $\overline{\eta}-\overline{\eta}'$ 
 are defined in Eq.(\ref{defmixingpara}). They read 
\begin{align*}
 \delta_{\overline{\eta}} &= \frac{c_\theta^2}{16 \pi^2 F_\pi^2}  A_0(m_K^2)
-\bigg[
-\frac{8\widetilde{c}_m \widetilde{c}_d \, (2m_K^2 + m_\pi^2)}{F_\pi^2 M_{S_1}^2}
- \frac{16 c_m c_d\, (m_K^2 - m_\pi^2)(c_\theta^2 + 2\sqrt2 c_\theta s_\theta) }{3 F_\pi^2 M_{S_8}^2}
  \bigg] \,,  \nn\\
 \delta_{\overline{\eta}'} &= \frac{s_\theta^2}{16 \pi^2 F_\pi^2}  A_0(m_K^2)
-\bigg[
-\frac{8\widetilde{c}_m \widetilde{c}_d \, (2m_K^2 + m_\pi^2)}{F_\pi^2 M_{S_1}^2}
- \frac{16 c_m c_d\, (m_K^2 - m_\pi^2)(s_\theta^2 - 2\sqrt2 c_\theta s_\theta) }{3 F_\pi^2 M_{S_8}^2}
  \bigg] \,, \nn \\
\delta_{k} &=   \frac{s_\theta c_\theta}{16 \pi^2 F_\pi^2}  A_0(m_K^2)
+ \frac{16 c_d\, c_m\, (m_K^2- m_\pi^2)
( \sqrt2 s_\theta^2 + c_\theta s_\theta - \sqrt2 c_\theta^2 ) }{3 F_\pi^2 M_{S_8}^2}  \,,\nn\\
\delta_{m_{\overline{\eta}}^2} &= -\frac{1}{16 \pi^2 F_\pi^2}\bigg\{
 -\frac{m_\pi^2(c_\theta-\sqrt2 s_\theta)^2}{2}A_0(m_\pi^2)
\nonumber \\&\qquad \qquad
+\frac{ c_\theta^2 m_\pi^2+2\sqrt2 c_\theta s_\theta(m_\pi^2-2m_K^2)
 -4s_\theta^2 m_K^2}{3}A_0(m_K^2)
\nonumber \\&\qquad \qquad
- \frac{A_0(m_\eta^2)}{18}\bigg[
c_\theta^4(16m_K^2-7m_\pi^2)+4\sqrt2 c_\theta^3 s_\theta(8m_K^2-5m_\pi^2)
\nonumber \\&\qquad \qquad \qquad
+12c_\theta^2 s_\theta^2(4m_K^2-m_\pi^2)+16\sqrt2 c_\theta s_\theta^3(m_K^2-m_\pi^2)
+2s_\theta^4(2m_K^2+m_\pi^2)\bigg]
\nonumber \\&\qquad \qquad
- \frac{(4m_K^2-m_\pi^2)(2c_\theta^4-2\sqrt2c_\theta^3 s_\theta-3c_\theta^2 s_\theta^2
+2\sqrt2c_\theta s_\theta^3+2s_\theta^4)}{18}A_0(m_{\eta'}^2) \,\bigg\}
\nonumber\\
&
\quad +\bigg\{
\frac{8\widetilde{c}_m^2 \,(2m_K^2 + m_\pi^2)\,
\big[ c_\theta^2(4m_K^2-m_\pi^2)+4\sqrt2 c_\theta s_\theta(m_K^2-m_\pi^2) + s_\theta^2 (2m_K^2 + m_\pi^2)
\big] }{3 F_\pi^2 M_{S_1}^2}
\nonumber\\
 &
\qquad\quad
+\frac{16\,c_m^2 \,(m_K^2 - m_\pi^2)\,
\big[ c_\theta^2(8m_K^2-5m_\pi^2)+2\sqrt2 c_\theta s_\theta(4m_K^2-m_\pi^2) + 4s_\theta^2 (m_K^2 - m_\pi^2)
\big] }{9 F_\pi^2 M_{S_8}^2}
  \bigg\}
\nonumber\\& \quad
+ \frac{2}{3}\Lambda_2 \bigg[ 2\sqrt2 s_\theta c_\theta (m_K^2-m_\pi^2) + s_\theta^2 (2m_K^2+m_\pi^2)  \bigg]\,,\nn\\
\delta_{m_{\overline{\eta}'}^2 } &=  -\frac{1}{16 \pi^2 F_\pi^2}\bigg\{
 -\frac{m_\pi^2(\sqrt2c_\theta + s_\theta)^2}{2 }A_0(m_\pi^2)
\nonumber \\&\qquad \qquad
+\frac{-4c_\theta^2 m_K^2 +2\sqrt2 c_\theta s_\theta(2m_K^2-m_\pi^2)
+s_\theta^2 m_\pi^2  }{3}A_0(m_K^2)
\nonumber \\&\qquad \qquad
- \frac{(4m_K^2-m_\pi^2)(2c_\theta^4-2\sqrt2c_\theta^3 s_\theta-3c_\theta^2 s_\theta^2
+2\sqrt2c_\theta s_\theta^3+2s_\theta^4)}{18 }A_0(m_{\eta}^2)
\nonumber \\&\qquad \qquad
- \frac{A_0(m_{\eta'}^2)}{18 }\bigg[
2c_\theta^4(2m_K^2+m_\pi^2)
-16\sqrt2 c_\theta^3 s_\theta(m_K^2-m_\pi^2)
\nonumber \\&\qquad \qquad \qquad
+12c_\theta^2 s_\theta^2(4m_K^2-m_\pi^2)-4\sqrt2 c_\theta s_\theta^3(8m_K^2-5m_\pi^2)
+s_\theta^4(16m_K^2-7m_\pi^2)\bigg]
\,\bigg\}
\nonumber\\ &
\quad +\bigg\{
\frac{8\widetilde{c}_m^2 \,(2m_K^2 + m_\pi^2)\,
\big[ s_\theta^2(4m_K^2-m_\pi^2)-4\sqrt2 c_\theta s_\theta(m_K^2-m_\pi^2) + c_\theta^2 (2m_K^2 + m_\pi^2)
\big] }{3 F_\pi^2 M_{S_1}^2}
\nonumber\\ &
\qquad\quad
+\frac{16\,c_m^2 \,(m_K^2 - m_\pi^2)\,
\big[ s_\theta^2(8m_K^2-5m_\pi^2)-2\sqrt2 c_\theta s_\theta(4m_K^2-m_\pi^2) + 4c_\theta^2 (m_K^2 - m_\pi^2)
\big] }{9 F_\pi^2 M_{S_8}^2}
  \bigg\}
\nonumber\\& \quad
+ \frac{2}{3}\Lambda_2 \bigg[ -2\sqrt2 s_\theta c_\theta (m_K^2-m_\pi^2) + c_\theta^2 (2m_K^2+m_\pi^2)  \bigg]\,,
\nn
\end{align*}
\begin{align}
\label{mix.del.1}
\delta_{m^2} &=-\frac{1}{16 \pi^2 F_\pi^2}\bigg\{ -\frac{m_\pi^2
(\sqrt2 c_\theta^2-c_\theta s_\theta-\sqrt2 s_\theta^2)}{2}A_0(m_\pi^2)
\nonumber \\& \qquad \qquad
+\frac{\sqrt2 c_\theta^2 (2m_K^2-m_\pi^2) + c_\theta s_\theta(4m_K^2+m_\pi^2)
-\sqrt2 s_\theta^2(2m_K^2-m_\pi^2) }{3}A_0(m_K^2)
\nonumber \\&\qquad \qquad
+ \frac{A_0(m_\eta^2)}{18}\bigg[
\sqrt2 c_\theta^4(8m_K^2-5m_\pi^2)+ c_\theta^3 s_\theta (8m_K^2+m_\pi^2)
\nonumber \\&\qquad \qquad \qquad
+3\sqrt2 c_\theta^2 s_\theta^2(m_\pi^2-4m_K^2)+4 c_\theta s_\theta^3(2m_\pi^2-5m_K^2)
+4\sqrt2s_\theta^4(m_\pi^2-m_K^2) \bigg]
\nonumber\\ &
\qquad\qquad
+ \frac{A_0(m_{\eta'}^2)}{18}\bigg[
4\sqrt2 c_\theta^4(m_K^2-m_\pi^2)+ 4c_\theta^3 s_\theta ( -5m_K^2+2m_\pi^2)
\nonumber \\&\qquad \qquad \qquad
+3\sqrt2 c_\theta^2 s_\theta^2(4m_K^2-m_\pi^2)+ c_\theta s_\theta^3(m_\pi^2+8m_K^2)
+\sqrt2s_\theta^4(5m_\pi^2-8m_K^2)\bigg]
 \bigg\}
\nonumber \\ &
+\bigg\{ \frac{16  c_m^2\, (m_K^2- m_\pi^2)\,(4m_K^2-m_\pi^2)
( \sqrt2 s_\theta^2 + c_\theta s_\theta - \sqrt2 c_\theta^2 ) }{9 F_\pi^2 M_{S_8}^2}
\nonumber \\&
\qquad
-\frac{16 \widetilde{c}_m^2\, (m_K^2- m_\pi^2)\,(2m_K^2+m_\pi^2)
( \sqrt2 c_\theta^2 - c_\theta s_\theta - \sqrt2 s_\theta^2 )  }{ 3F_\pi^2 M_{S_1}^2}
 \bigg\} 
\nonumber\\& \quad
- \frac{2}{3}\Lambda_2 \bigg[ \sqrt2 c_\theta^2 ( m_K^2 -m_\pi^2)
+ s_\theta c_\theta ( 2m_K^2+ m_\pi^2) - \sqrt2 s_\theta^2 ( m_K^2-m_\pi^2)  \bigg]\,.
\end{align}

\section{Renormalization of pion decay constant}
\def\theequation{\Alph{section}.\arabic{equation}}
\setcounter{equation}{0}
\label{app.C}

The definition of pseudoscalar weak  decay constant is
\begin{eqnarray}
\langle 0| A_\mu^a |\phi^b(p) \rangle = i F_{\phi^b}\, p_\mu \,\delta^{ab}\,,
\end{eqnarray}
where the axial-vector current is $A_\mu^a = \bar{q}\gamma_\mu \gamma_5 \frac{\lambda^a}{2}q  $.

Throughout this work we have expressed $F$ in  the chiral limit in terms of the physical pion decay constant 
$F_\pi$. The corresponding expression coincides with the one in $SU(3)$ $\chi$PT \cite{gasserleu,kaiser98}, once resonance saturation 
of the ${\cal O}(p^4)$ $\chi$PT counterterms is assumed \cite{ecker89npb}. It reads  
\begin{eqnarray}
\label{ffpi}
F_{\pi} &=&   F \bigg\{ 1 +
\frac{1}{16\pi^2 F_\pi^2}\bigg[   A_0(m_\pi^2) + \frac{1}{2} A_0(m_K^2) \bigg]
\nonumber \\ &&
\qquad\qquad
+\bigg[ \frac{4 \widetilde{c}_d \,\widetilde{c}_m (m_\pi^2+2m_K^2)}{F_\pi^2 M_{S_1}^2}
- \frac{8 c_d\, c_m\, (m_K^2-m_\pi^2)}{3 F_\pi^2 M_{S_8}^2} \bigg] \bigg\}\,.
\end{eqnarray}

 \section{Scattering amplitudes}
\def\theequation{\Alph{section}.\arabic{equation}}
\setcounter{equation}{0}
\label{app.D}

By using crossing symmetry and isospin symmetry, as we discussed  in Section \ref{pwamp}, 
all the meson-meson scattering amplitudes with well defined isospin and angular momentum 
in $U(3)$ $\chi$PT can be reduced to the calculation of 16 independent processes.

Due to the much lengthy expressions of the one-loop $U(3)$ $\chi$PT scattering amplitudes, 
we only provide here the analytical expressions for the tree level 
results at leading order, denoted with the superscript $(2)$.  
For the remaining parts, comprising  
the ones from the loops and resonance exchanges, one can download the Mathematica file 
from http://www.um.es/oller/u3FullAmp16.nb, that can also be provided by the authors under request.

\begin{enumerate}

\item $\pi^+ \pi^- \to \pi^0 \pi^0$
\begin{eqnarray}
T_{\pi^+ \pi^- \to \pi^0 \pi^0}^{\rm (2)}&=& \frac{s-m_\pi^2}{F_\pi^2}\,.
\end{eqnarray}

\item $K^+ \pi^+ \to K^+ \pi^+$
\begin{eqnarray}
T_{K^+ \pi^+ \to K^+ \pi^+}^{\rm (2)}&=& \frac{m_\pi^2 + m_K^2 - s}{2F_\pi^2}\,.
\end{eqnarray}

\item $K^+ K^- \to K^0 \bar{K}^0$
\begin{eqnarray}
T_{K^+ K^- \to K^0 \bar{K}^0}^{\rm (2)}&=& \frac{s + t - 2m_K^2}{2F_\pi^2}\,.
\end{eqnarray}

\item $\pi^0\pi^0 \to \eta\eta$
\begin{eqnarray}
T_{\pi^0\pi^0 \to \eta\eta}^{\rm (2)}&=& \frac{m_\pi^2 (c_\theta- \sqrt2 s_\theta)^2}{3F_\pi^2}\,.
\end{eqnarray}

\item $\pi^0\pi^0 \to \eta\eta'$
\begin{eqnarray}
T_{\pi^0\pi^0 \to \eta\eta'}^{\rm (2)}&=& \frac{m_\pi^2 (\sqrt2 c_\theta^2 - c_\theta s_\theta -\sqrt2 s_\theta^2)}{3F_\pi^2}\,.
\end{eqnarray}

\item $\pi^0\pi^0 \to \eta'\eta'$
\begin{eqnarray}
T_{\pi^0\pi^0 \to \eta'\eta'}^{\rm (2)}&=& \frac{m_\pi^2 (\sqrt2 c_\theta + s_\theta)^2 }{3F_\pi^2}\,.
\end{eqnarray}

\item $K^0 \bar{K}^0 \to \eta\eta$
\begin{eqnarray}
T_{K^0 \bar{K}^0\to \eta\eta}^{\rm (2)}&=& \frac{c_\theta^2(9 s - 6 m_\eta^2 - 2m_\pi^2 )
 +4\sqrt2 c_\theta s_\theta (2m_K^2 -m_\pi^2) + 8 s_\theta^2 m_K^2 }{12F_\pi^2}\,. \nonumber \\
\end{eqnarray}

\item $K^0 \bar{K}^0 \to \eta\eta'$
\begin{eqnarray}
&& T_{K^0 \bar{K}^0\to \eta\eta'}^{\rm (2)}= \nonumber \\&& 
\frac{2\sqrt2 c_\theta^2( m_\pi^2-2m_K^2 )
- c_\theta s_\theta (3m_\eta^2 + 3m_{\eta'}^2+ 8m_K^2 + 2m_\pi^2 - 9s) + 2\sqrt2 s_\theta^2 (2m_K^2 - m_\pi^2)}{12F_\pi^2}\,. 
\nonumber \\
\end{eqnarray}

\item $K^0 \bar{K}^0 \to \eta'\eta'$
\begin{eqnarray}
T_{K^0 \bar{K}^0\to \eta'\eta'}^{\rm (2)}&=& \frac{ 8 c_\theta^2 m_K^2
-4\sqrt2 c_\theta s_\theta (2m_K^2 -m_\pi^2) +  s_\theta^2(9 s - 6 m_{\eta'}^2 - 2m_\pi^2 ) }{12F_\pi^2}\,. \nonumber \\
\end{eqnarray}

\item $\eta\eta \to \eta\eta$
\begin{eqnarray}
 && T_{\eta\eta \to \eta\eta}^{\rm (2)}= \frac{1}{9F_\pi^2} \bigg[ 
c_\theta^4(16m_K^2 -7m_\pi^2) + 4\sqrt2 c_\theta^3 s_\theta (8 m_K^2 - 5 m_\pi^2)
\nonumber \\ &&
 +12c_\theta^2 s_\theta^2 (4m_K^2-m_\pi^2)
+16\sqrt2 c_\theta s_\theta^3 (m_K^2 - m_\pi^2)+ 2s_\theta^4(2m_K^2+m_\pi^2)   \bigg] \,. \nonumber \\
\end{eqnarray}

\item $\eta\eta \to \eta\eta'$
\begin{eqnarray}
 && T_{\eta\eta \to \eta\eta'}^{\rm (2)}= \frac{1}{9F_\pi^2} \bigg[ 
\sqrt2 c_\theta^4( -8m_K^2 + 5m_\pi^2) - c_\theta^3 s_\theta ( 8 m_K^2 + m_\pi^2)
\nonumber \\ &&
 +3\sqrt2 c_\theta^2 s_\theta^2 (4m_K^2-m_\pi^2)
+4 c_\theta s_\theta^3 (5m_K^2 - 2m_\pi^2)+ 4\sqrt2 s_\theta^4( m_K^2-m_\pi^2)   \bigg] \,. \nonumber \\
\end{eqnarray}

\item $\eta\eta \to \eta'\eta'$
\begin{eqnarray}
 && T_{\eta\eta \to \eta'\eta'}^{\rm (2)}= 
\frac{( 4m_K^2-m_\pi^2)(2c_\theta^4-2\sqrt2 c_\theta^3 s_\theta -3 c_\theta^2 s_\theta^2 +2\sqrt2 c_\theta s_\theta^3 
 + 2 s_\theta^4) }{9F_\pi^2} \,. \nonumber \\
\end{eqnarray}

\item $\eta\eta' \to \eta'\eta'$
\begin{eqnarray}
 && T_{\eta\eta' \to \eta'\eta'}^{\rm (2)}= \frac{1}{9F_\pi^2} \bigg[ 
4 \sqrt2 c_\theta^4( -m_K^2 + m_\pi^2) +4 c_\theta^3 s_\theta ( 5 m_K^2 - 2 m_\pi^2)
\nonumber \\ &&
 +3\sqrt2 c_\theta^2 s_\theta^2 ( - 4m_K^2 + m_\pi^2)
- c_\theta s_\theta^3 (8m_K^2 + m_\pi^2)+ \sqrt2 s_\theta^4( 8m_K^2-5m_\pi^2)   \bigg] \,. \nonumber \\
\end{eqnarray}

\item $\eta'\eta' \to \eta'\eta'$
\begin{eqnarray}
 && T_{\eta'\eta' \to \eta'\eta'}^{\rm (2)}= \frac{1}{9F_\pi^2} \bigg[ 
2 c_\theta^4( 2m_K^2 + m_\pi^2) -16\sqrt2 c_\theta^3 s_\theta (  m_K^2 - m_\pi^2)
\nonumber \\ &&
 +12\sqrt2 c_\theta^2 s_\theta^2 ( 4m_K^2 - m_\pi^2)
- 4\sqrt2 c_\theta s_\theta^3 (8m_K^2 - 5m_\pi^2)+ s_\theta^4( 16m_K^2-7m_\pi^2)   \bigg] \,. \nonumber \\
\end{eqnarray}

\item $K^+ \pi^0 \to K^+\eta$
\begin{eqnarray}
T_{K^+ \pi^0 \to K^+\eta}^{\rm (2)}&=& -\frac{ c_\theta ( -9 t +3 m_\eta^2 +8m_K^2 + m_\pi^2 )
+ 2\sqrt2  s_\theta ( 2m_K^2 + m_\pi^2 )}{12\sqrt3 F_\pi^2}\,. \nonumber \\
\end{eqnarray}

\item $K^+ \pi^0 \to K^+\eta'$
\begin{eqnarray}
T_{K^+ \pi^0 \to K^+\eta'}^{\rm (2)}&=& \frac{ 2\sqrt2  c_\theta ( 2m_K^2 + m_\pi^2 )
 -s_\theta ( -9 t +3 m_{\eta'}^2 +8m_K^2 + m_\pi^2 )
}{12\sqrt3 F_\pi^2}\,. \nonumber \\
\end{eqnarray}

\end{enumerate}

\section{Numerical inputs}\label{appendix-inputs}

In the numerical discussion, we take the average values for the masses 
of the charged and neutral pions and kaons.  The values taken are summarized below  in units of MeV: 
\begin{eqnarray}
&& m_{\pi} = 137.3\,, \quad m_{K} = 495.6\,, \quad m_\eta = 547.9\,,\quad
m_{\eta'} = 957.7\,, \nonumber \\ &&
F_\pi = 92.4\,, \quad\,\,\, \mu = 770.0\,, \quad \,\,\,\, M_\phi = 1026.0, \quad
M_\omega = 783.0\,.
\end{eqnarray}
The value for $M_\phi$ is adjusted so as to reproduce the pole position of the $\phi(1020)$ resonance in Table~\ref{tab:pole}  according to 
the PDG \cite{pdg}.




\begin{thebibliography}{30}
\bibitem{pdg} K. Nakamura, {\it et al.}, (Particle Data Group), J. Phys. G 37 (2010) 075021.  
\bibitem{chewbook} G.~F.~Chew, ``S-matrix Theory of Strong Interactions", W.~A.~Benjamin, Inc., New York, 1961.
\bibitem{martin}A.~D.~Martin and T.~D.~Spearman, ``Elementary Particle Theory", North-Holland Publishing Company, Amsterdam, 1970.
\bibitem{eden} R.~J.~Eden, P.~V.~Landshoff, D.~I.~Olive and J.~C.~Polkinghorne, ``The Analytic S-Matrix", Cambridge University 
Press, Cambridge, 1966.
\bibitem{truong}T.~N.~Truong, Phys. Rev. Lett. {\bf 61} (1988) 2526.
\bibitem{dobado}A.~Dobado, M.~J.~Herrero and T.~N.~Truong, Phys. Lett. B {\bf 235} (1990) 134.
\bibitem{dobado97} A.~Dobado and J.~R.~Pel\'aez, Phys.~Rev.~{\bf D56} (1997) 3057. 
\bibitem{pavon}J.~Nieves, M.~Pav\'on Valderrama and E.~Ruiz Arriola, Phys. Rev. D {\bf 65} (2002) 036002.
\bibitem{oller97} J.~Oller and E.~Oset, Nucl.~Phys.~{\bf A620} (1997) 438; (E)$-ibid$  A {\bf 652}(1999) 407.
\bibitem{oller99prd} J.~A.~Oller and E.~Oset, Phys.~Rev.~{\bf D60} (1999) 074023.
\bibitem{chew60} G.~F.~Chew, S.~Mandelstam, Phys.~Rev.~{\bf 119} (1960) 467. 
\bibitem{oller08prl} M.~Albaladejo and J.~A.~Oller, Phys.~Rev.~Lett.~{\bf101} (2008) 252002.
\bibitem{igi}Keigi~Igi and Ken-ichi~Hisaka, Phys. Rev. D {\bf 59} (1999) 034005.
\bibitem{leutwyler06} I.~Caprini, G.~Colangelo and H.~Leutwyler, Phys.~Rev.~Lett.~{\bf 96} (2006) 132001.  
\bibitem{descotes06} S.~Descotes-Genon and B.~Moussallam, Eur.~Phys.~J.~{\bf C48} (2006) 553.  
\bibitem{madrid}R. Kaminski, J.~R.~Pel\'aez and F.~J.~Yndur\'ain, Phys. Rev. D {\bf 77} (2008) 054015; R.~Garc\'{\i}a-Mart\'{\i}n, 
R.~Kaminski, J.~R.~Pel\'aez, J. Ruiz de Elvira and F.~J.~Yndur\'ain, arXiv: 1102.2183.
\bibitem{zheng01} Z.~Xiao and H.~Q.~Zheng, Nucl.~Phys.~{\bf A695} (2001) 273. 
\bibitem{zheng04} H.~Q.~Zheng, Z.~Y.~Zhou, G.~Y.~Qin, Z.~Xiao, J.~J.~Wang and N.~Wu, Nucl.~Phys.~{\bf A733} (2004) 235. 
\bibitem{pelaez04}J.~R.~Pel\'aez, Phys.~Rev.~Lett.{\bf 92} (2004) 102001; 
 J.~R.~Pel\'aez, Mod.~Phys.~Lett.~{\bf A19} (2004) 2879.  
\bibitem{pelaez06prl} J.~R.~Pel\'aez and G.~Rios, Phys.~Rev.~Lett. {\bf 97} (2006) 242002.  
\bibitem{sun07}Z.~X.~Sun, L.~Y.~Xiao, Z.~Xiao and H.~Q.~Zheng, Mod. Phys. Lett. A {\bf 22} (2007) 711.
\bibitem{guo}L.~Y.~Xiao, Z.-H. Guo and H.~Q.~Zheng, Int. J. Mod. Phys. A {\bf 22} (2007) 4603.
\bibitem{guo2}Z.-H.~Guo, J.~J.~Sanz-Cillero and H.-Q.~Zheng, JHEP {\bf 06} (2007) 030. 
\bibitem{arriola1}J.~Nieves and E.~Ruiz Arriola, Phys. Rev. D {\bf 80} (2009) 045023.
\bibitem{gasserleu}J.~Gasser and H.~Leutwyler, Ann. Phys. {\bf 158} (1984) 142; Nucl. Phys. B {\bf 250} (1985) 465.
\bibitem{broniowski} E. Ruiz Arriola and W. Broniowski, Phys. Rev. D {\bf 81} (2010) 054009.
\bibitem{bardeen}W.~A.~Bardeen, Phys. Rev. {\bf 184} (1969) 1848.
\bibitem{fuji}K.~Fujikawa, Phys. Rev. D {\bf 21} (1980) 2848.
\bibitem{adler}S.~L.~Adler and W.~A.~Bardeen, Phys. Rev. {\bf 182} (1969) 1517.
\bibitem{wein23}G.'t Hooft, Phys. Rev. D {\bf 14} (1976) 3432; Phys. Rep. {\bf 142} (1986) 357.
\bibitem{ua1innc}G.'t Hooft, Nucl.~Phys.~{\bf B72} (1974) 461;
S.~Coleman and E.~Witten, Phys.~Rev.~Lett.~{\bf 45} (1980) 100; G.~Veneziano, Phys.~Lett.~{\bf B95} (1980) 90.
\bibitem{wit79} E.~Witten, Nucl.~Phys.~{\bf B156} (1979) 269.
\bibitem{oriua}P.~Di~Vecchia and G.~Veneziano, Nucl. Phys. B {\bf 171} (1980) 253; 
C. Rosenzweig, J. Schechter and T. Trahem, Phys. Rev. D {\bf 21} (1980) 3388;
E.~Witten, Ann. Phys. {\bf 128} (1980) 363.
\bibitem{feldmann} T.~Feldmann, Int.~J.~Mod.~Phys.~{\bf A15} (2000) 159.
\bibitem{borasoy02npa} N.~Beisert and B.~Borasoy, Nucl.~Phys. ~{\bf A705} (2002) 433.
\bibitem{becher99epjc} T.~Becher and H.~Leutwyler, Eur.~Phys.~J.~{\bf C9} (1999) 643.
\bibitem{alarcon}J.~M.~Alarc\'on, J.~Mart\'{\i}n Camalich, J.~A.~Oller and L.~Alvarez-Ruso, Phys. Rev. C {\bf 83} (2011) 055205.
\bibitem{kaiser98} R.~Kaiser and H.~Leutwyler, Proceedings of ``Nonperturbative Methods in Quantum Field Theory". Edited by A. W.~Schrweiber, 
A.~G.~Williams and A.~W.~Thomas. World Scientific, Singapore, 1998.  
\bibitem{herrera97}P.~Herrera-Siklody, J.~I.~Latorre, P.~Pascual and J.~Taron, Nucl.~Phys.~{\bf B497} (1997) 345.
\bibitem{kaiser00}R.~Kaiser and H.~Leutwyler, Eur.~Phys.~J.~{\bf C17} (2000)623.
\bibitem{u3chpteta}P.~Herrera-Siklody, J.~I.~Latorre, P.~Pascual and J.~Taron, Phys.~Lett.~{\bf B419} (1998) 326.
\bibitem{mongolia} Z.~H.~Guo, J.~Prades and J.~A.~Oller, Nucl.\ Phys.\ Proc.\ Suppl.\ B {\bf 207-208} (2010) 184.
\bibitem{borasoy03prd} N.~Beisert and B.~Borasoy, Phys.~ReV. ~{\bf D67} (2003) 074007.
\bibitem{zhou11prd}Z.~Y.~Zhou and Z.Xiao, Phys.~Rev.~{\bf D83} (2011) 014010.
\bibitem{jamin}M.~Jamin, J.~A.~Oller and A.~Pich, Nucl. Phys. B {\bf 587} (2000) 331.
\bibitem{kubis09}B.~Kubis and S.~P.~Schneider, Eur. Phys. J. C {\bf 62} (2009) 511.
\bibitem{escribano10} R.~Escribano, P.~Masjuan and J.~J.~Sanz-Cillero, JHEP {\bf 1105} (2011) 094.  
\bibitem{ecker89npb} G.~Ecker, J.~Gasser, A.~Pich, E.~de~Rafael, Nucl.~Phys.~{\bf B321} (1989) 311. 
\bibitem{ecker89plb} G.~Ecker, J.~Gasser, H.~Leutwyler, A.~Pich, E.~de~Rafael, Phys.~Lett.~{\bf B223} (1989) 425.
\bibitem{callen} S.~R.~Coleman, J.~Wess and B.~Zumino, Phys. Rev. {\bf 177} (1969) 2239; C.~G.~Callan, S.~Coleman, J.~Wess and B.~Zumino, Phys. Rev. {\bf 177} (1969) 2247. 
\bibitem{cirigliano06npb} V.~Cirigliano, {\it et al.}, Nucl.~Phys.~{\bf B753} (2006) 139.
\bibitem{morgan} B.~R.~Martin, D.~Morgan and G.~Shaw, ``Pion-Pion Interactions in Particle Physics", Academic Press Inc., New York, 1976.
\bibitem{guerrero}F.~Guerrero and J.~A.~Oller, Nucl. Phys. B {\bf 537} (1999) 459; (E)~J.~A.~Oller, Nucl. Phys. B {\bf 602} (2001) 641.
\bibitem{prdlong} J.~A.~Oller, E.~Oset and J.~R.~Pel\'aez, Phys. Rev. D {\bf 59} (1999) 074001; (E)-$ibid$ {\bf 60} (1999) 074023; (E)-$ibid$ {\bf 75} (2007) 099903.
\bibitem{prlshort} J.~A.~Oller, E.~Oset and J.~R.~Pel\'aez, Phys. Rev. Lett. {\bf 16} (1998) 3452.
\bibitem{plbkn}J.~A.~Oller and U.~G.~Meissner,
  Phys.\ Lett.\  B {\bf 500} (2001) 263.
\bibitem{higgs} J.~A.~Oller, Phys. Lett. B {\bf 477} (2000) 187.
\bibitem{lacour}A.~Lacour, J.~A.~Oller and U.-G.~Mei{\ss}ner, Ann. Phys. {\bf 326} (2011) 241.
\bibitem{adlerzero}S.~L.~Adler, Phys. Rev. {\bf 139} (1965) B1638.
\bibitem{cdd}L.~Castillejo, R.~H.~Dalitz and F.~J.~Dyson, Phys. Rev. {\bf 101} (1956) 453.
\bibitem{rios} A.~G\'omez Nicola, J.~R.~Pel\'aez and G.~R\'{\i}os, Phys. Rev. D {\bf 77} (2008) 056006.
\bibitem{pennington} M. Boglione and M.~R.~Pennington, Z. Phys. C {\bf 75} (1997) 113.
\bibitem{frogratt77npb} C.~D.~Frogratt and J.~L.~Petersen, Nucl.~Phys.~{\bf B129} (1977) 89. 
\bibitem{na48} J.~R.~Batley {\it et al.} (NA48/2 Collaboration), Eur. Phys. J. C {\bf 54} (2008) 411.
\bibitem{ochs74} W.~Ochs, Thesis, University of Munich , 1974.
\bibitem{refpipiphase} 
B.~Hyams, {\it et al.}, Nucl.~Phys.~{\bf B64} (1973) 134; 
P.~Estrabooks, {\it et al.}, AIP Conf.~Proc.~{\bf 13} (1973) 37;
G.~Grayer, {\it et al.}, {\it Proceedings of the 3rd Philadephia Conference on Experimental 
Meson Spectroscopy}, Philadephia, 1972 (American Institute of Physics, New York, 1972), p.5;
S.~D.~Protopopescu and M.~Alson-Garnjost, Phys.~ReV.~{\bf D7} (1973) 1279.
\bibitem{kaminski97zpc} R.~Kaminski, L.~Lesniak and K.~Rybicki, Z.~Phys.~{\bf C74} (1997) 79.
\bibitem{cohen80prd} D.~Cohen, {\it et al.}, Phys.~ReV.~{\bf D22} (1980) 2595.
\bibitem{martin79npb} A.~D.~Martin and E.~N.~Ozmutlu, 
Nucl.~Phys.~{\bf B158} (1979) 520.
\bibitem{etkin} A. Etkin {\it et al.}, Phys. Rev.  {\bf D25} (1982) 1786.
\bibitem{mercer71npb} A.~Mercer, {\it et al.}, Nucl.~Phys.~{\bf B32} (1971) 381.
\bibitem{estrabooks78npb} P.~Estabrooks, {\it et al.}, Nucl.~Phys.~{\bf B133} (1978) 490.
\bibitem{bingham72npb} H.~H.~Bingham, {\it et al.}, Nucl.~Phys.~{\bf B41} (1972) 1.
\bibitem{aston88npb} D.~Aston, {\it et al.}, Nucl.~Phys.~{\bf B296} (1988) 493.
\bibitem{armstrong84zpc} T.~A.~Armstrong, {\it et al.}, Z.~Phys.~{\bf C25} (1984) 91.
\bibitem{hoogland77npb} W.~Hoogland, {\it et al.}, Nucl.~Phys.~{\bf B126} (1977) 109.
\bibitem{losty74npb} M.~J.~Losty, {\it et al.}, Nucl.~Phys.~{\bf B69} (1974) 185.
\bibitem{bakker70npb} A.~M.~Bakker, {\it et al.}, Nucl.~Phys.~{\bf B24} (1970) 211.
\bibitem{cho70plb} Y.~Cho, {\it et al.}, Phys.~Lett.~{\bf B32} (1970) 409.
\bibitem{lambda} D.~Jido, J.~A.~Oller, E.~Oset, A.~Ramos and U.-G.~Mei{\ss}ner, Nucl. Phys. A {\bf 725} (2003) 181.
\bibitem{lindenbaum92plb} S.~J.~Lindenbaum and R.~S.~Longacre, 
Phys.~Lett.~{\bf B274} (1992) 472.
\bibitem{estrabooks74npb} P.~Estrabooks and A.~D.~Martin, Nucl.~Phys.~{\bf B79} (1974) 301.
\bibitem{guo09prd} Z.~H.~Guo, J.~J.~Sanz-Cillero, Phys.~Rev.~{\bf D79} (2009) 096006.
\bibitem{eden64pr} R.~J.~Eden and J.~R.~Taylor, Phys.~ReV.~{\bf 133} (1964) B1575. 
\bibitem{ruben76} R. Garc\'{\i}a-Mart\'{\i}n, J.~R.~Pel\'aez and F.~J.~Yndur\'ain, Phys. Rev. D {\bf 76} (2007) 074034.
\bibitem{narison22} G.~Mennessier, S.~Narison and X.-G.~Wang, Phys. Lett. B~{\bf 688} (2010) 59.
\bibitem{mixing}
J.~A.~Oller,  Nucl.\ Phys.\  A {\bf 727} (2003) 353.
\bibitem{SNVeneziano89}S.~Narison and G.~Veneziano, Int. J. Mod. Phys. A {\bf 4} (1989) 2751.
\bibitem{SN98}S.~Narison, Nucl. Phys. B {\bf 509} (1998) 312. 
\bibitem{isgur}J.~Weinstein and N.~Isgur, Phys. Rev. Lett. {\bf 48} (1982) 659; Phys. Rev. D {\bf 27} (1983) 588; Phys. Rev. D {\bf 41} (1990) 2236.
\bibitem{wein} J.~Weinstein,Phys. Rev. D {\bf 47} (1993) 911.
\bibitem{speth}G.~Janssen, B.~C.~Pearce, K.~Holinde and J.~Speth, Phys. Rev. D {\bf 52} (1995) 2690.
\bibitem{alar1} L.~Alvarez-Ruso, J.~A.~Oller and J.~M.~Alarc\'on, Phys. Rev. D {\bf 80} (2009) 045011; {\it ibid} {\bf 82} (2010) 094028.
\bibitem{weinberg75} S.~Weinberg, Phys. Rev. D {\bf 11} (1975) 3583.
\bibitem{pelaezconf}J.~R.~Pel\'aez, hep-ph/0509284; J.~R.~Pel\'aez and G.~R\'{\i}os, arXiv:0905.4689 [hep-ph].
\bibitem{sun05}Z.~X.~Sun, L.~Y.~Xiao, Z.~G.~Xiao, H.~Q.~Zheng and Z.~Y.~Zhou, AIP Conf. Proc. {\bf 756} (2005) 324.
\bibitem{peris95plb} S.~Peris and E.~de Rafael, Phys.~Lett.~B {\bf 348} (1995) 539.
\bibitem{ksfr}K.~Kawarabayashi and M.~Suzuki, Phys. Rev. Lett. {\bf 16} (1966) 266; Riazuddin and Fayyazuddin, Phys. Rev. {\bf 147} (1966) 1071.
\bibitem{pich2011}A.~Pich, I.~Rosell and J.~J.~Sanz-Cillero, JHEP {\bf 1102} (2011) 109. 
\bibitem{sakurai}J.~J.~Sakurai, ``Currents and Mesons". University of Chicago Press, Chicago, 1969.
\bibitem{bij97}J.~Bijnens {\it et al.}, Nucl. Phys. B {\bf 508} (1997) 263.
\bibitem{jaminff} M.~Jamin, J.~A.~Oller and A.~Pich,  Nucl.\ Phys.\  B {\bf 622} (2002) 279.
\bibitem{localduality} J.~Ruiz~Elvira, J.~R.~Pel\'aez, M.~R.~Pennington and D.~J.~Wilson, arXiv:1009.6204 [hep-ph].
\end{thebibliography}
 \end{document}